\documentclass[aps,twocolumn,amsmath,nofootinbib,showpacs,showkeys]{revtex4-1}

\usepackage{epsfig}
\usepackage{natbib}

\begin{document}

\title{Renormalization of chiral two pion exchange NN interactions 
  with \\ $\Delta$-excitations: correlations in the partial wave
  expansion}

\author{M. Pavon
  Valderrama}\email{m.pavon.valderrama@fz-juelich.de}
\altaffiliation[present address: ]{Departamento de F\'{\i}sica Te\'orica and
Instituto de F\'{\i}sica Corpuscular (IFIC),
Institutos de Investigaci\'on de Paterna, Aptd. 22085, E-46071 Valencia, Spain}
\affiliation{Institut f\"ur Kernphysik and J\"ulich Center for Hadron
  Physics, Forschungszentrum J\"ulich, 52425 J\"ulich, Germany}
\author{E. Ruiz Arriola}\email{earriola@ugr.es}
\affiliation{Departamento de F\'{\i}sica At\'{o}mica, Molecular y
  Nuclear, Universidad de Granada, E-18071 Granada, Spain.}

\date{\today}

\begin{abstract} 
\rule{0ex}{3ex}
In this work we consider the renormalization of the chiral two-pion exchange
potential with explicit $\Delta$-excitations for nucleon-nucleon scattering
at next-to-leading (${\rm NLO}$) and next-to-next-to-leading order
(${\rm N^2LO}$).
Due to the singular nature of the chiral potentials,
correlations between different partial waves are generated.
In particular we show that two-body scattering by a short distance
power like singular attractive interaction can be renormalized in all
partial waves with a single counterterm, provided the singularities 
are identical.
A parallel statement holds in the presence of tensor interactions when
the eigenpotentials in the coupled channel problem also coincide.
While this construction reduces the total number of counterterms to
eleven in the case of nucleon-nucleon scattering with chiral two-pion
exchange interactions with $\Delta$ degrees of freedom, the
differences in the scattering phases as compared to the case with
the uncorrelated partial wave renormalization become smaller as the angular
momentum is increased in the elastic scattering region.

\end{abstract}


\maketitle

\section{Introduction}
\label{sec:intro}

The basic and fundamental problem of nuclear physics is the
determination of the nucleon-nucleon (NN) interaction~\cite{Brown:76}.
Field theoretical approaches to the nuclear force state
that the NN potential can be expressed as a sum of
increasingly heavy meson exchange contributions~(for reviews see
e.g. \cite{Machleidt:1987hj,Machleidt:1989tm} and references therein).
The resulting high-quality potentials describe neutron-proton and
proton-proton scattering data with a
$\chi^2 /{\rm d.o.f. \le 1}$~\cite{Stoks:1993tb,Stoks:1994wp,Wiringa:1994wb,Machleidt:2000ge}.
They all include at large distances
the charge dependent one pion exchange (OPE) potential 
and typically need of the order of 40 parameters
for parametrizing the shorter range components of the interaction.
On the contrary, quantum chromodynamics (QCD),
the underlying fundamental theory of the strong interaction,
requires only two parameters in the isospin symmetric limit:
$\Lambda_{\rm QCD}$ and the average up and down quark masses.
These fundamental QCD parameters can be traded for experimentally
accessible observables such as $f_\pi$,
the pion weak decay constant, and $m_\pi$, the averaged pion mass.
Obviously, the large number of parameters needed in phenomenological approaches
arises within very specific schemes and functional forms.
It is not clear whether this number of parameters can effectively be reduced
by invoking relevant QCD features
while maintaining the quality of the description at the same time.
Ultimately, lattice {\it ab initio} calculations of the NN potential,
for which incipient results already exist~\cite{Ishii:2006ec,Aoki:2009ji},
will eventually solve the problem.

In the present paper we deal with a situation where a reduction of parameters
arises within the context of the renormalization of the chiral potentials
deduced from the effective field theory (EFT) approach suggested by
Weinberg~\cite{Weinberg:1990rz} two decades ago (for comprehensive
reviews see Refs.~\cite{Bedaque:2002mn,Epelbaum:2008ga}).
These chiral potentials turn out to be singular interactions
which exhibit an inverse power law  behaviour $\sim 1/r^n$
at distances below the pion Compton wavelength, $m_{\pi} r \ll 1$.
In case they are attractive the resulting amplitudes are sensitive to
short distance physics and require renormalization.
As will be shown,
the renormalizability of singular attractive potentials can be translated
into a mathematical short distance constraint on the scattering amplitude
for different partial waves.
In the simplifying case of two-body scattering by a central attractive
singular potential only one counterterm is needed
in order to renormalize all the partial waves.
In the more complex case of NN scattering, where spin, isospin dependence
and tensor forces are present, this number can rise to eleven counterterms
provided that certain conditions are met.
We review below the EFT approach from the perspective of the number of
parameters in a way that our results can easily be displayed.

The main appeal of the EFT idea in nuclear physics lies
in the promise of a model independent approach
where the long and short range contributions to observables
can be disentangled with the aid of a sensible hierarchy,
eluding the ubiquitous problem of fine tunings.
This feature is explicitly displayed through the introduction of counterterms
in the effective Lagrangian, 
which encode the underlying but unresolved short distance physics and
are organized according to a power counting.
The standard EFT formulation of the nuclear forces
exploits the spontaneous breakdown of chiral symmetry,
which requires derivative couplings for the pion.
This implies that higher pion exchanges are power suppressed at
momenta which are small compared with the chiral scale
$\Lambda_\chi \sim (4 \pi f_\pi, M_N) \sim 1 {\rm GeV}$.
On the contrary, at high virtual momenta, pion exchanges become large,
eventually requiring a suitable renormalization through the
introduction of counterterms.
Furthermore, the non-perturbative nature of the NN interaction makes
using a sensible resummation of diagrams mandatory.
A simple and effective method already suggested
by Weinberg~\cite{Weinberg:1990rz,Weinberg:1991um}
and implemented for the first time by Ray, Ord\'o\~{n}ez
and van Kolck~\cite{Ordonez:1993tn,Ordonez:1995rz}
proceeds in two steps:
first, one deduces a chiral nuclear potential and afterwards solves
the corresponding Lippmann-Schwinger equation in momentum space,
or equivalently, the Schr\"odinger equation in coordinate space.
This scheme complies to the familiar and widely accepted concept of nuclear
potential, which in the chiral case can be organized as an expansion
in powers of
$Q$~\cite{Kaiser:1997mw,Kaiser:1998wa,Epelbaum:1998ka,Rentmeester:1999vw,Friar:1999sj,Kaiser:1999ff,Kaiser:1999jg,Kaiser:2001at,Kaiser:2001pc,Kaiser:2001dm,Entem:2002sf}
\begin{eqnarray}
\label{eq:chi-pot}
V (r) &=& V_{\chi}^{(0)}(r) + V_{\chi}^{(2)}(r) + V_{\chi}^{(3)}(r) + V_{\chi}^{(4)}(r) \nonumber \\ &+& 
{\mathcal O}(Q^5) \, , 
\end{eqnarray}
where $Q$ represents either the pion mass or the momentum of the nucleons
(or additionally the nucleon-$\Delta$ splitting in case we include
the $\Delta$-excitation as an explicit degree of freedom).
Within this scheme and making use of finite cut-offs
at ${\rm N^3LO}$~\cite{Entem:2003ft,Epelbaum:2004fk},
the number of parameters becomes comparable with that of the phenomenological
potentials even though the long distance behaviour is determined
by chiral symmetry.

A stringent constraint follows from the natural requirement of short
distance insensitivity: physics not explicitly taken into account
should be under control by fixing a sufficient amount of low energy
parameters.
Such a condition represents the basis of the renormalization process
as understood in the present work~\footnote{
By renormalization we specifically mean the existence of well-defined
scattering amplitudes when the cut-off is removed.
This condition allows to identify all the short distance operators needed
to remove the cut-off dependence, as done, for example,
in Ref.~\cite{Nogga:2005hy} for the OPE case.
Once these counterterms are included in the computation approximate cut-off
independence is assured and consequently there is no problem
in keeping a finite cut-off.
For a different view on renormalization within an effective field theory
context, see Lepage~\cite{Lepage:1997cs} and the related discussions
of Refs.~\cite{Epelbaum:2006pt,Epelbaum:2009sd}.
}.
To fulfill this goal it is necessary to achieve approximate cut-off
independence over a certain cut-off region.
This immediately raises the question of
what cut-off values can be regarded as natural
and how much the {\it a priori} arbitrary cut-off can be varied.
The shortest de Broglie wavelength probed
in elastic NN scattering below pion production threshold
is $\lambda_{\rm min} \sim 0.5 {\rm fm}$, and thus we might expect stable
results for similar short distance cut-offs.
Otherwise the cut-off becomes an essential parameter of the theory.
This particularly applies when the cut-off must be fine tuned to physical
observables, a situation which actually takes place for specific power
counting schemes.

By naive power counting one expects chiral potentials to be singular
for $r \ll 1/m_\pi$~\footnote{The operator product expansion for six
  quark operators predict the functional form of the NN potential at
  short distance, which turns out to be a little weaker than $1/r^2$
  and repulsive~\cite{Aoki:2009pi,Aoki:2010kx}.  This shows that the
  NN potential computed on the lattice is regular and hence might
  predict uniquely the NN
  scattering data as well as the deuteron properties and also that
  this short distance dependence is quite different from the chiral
  potentials.}  ,
\begin{eqnarray}
V (r) \to \frac{1}{\Lambda_\chi^{\nu+2} r^{\nu+3}} \, ,
\label{eq:sing-chi}
\end{eqnarray}
where $\nu$ represents the order in the chiral expansion~\footnote{
  The inclusion of static degrees of freedom,
  such as the $\Delta$ isobar excitation in the small scale expansion,
  can change the expected power law behaviour of the potential.},
see Eq.~(\ref{eq:chi-pot}).
While the resummations implied by solving the wave equation mix up
the chiral power counting,
they also enable finding non-perturbative new features
when the short distance cut-off $r_c$ becomes much smaller
than any other long distance length scale.
There arises the possibility of finding the adequate number of counterterms
which is compatible with the power counting of the long range potential
and of obtaining a finite and unique limit
for the corresponding scattering amplitude.
In this regard several studies have found that
the original Weinberg power counting is inconsistent with 
renormalizability~\cite{Nogga:2005hy,Valderrama:2005wv,PavonValderrama:2005uj}.
This unexpected result has suggested several alternative approaches
and heated debates questioning the particular power counting,
the renormalization process itself or the correctness of the
non-perturbative resummation.
We will not ponder on the pros and cons of any particular approach
as this has been already done from several viewpoints in
Refs.~\cite{Nogga:2005hy,Valderrama:2005wv,Birse:2005um,Epelbaum:2006pt,Epelbaum:2009sd}.
At present it is unclear what aspects of the original EFT framework
will ultimately prevail or be universally accepted
by the nuclear physics community
(in this regard, see the related discussion in Ref.~\cite{Machleidt:2010kb}).
Rather than considering the problem solved,
we think that further work is still needed to settle the issue.

In previous works~\cite{PavonValderrama:2005gu,Valderrama:2005wv,PavonValderrama:2005uj,PavonValderrama:2007nu,Entem:2007jg}
we have shown that the necessary minimum number of counterterms 
renormalizing a singular interaction can in fact be determined {\it a priori}
from the behaviour of the configuration-space potential near the origin
(for an earlier coordinate space treatment, see Ref.~\cite{Beane:2000wh}).
On a more numerical basis,
similar observations have been made in momentum space
either using finite cut-offs~\cite{Nogga:2005hy,Entem:2007jg}
or subtractive methods~\cite{Yang:2007hb,Yang:2009kx,Yang:2009pn,Yang:2009fm}.
The main result is that in any uncoupled partial wave where an attractive
singular potential is present, a counterterm is needed in order to
renormalize this particular partial wave.
An interesting corollary is that
finite cut-off effects are less important
the more singular the potential. 
Indeed, at order $\nu $ in the chiral counting the potential behaves as
$ 1/\Lambda_\chi^{\nu+2} r^{3+\nu}$,
generating finite cut-off corrections $\delta_\Lambda (k)$
to the renormalized phase shifts $\delta_\infty (k)$ which scale as
\begin{eqnarray}
\delta_\infty (k) - \delta_\Lambda (k) ={\cal O} (\Lambda^{-5/2-\nu/2}) \, ,
\end{eqnarray}
for large enough cut-offs~\cite{PavonValderrama:2007nu,Entem:2007jg},
meaning in particular that cut-off independence is achieved
in this case.
On the contrary, if the potential is singular and repulsive
the effect of counterterms becomes negligible
for small enough cut-off radii.

A particularly problematic consequence of the non-perturbative
treatment of singular potentials is that the renormalization
of two body scattering by an attractive singular central interaction
requires an infinite number of counterterms,
one per each partial wave.
The unlimited proliferation of counterterms when renormalizing singular
interactions has been, among others, an argument against
removing the cut-off, as the resulting effective field theory
will be unable to predict observables~\cite{Epelbaum:2006pt}.
This problem can be cured in perturbative power countings,
like the one proposed by Kaplan, Savage and Wise~\cite{Kaplan:1998tg,Kaplan:1996xu},
where the proliferation of counterterms is naturally limited
by the order of the approximation.
Unfortunately the singularity in the tensor forces makes the previous
proposal poorly convergent in the $^3S_1-{}^3D_1$ channel~\cite{Fleming:1999ee}
(see however~\cite{ Beane:2008bt} for a renewed formulation).
In the modified Weinberg proposal of Nogga, Timmermans and
van Kolck~\cite{Nogga:2005hy},
OPE is iterated in low angular momentum waves while treated perturbatively
in sufficiently peripheral waves (usually $l > 2$).
This choice naturally limits the number of necessary counterterms and,
although it has has been criticized as arbitrary
in Ref.~\cite{Epelbaum:2006pt},
it is sustained by the perturbative analysis of Ref.~\cite{Birse:2005um}.
Higher order corrections are treated in perturbation theory and
the corresponding (finite) number of counterterms is determined
by imposing cut-off independence on the results,
generating convergent amplitudes
for the central waves with the ${\rm NLO}$ and ${\rm N^2LO}$
chiral potentials~\cite{Valderrama:2009ei}.
However, as we will show, there is a non-perturbative way of
also obtaining a finite number of counterterms. 

In a recent paper~\cite{Valderrama:2008kj} we have analyzed
the role of $\Delta$ degrees of freedom for the central waves and the deuteron
with the chiral $\Delta$ potentials of Refs.~\cite{Kaiser:1998wa,Krebs:2007rh}
with a reasonable phenomenological success.
This particular potential furnishes {\it simultaneously }
the theoretical requirements of renormalizability~\footnote{
  The divergence structure of this potential is identical to that of
  the chiral quark model in the Born-Oppenheimer
  approximation~\cite{Arriola:2009vp}, which being second order
  perturbation theory provides only attractive and singular
  potentials.}
and power counting.
Actually, convergence is achieved for reasonable cut-offs of the order of
$r_c \sim 0.5 {\rm fm}$,
that is, scales comparable with the shortest wavelength probed in NN
elastic scattering below pion production threshold~\footnote{
  This is a purely coordinate space
  argument where the cut-off in the potential has been removed . In
  momentum space the corresponding cut-off $\Lambda \sim \sqrt{m_\pi
    M_N} \sim 350 {\rm MeV}$ implies also a regularization of the
  potential and an effective quenching of the $g_{\pi NN}$ coupling
  constant.  This might be one of the reasons why momentum space calculations
  renormalizing the LS equation~\cite{Entem:2007jg,Valderrama:2007ja}
  require much larger cut-offs $\Lambda \sim 1-4 {\rm GeV}$ than
  naively expected.}.
Moreover, as discussed in Ref.~\cite{Machleidt:2009bh},
it may provide a good starting point for nuclear matter calculations as
it has a rather small $D$-state probability,
implying a sufficiently small wound integral which insures better convergence
properties for the few body correlations and the nuclear many body problem.

In the present work we analyze further the TPE potential with
$\Delta$-excitations~\cite{Kaiser:1998wa,Krebs:2007rh},
addressing the calculation of non-central partial waves.
We show how the number of counterterms can be made finite by implementing a
renormalization prescription correlating an infinite number of partial waves.
Thus, this is a compelling example where, contrary to naive expectations,
singular potentials may be consistently renormalized with
a single common counterterm for {\it all} partial waves.
The idea behind such a procedure is quite simple:
if the potential has an inverse power law behaviour at short distances,
$V(r) \sim C / r^n$,
with a coefficient $C$ independent of energy and angular momentum,
we expect all the reduced wave functions of the system to behave the same way
at small enough radii, regardless of the energy or the angular momentum,
as the contribution from these two factors will become negligible
in comparison with the strength of the potential.
As a trivial consequence,
all partial waves can be related to the zero energy s-wave.
The issue is analyzed in detail in Section~\ref{sec:central}
both for regular and singular potentials. 
While this becomes a relevant observation for uncoupled channels,
the tensor force requires a suitable generalization of the result
for coupled channels, which is discussed in Section~\ref{sec:tensor}.
Surprisingly,
the potentials computed in Refs.~\cite{Kaiser:1998wa,Krebs:2007rh}
including TPE with $\Delta$ excitation do fulfill the necessary
mathematical conditions to link partial waves with different
angular momenta (see Section~\ref{sec:np-scatt}).
Actually, we can estimate the finite cut-off error induced by these angular
momentum correlations and which are exclusive of singular potentials.
Other potentials do not automatically comply with these requirements,
so the question on the consistency of the partial waves correlations
is not independent of the potential and indirectly
on the power counting invoked to compute it.
Of course, mathematical consistency does not necessarily mean
phenomenological success, and we test our proposal against
the widely accepted partial wave analysis (PWA)
of the Nijmegen group in Section~\ref{sec:num-res}.
We see that actually there is no big difference between using the
finite number of counterterms or renormalizing independently wave by
wave, suggesting that improvements might be sought in the TPE chiral
potentials as well. For completeness we address the problem of the
familiar OPE in Appendix~\ref{app:OPE} . 
Finally, in Section~\ref{sec:concl} we summarize our main results
and present our main conclusions and outlook for further work.

As in previous works~\cite{PavonValderrama:2005gu,Valderrama:2005wv,PavonValderrama:2005uj,PavonValderrama:2007nu,Entem:2007jg}
we use extensively the coordinate space formulation 
which greatly simplifies the treatment and
allows handy analytical calculations.
In this approach, contact operators are treated implicitly
via boundary conditions in coordinate space.
We do not follow an {\it a priori} power counting for the contact operators,
but rather deduce the short range operator structure from the condition
of cut-off independence and assuming
that the long range potential is to be fully iterated.
The equivalence to momentum space renormalization has been discussed
in detail for scattering states in Ref.~\cite{Entem:2007jg}
and the deuteron in Ref.~\cite{Valderrama:2007ja}.

\section{Central Delta-Shell Potentials and the Partial Wave Expansion}
\label{sec:central}

As a preparation we will consider first the simplest two-body
scattering problem described by a central potential $V$, which for the
$l$-wave reads
\begin{eqnarray}
\label{eq:schroe-regul}
- u_{k,l}'' + \left[ 2\mu\, V(r) + \frac{l (l+1)}{r^2}
\right]\,u_{k,l}(r) &=& k^2 u_{k,l}(r) \, ,
\end{eqnarray}
where $u_{k,l}$ is the reduced wave function, $\mu$ the reduced mass
of the system, and $k = \sqrt{2\mu E}$ is the center of mass momentum.
The asymptotic long distance boundary condition is taken as 
\begin{eqnarray}
\label{eq:ukl-ps}
u_{k,l} (r) \to \sin \left[ k r - \frac{l\pi}{2} + \delta_l(k) \right]  \, ,
\end{eqnarray}
where $\delta_{l}(k)$ is the corresponding phase shift.
We assume that $V(r)$ can be decomposed as the sum of a finite range potential
$V_F$ (bounded by an exponential fall-off $\sim e^{-m r}$) 
and a contact range interaction $V_C$
\begin{eqnarray}
V(r) = V_F(r; r_c) + V_C(r; r_c) \, ,
\end{eqnarray}
where we have added the auxiliary cut-off scale $r_c$, which will be
needed in order to regularize the contact range interaction.
For convenience we have also regularized the finite range potential
in the following way
\begin{eqnarray}
\label{eq:V_F_reg}
V_F(r; r_c) = V_F(r)\,\theta(r - r_c) \, ,
\end{eqnarray}
which means that the short range components of the finite range potential
are effectively absorbed in the contact potential $V_C$.
For the contact potential we only consider for definiteness
the case in which $V_C$ is a delta-shell interaction
\begin{eqnarray}
V_C(r; r_c) = \frac{C_0(r_c)}{4 \pi r^2}\,\delta (r - r_c) \, ,
\end{eqnarray}
where $C_0$ does not depend on energy, and no higher derivatives
of the delta function are considered. 
This is the simplest possible contact interaction and it actually
becomes equivalent to a short distance boundary condition.
We analyze below what can be obtained with such an interaction when
the cut-off $r_c$ is removed, both in the case of regular and singular
interactions.

\subsection{Delta-Shell Potentials and Regular Interactions}

As mentioned, the two-body scattering problem can be described by the
corresponding reduced Schr\"odinger equation,
Eq.~(\ref{eq:schroe-regul}).
For radii below the cut-off $r_c$, there is no potential
(due to the specific regularization employed for the finite range piece
of the potential, see Eq.~(\ref{eq:V_F_reg}))
and the solution for the wave function is simply
\begin{eqnarray}
\label{eq:uk_reg}
u_{k,l}(r) = const \times r^{l+1} \quad \mbox{for $r < r_c$} \, ,
\end{eqnarray}
where the regular solution has been chosen.

The solution for radii above the cut-off depends (i) on the size of the cut-off
with respect to the range of $V_F$ and (ii) on whether $V_F$ is a regular or
singular interaction.
For the present discussion, we will assume that the cut-off
is much smaller than the range of $V_F$, which we will call $a_F$,
$r_c \ll a_F$,
and that the finite range potential is a regular one,
$\lim_{r \to 0} r^2\,V_F(r) = 0$.
Under these conditions the reduced wave function can be written as a linear
combination of a regular and irregular solution for $r > r_c$, i.e.
\begin{eqnarray}
u_{k,l}(r) = a_l\, u_{k,l}^{({\rm reg})}(r) + b_l\, u_{k,l}^{({\rm irr})}(r)
\quad \mbox{for $r > r_c$} \, ,
\end{eqnarray}
where the superscripts $^{({\rm reg})}$ and $^{({\rm irr})}$ denote the
regular and irregular solutions, respectively.

For small radii, say $r_c < r \ll a_F$, the behaviour of the regular
and irregular wave functions is given by
\begin{eqnarray}
u_{k,l}^{({\rm reg})}(r) &\sim& r^{l+1} \, , \\
u_{k,l}^{({\rm irr})}(r) &\sim& \frac{1}{r^l} \, ,
\end{eqnarray}
where corrections depending on the presence of the potential $V_F$ or
the finite momentum $k$ do not appear until higher relative powers of
$r$ are considered.

The effect of the delta shell potential in the Schr\"odinger equation,
Eq.~(\ref{eq:schroe-regul}), is to generate a discontinuity in the first
derivative of the reduced wave function at $r = r_c$.
The previous statement can be summarized in the following
relation between the logarithmic derivatives of the wave functions
for $r < r_c$ and $r > r_c$
\begin{eqnarray}
\label{eq:C0-general-running}
\frac{2\mu\,C_0(r_c)}{4 \pi r_c^2} &=& 
\frac{a_l(r_c)\, {u_{k,l}^{({\rm reg})}}'(r_c) + 
b_l(r_c)\, {u_{k,l}^{({\rm irr})}}'(r_c)}
{a_l(r_c)\, u_{k,l}^{({\rm reg})}(r_c) +
b_l(r_c)\, u_{k,l}^{({\rm irr})}(r_c)} \nonumber \\ &-& \frac{l+1}{r_c} \, .
\end{eqnarray}
From this expression it can be checked that if we want the effect of the delta
function to be nontrivial, we need $C_0(r_c)$ to be a running
coupling constant.
In fact, for a constant value of $C_0$ one finds in the $r_c \to 0$ limit that
\begin{eqnarray}
\label{eq:lin-C0-constant}
\frac{b_l(r_c)}{a_l(r_c)} \to - 
\frac{u_{k,l}^{({\rm reg})}(r_c)}{u_{k,l}^{({\rm irr})}(r_c)} \simeq - 
r_c^{2l + 1} \to 0 \, ,
\end{eqnarray}
meaning that the regular solution is effectively chosen as $|b_l| \ll |a_l|$.
Therefore, in order to avoid a trivial or irrelevant contact
interaction one needs that $C_0(r_c)$ evolves with $r_c$ in a very
specific way, a dependence that can be obtained by solving
Eq.~(\ref{eq:C0-general-running}) for a given $b_l / a_l$ value.

The running of $C_0(r_c)$ is so strongly determined by the scaling properties
of the regular and irregular wave functions near the origin that,
if $C_0(r_c)$ is set to be non-trivial in a given partial wave,
it will become trivial in all the other waves.
This can be checked as follows. If we fit $C_0(r_c)$ to reproduce $b_l
/ a_l$ in the partial wave $l = l_1$ and call this counterterm
$C_0^{(l_1)}(r_c)$, its exact value can be obtained from solving
Eq.~(\ref{eq:C0-general-running}) for $l = l_1$.
Using now the counterterm $C_0^{(l_1)}(r_c)$ for computing the linear
combination of solutions for $l = l_2 (\neq l_1)$, we get the following
\begin{eqnarray}
\frac{b_{l_2}}{a_{l_2}} \sim r_c^{2 l_2 + 1}
\quad \mbox{for $l_2 \neq l_1$} \, ,
\end{eqnarray}
which is just the same scaling as the corresponding one for a constant
counterterm, Eq.~(\ref{eq:lin-C0-constant}).
Therefore we can take the simplification that a given counterterm only
acts on a determined partial wave when the cut-off is removed, as is
usually assumed. This means {\it de facto} a total independence of
non-trivial counterterms for any other partial wave. Note that only a
trivial counterterm produces a short distance interaction {\it common}
to all partial waves.

\subsection{Delta Shell Potentials and Attractive Singular Interactions}

As we have seen, in order to have a non-trivial effect, the running of
the counterterm depends on the scaling properties of the wave function
near the origin, $r \to 0$.
Given the fact that for regular potentials the scaling is different
for each partial wave, the result
is that only one partial wave will be affected by a given counterterm.
On the contrary, as we will see, for attractive singular potentials
the scaling does not depend on the angular momentum.
Therefore, the scaling is independent of the partial wave chosen, and a given
counterterm will affect all the partial waves simultaneously.
This is our key observation which we will extend to tensor forces in
Section~\ref{sec:tensor} and put forward below for the relevant case
of TPE chiral NN interactions with $\Delta$-excitations in
Section~\ref{sec:np-scatt}.

Indeed, if we consider the behaviour of the reduced wave function for
a power-law attractive singular potential which for short enough
distances behaves as
\begin{eqnarray}
{2 \mu}\,V_F(r) \to - \frac{R_F^{n-2}}{r^n} \, ,
\end{eqnarray}
where $n > 2$ and $R_F$ is some given length scale which sets the range of
the power-law behaviour of $V_F$.
This new scale $R_F$ may not coincide with the generic range $a_F$ of
the potential as several lower energy scales may be present in the
system~\footnote{Actually in the chiral NN potential this is mostly
  the case where $a_F \sim 1/m_\pi $ and $R_F \sim 1/f_\pi$.
}.
For a potential like the previous one and for distances $r_c < r \ll R_F$,
the reduced wave function can be described by the WKB approximation
since the de Broglie wavelength is slowly varying,  
\begin{eqnarray}
R_F \frac{d}{dr} \frac1{\sqrt{2\mu\,(E - V_F(r))}}
\sim \frac{n}{2} (r/R_F)^{n/2-1} \ll 1  \, ,
\end{eqnarray}
yielding the short distance behaviour 
\begin{eqnarray}
\label{eq:uk-uv}
u_{k,l} (r) &\simeq&  A_l\,{\left( \frac{r}{R_F} \right)}^{n/4}\,
\sin{\left[\frac{2}{n-2}\,{\left(\frac{R_F}{r}\right)}^{n/2 - 1} + 
\varphi_l(k) \right]} \nonumber \\ 
&& \mbox{for \, \, $R_F \gg r > r_c$} \, ,
\end{eqnarray}
where $A_l$ is some normalization constant and $\varphi_l(k)$ is a
short distance phase which in principle depends on the angular momentum
and the energy.
For $r < r_c$,  the reduced wave function $u_{k,l}$ will show the expected
$r^{l+1}$ behaviour, see Eq.~(\ref{eq:uk_reg}).

Taking into account the behaviour of the wave function around the cut-off, 
we can rewrite the equation that describes the running of $C_0$ 
for the case of singular interactions as
\begin{eqnarray}
\frac{2\mu\,C_0(r_c)}{4 \pi r_c^2} &=& -\frac{2}{R}\,
{\left( \frac{R_F}{r_c} \right)}^{n/2}\, \nonumber \\ &\times& 
\cot{\left[\frac{2}{n-2}\,{\left(\frac{R_F}{r_c}\right)}^{n/2 - 1} + 
\varphi_l(k) \right]} \nonumber \\ &-& \frac{l+1}{r_c} \, .
\label{eq:C0-singular-running}
\end{eqnarray}
As can be immediately realized, for $r_c \to 0$ the explicit
$l$-dependent term stemming from the behaviour of the wave function
for $r < r_c$ can be dropped, leading to the following ultraviolet
behaviour for $C_0$
\begin{eqnarray}
\frac{2\mu\,C_0(r_c)}{4 \pi r_c^2} &\to& -\frac{2}{R_F}\,
{\left( \frac{R}{r_c} \right)}^{n/2}\, \nonumber \\ &\times&
\cot{\left[\frac{2}{n-2}\,{\left(\frac{R_F}{r_c}\right)}^{n/2 - 1} + 
\varphi_l(k) \right]} \, , \nonumber \\ 
\label{eq:C0-singular-limit}
\end{eqnarray}
which {\it does not depend explicitly on angular momentum}.
In particular the previous equation means that for $r_c \to 0$
we have the following identifications 
\begin{eqnarray}
\varphi_l(k_1) = \varphi_l(k_2) \quad \mbox{and} \quad
\varphi_{l_1}(k) = \varphi_{l_2}(k) \, .
\end{eqnarray}
In other words, the short distance phase is independent of angular
momentum or energy.
In the more general case that we accept energy dependent counterterms $C_k$ or
higher derivatives of the delta, we can obtain an energy dependent
semi-classical phase $\varphi_l(k_1) \neq \varphi_l(k_2)$,
but the angular momentum independence will hold~\footnote{
The $n$-th derivative of the delta function potential generates a discontinuity
on the $(n+1)$-th derivative of the wave function.
As long as the behaviour of the wave function is given by Eq.~(\ref{eq:uk-uv}),
its $(n+1)$-th derivative will be angular momentum independent.
}.
The only way in which one can break the condition 
$\varphi_{l_1}(k) = \varphi_{l_2}(k)$ is by accepting terms explicitly
depending on the angular momentum in the contact range interaction.

The previous results can be efficiently cast in the language of short
distance boundary conditions as follows
\begin{eqnarray}
\left. \frac{u_{k,l_1}'}{u_{k,l_1}} \right|_{r = r_c + \epsilon} =
\left. \frac{u_{k,l_2}'}{u_{k,l_2}} \right|_{r = r_c + \epsilon} \, ,
\end{eqnarray}
that is, the logarithmic derivative at the cut-off radius of the reduced wave
function does not depend on the angular momentum.
This is the form in which we will effectively implement the condition
$\varphi_{l_1}(k) = \varphi_{l_2}(k)$.

It is important to notice that the angular momentum independence of
the behaviour of the $l$-wave reduced wave function near the origin
is only realized when the cut-off is small enough, so the behaviour
described in Eq.~(\ref{eq:uk-uv}) is valid.
In general Eq.~(\ref{eq:uk-uv}) will be applicable for a radius such
that the WKB approximation holds, and this radius will be smaller for
higher partial waves.
If we denote the previous radius as $r^{\rm WKB}_l$, we will generally have
$r^{\rm WKB}_{l_1} > r^{\rm WKB}_{l_2}$ for $l_1 < l_2$.
This means that for a given cut-off radius there will be a critical value
of the angular momentum for which $r^{\rm WKB}_{l_c} > r_c$, and
therefore the condition of angular momentum independence should
only be used for $l < l_c$, i.e.
\begin{eqnarray}
\left. \frac{u_{k,0}'}{u_{k,0}} \right|_{r = r_c + \epsilon} =
\left. \frac{u_{k,1}'}{u_{k,1}} \right|_{r = r_c + \epsilon} =
\dots =
\left. \frac{u_{k,l_c-1}'}{u_{k,l_c-1}} \right|_{r = r_c + \epsilon} \, ,
\end{eqnarray}
while for higher angular momenta the behaviour of the reduced wave function
roughly corresponds to what is to be expected for a regular potential, and
it may be possible to make a perturbative treatment~\footnote{
The ultraviolet or WKB behaviour described by Eq.~(\ref{eq:uk-uv}) sets in
once there is a deeply bound state in the system, or equivalently,
once the zero energy wave function has reached its first zero
(the finite energy wave function will contain anyway zeros due to the
$\sin{(k r + \delta_l - l \pi/2)}$ behaviour at large distances).
This means that for radii bigger than $r_{l}^{\rm WKB}$ or $r_{l}^{\rm bound}$
the perturbative expansion is expected to converge.}
as suggested in Refs.~\cite{Birse:2005um,Long:2007vp}.

\section{The Inclusion of the Tensor Force}
\label{sec:tensor}

The previous analysis about the connection between the short range physics
in different partial waves is only valid for the uncoupled channel case.
If a tensor force is present, as in the nucleon-nucleon interaction, 
we will need to account for the induced coupled channel structure happening
in spin triplet channels in our analysis.
For simplicity, we will assume a two-body system in which the finite range
piece of the potential only contains a non-tensor (central) and a tensor piece
\begin{eqnarray}
V_F(\vec{r}) = V_{NT}(r) + S_{12}(\hat{r})\,V_T(r) \, .
\end{eqnarray}
with $S_{12}(\hat{r}) = 3\,\vec \sigma_1 \cdot \hat r \vec \sigma_2 \cdot \hat
r - \vec\sigma_1 \cdot \vec \sigma_2$ ($\vec{\sigma}_{1}$ and
$\vec{\sigma}_{2}$ are the spin operators of particle $1$ and $2$),
and where we do not specify any additional operator structure of the tensor
and non-tensor pieces of the potential (like, for example,
spin or isospin dependence).
The behaviour of the non-tensor and tensor piece at short enough
distances is given by
\begin{eqnarray}
V_{NT}(r) &\to& \frac{C_{NT}}{r^n} \, , \\
V_{T}(r) &\to& \frac{C_{T}}{r^n} \, ,
\end{eqnarray}
where we have assumed that they have the same power-law divergent behaviour
at short distances.
We have not yet determined whether the potentials are attractive or repulsive.

The tensor force will couple spin triplet channels for which $l = j \pm 1$,
being the corresponding reduced Schr\"odinger equation for $r > r_c$
\begin{eqnarray}
- u_j'' + U_{11} u_j + U_{12} w_j &=& k^2 u_j
\, , \\
- w_j'' + U_{12} u_j + U_{22} w_j  &=& k^2 w_j \, .
\end{eqnarray}
where 
\begin{eqnarray}
U_{11} &=& 2\mu\,V_{NT} - 2\mu\,\frac{2j-2}{2j+1}\,V_T + 
\frac{j(j-1)}{r^2} \, , \\ 
U_{12} &=&  6\,\frac{\sqrt{j(j+1)}}{2j+1} \, , \\ 
U_{22} &=&  2\mu\,V_{NT} - 2\mu\,\frac{2j+4}{2j+1}\,V_T + 
\frac{(j+1)(j+2)}{r^2} \, . \nonumber \\ 
\end{eqnarray}
The previous Schr\"odinger equation has four linearly independent
solutions, of which only two of them are regular and therefore
physically acceptable.
The coupled channel equations can be efficiently cast into the following
compact notation
\begin{eqnarray}
- {\bf u}'' + \left[ 2\mu\,{\bf V} + \frac{{\bf L}^2}{r^2} \right]\,{\bf u} =
k^2\,{\bf u}
\end{eqnarray}
where the wave function is now a matrix
\begin{eqnarray}
{\bf u} = 
\begin{pmatrix}
u_j^{(a)} & u_j^{(b)} \\
w_j^{(a)} & w_j^{(b)}
\end{pmatrix} \, ,
\end{eqnarray}
with the $^{(a)}$ and $^{(b)}$ superscripts representing the two linearly
independent asymptotic ($r \to \infty$) solutions of the system,
and where the potential and the angular momentum operator are
also $2 \times 2$ matrices:
\begin{eqnarray}
{\bf V} &=& {\bf 1}\,V_{NT}(r) + {\bf S}_{12}^j\,V_T(r) \, , \\
{\bf L}^2 &=&
\begin{pmatrix}  
j(j-1) & 0 \\ 
0 & (j+1)(j+2)
\end{pmatrix}
\, .
\end{eqnarray}
In the previous definition of the potential in matrix form, ${\bf 1}$
represents the identity and ${\bf S}_{12}^j$ the tensor operator, i.e.
\begin{eqnarray}
{\bf 1} &=& 
\begin{pmatrix}  
1 & 0 \\ 
0 & 1
\end{pmatrix} \, , \\
{\bf S}_{12}^j &=& \frac{1}{2j + 1}
\begin{pmatrix}  
-2(j-1) & 6\sqrt{j(j+1)} \\ 
6\sqrt{j(j+1)} & -2(j+2)
\end{pmatrix}
\, .
\end{eqnarray}
The tensor operator can be diagonalized with the following rotation
\begin{eqnarray}
\label{eq:tensor-rotation}
{\bf R}_j = \frac{1}{\sqrt{2j+1}}
\begin{pmatrix}  
\sqrt{j+1} & \sqrt{j} \\ 
-\sqrt{j} & \sqrt{j+1}
\end{pmatrix}
\, , 
\end{eqnarray}
leading to
\begin{eqnarray}
\label{eq:S12-rotated}
{{\bf S}_{12}^j\,}_{, \rm D} &=&
{\bf R}_j {\bf S}_{12}^j {\bf R}_j^{\rm T} = 
\begin{pmatrix}  
2 & 0 \\ 
0 & -4
\end{pmatrix}
\, , \\
{\bf L}_{j,\rm D}^2 &=&
{\bf R}_j {\bf L}^2 {\bf R}_j^{\rm T} =
\begin{pmatrix}  
j(j+1) & 2\sqrt{j(j+1)} \\ 
2\sqrt{j(j+1)} & j(j+1) + 2
\end{pmatrix} 
\, , \nonumber \\
\end{eqnarray}
where the $_{\rm D}$ subscript indicates that the corresponding quantities
are defined in the {\it diagonal} basis.
In this basis, the reduced Schr\"odinger equation is written
in the following way
\begin{eqnarray}
- {\bf v}_j'' &+& 2\mu\,\left[ {\bf V}_{j, \rm D} +
\frac{{\bf L}^2_{j, \rm D}}{r^2} +
\frac{2\mu \, {\bf C}_{\rm D}(r_c)}{4 \pi r_c^2}\,\delta(r - r_c)
\right]\,{\bf v}_j \nonumber \\ &=&
k^2\,{\bf v}_j \, ,
\end{eqnarray}
where ${\bf v}_j = {\bf R}_j\,{\bf u}_j$ is the rotated wave function, 
${\bf V}_{j, \rm D} = {\bf R}_j\,{\bf V}\,{\bf R}_j^{\rm T}$ represents
the potential in the diagonal basis, and the contact interaction has
been explicitly introduced.
At short enough distances, ${\bf V}_{j, \rm D}$ behaves in the following way
\begin{eqnarray}
{\bf V}_{j,\rm D}(r) &\to& \frac{1}{r^n}
\begin{pmatrix}  
C_{NT} + 2\,C_T & 0 \\ 
0 & C_{NT} -4\,C_T
\end{pmatrix}
\, ,
\end{eqnarray}
which means that depending on the signs and relative values of the
van der Waals coefficients of the non-tensor and tensor piece,
$C_{NT}$ and $C_T$,
the diagonalized potential may be attractive-attractive, attractive-repulsive
or repulsive-repulsive.
Only two of these situations, namely the attractive-attractive and
attractive-repulsive case, admit
counterterms~\cite{PavonValderrama:2005wv,PavonValderrama:2005uj}, and
are therefore of interest from the point of view of renormalization. 

\subsection{Attractive-Attractive Case}

Situations where both eigenchannels are attractive are the easiest to
handle.
If a singular power-law potential is assumed,
the behaviour of the reduced wave function for $r < r_c$ can be safely ignored,
yielding the following relation 
\begin{eqnarray}
\label{eq:delta_shell_coupled}
{{\bf v}_j}'(r_c) {{\bf v}_j (r_c)\,}^{-1} =
\frac{2\mu}{4 \pi r_c^2}\,{\bf C}_{\rm D}(r_c) \, ,
\end{eqnarray}
meaning that there are three free parameters in this case
(due to ${\bf C}_{\rm D}$ being real and symmetric),
in agreement with previous analysis of singular potentials in coupled
channels~\cite{PavonValderrama:2005wv,PavonValderrama:2005uj}.
The relation between the short range wave functions of channels with
different total angular momentum is therefore
\begin{eqnarray}
{\bf v}^{\rm T}_{j_2} {\bf v}_{j_1}' = 
{{\bf v}^{\rm T}_{j_2}\,}' {\bf v}_{j_1} \, ,
\end{eqnarray}
where we have made use of ${\bf C}_{\rm D} = {\bf C}_{\rm D}^{\rm T}$.
This relation is invariant with respect to the relative normalization of the
two linearly independent solutions of each partial wave, which were
previously denoted with the $^{(a)}$ and $^{(b)}$ superscripts,
and also on the set of linearly independent solutions chosen,
as can be easily checked.
It should be noted too that the previous relation is reminiscent of
the coupled-channel version of the two-potential formula
of Ref.~\cite{PavonValderrama:2009nn}.

A problem arises in relating the $j=0$ triplet state with other coupled
channels, as the $^3P_0$ wave is effectively an uncoupled state,
\begin{eqnarray}
{\bf v}_{j = 0}(r) &=& 
\begin{pmatrix}  
0 & 0 \\ 
0 & v_{^3P_0}(r)
\end{pmatrix}
\, .
\end{eqnarray}
Therefore, the previous representation of the short distance potential,
i.e. the coupled-channel delta-shell of Eq.~(\ref{eq:delta_shell_coupled}),
cannot be a correct representation of the short range physics of
the $^3P_0$ channel.
As a consequence the $^3P_0$ channel cannot be {\it unambiguously}
related with the other coupled channels.
It is possible however to obtain convergent amplitudes by
relating the $^3P_0$ wave function with any of the ``lower''
{\it eigen} wave functions of the reference wave function ${\bf v}_j$,
\begin{eqnarray}
\frac{v_{{}^3P_0}'(r_c)}{v_{{}^3P_0}(r_c)} =
\frac{({\bf v}_j)_{21}'(r_c)}{({\bf v}_j)_{21}(r_c)} \quad \mbox{ or } \quad
\frac{v_{{}^3P_0}'(r_c)}{v_{{}^3P_0}(r_c)} =
\frac{({\bf v}_j)_{22}'(r_c)}{({\bf v}_j)_{22}(r_c)} \, , \nonumber \\
\end{eqnarray}
where ${}_{21}$ and ${}_{22}$ are the corresponding indices in the rotated
wave function matrix $v_2$.
Both possibilities yield a renormalizable phase for the $^3P_0$ wave,
but not the same one: the phase shift depends on which of the previous
two equations is used, indicating the presence of model
dependence in any of these choices.
Therefore, the only way to avoid model dependence is to treat the $^3P_0$ wave
as an independent wave in the attractive-attractive case.

\subsection{Attractive-Repulsive Case}

If one of the eigenchannels is attractive and the other is repulsive, then
we have that Eq.~(\ref{eq:delta_shell_coupled}) can only be applied in the
attractive eigenchannel.
In particular, the delta-shell coupling matrix ${\bf C}_{\rm D}(r_c)$
takes the simplifying form
\begin{eqnarray}
{{\bf C}_{\rm D}(r_c)}_{BC} = C_A(r_c)\,\delta_{AB} \delta_{AC} \, , 
\end{eqnarray}
where the labels $B, C = 1, 2$ are matrix indices,
and $A$ represents the index associated
with the attractive solution.
That is, only one counterterm is needed in order to renormalize an
attractive-repulsive coupled channel,
as delta-shell contributions become trivial in the $r_c \to 0$ limit,
except if they happen in the $AA$ subchannel.
Combining the previous result with the boundary condition
induced by the delta-shell potential
\begin{eqnarray}
\label{eq:delta-shell-coupled-bc}
{{\bf v}_j}'(r_c) =
\frac{2\mu}{4 \pi r_c^2}\,{\bf C}_{\rm D}(r_c)\,{{\bf v}_j (r_c)}
\, ,
\end{eqnarray}
the following renormalization conditions are obtained
\begin{eqnarray}
\label{eq:AR-corr-1}
\frac{{\left( {{\bf v}_{j_1}}'(r_c)\,\right)}_{AA}}
{{\left( {{\bf v}_{j_1}}(r_c)\,\right)}_{AA}} &=&
\frac{{\left( {{\bf v}_{j_2}}'(r_c)\,\right)}_{AA}}
{{\left( {{\bf v}_{j_2}}(r_c)\,\right)}_{AA}} \, ,  \\
\label{eq:AR-corr-2}
\frac{{\left( {{\bf v}_{j_1}}'(r_c)\,\right)}_{AR}}
{{\left( {{\bf v}_{j_1}}(r_c)\,\right)}_{AR}} 
&=&
\frac{{\left( {{\bf v}_{j_2}}'(r_c)\,\right)}_{AR}}
{{\left( {{\bf v}_{j_2}}(r_c)\,\right)}_{AR}} \, , \\
\label{eq:AR-corr-3}
{\left( {{\bf v}_{j_1}}'(r_c)\,\right)}_{RA} &=& 
{\left( {{\bf v}_{j_2}}'(r_c)\,\right)}_{RA} = 0 \, , \\
\label{eq:AR-corr-4}
{\left( {{\bf v}_{j_1}}'(r_c)\,\right)}_{RR} &=& 
{\left( {{\bf v}_{j_2}}'(r_c)\,\right)}_{RR} = 
0 \, ,
\end{eqnarray}
where $R$ is the index of the repulsive eigenchannel.
The first two equations relate the attractive eigenchannels of different
partial waves, although they are redundant.
From the form of the delta-shell coupling matrix ${\bf C}_{\rm D}(r_c)$,
together with Eq.~(\ref{eq:delta-shell-coupled-bc}),
one obtains that
\begin{eqnarray}
\frac{{\left( {{\bf v}_{j}}'(r_c)\,\right)}_{AA}}
{{\left( {{\bf v}_{j}}(r_c)\,\right)}_{AA}} =
\frac{{\left( {{\bf v}_{j}}'(r_c)\,\right)}_{AR}}
{{\left( {{\bf v}_{j}}(r_c)\,\right)}_{AR}} \, ,
\end{eqnarray}
implying the equivalence of Eqs.~(\ref{eq:AR-corr-1}) and (\ref{eq:AR-corr-2}).
The last two equations, (\ref{eq:AR-corr-3}) and (\ref{eq:AR-corr-4}),
are just regularization conditions for the repulsive eigenchannels.
Contrary to what happened in the attractive-attractive case,
in the attractive-repulsive case the $j=0$ $^3P_0$ wave
(if attractive) can be directly related with the other coupled triplet waves.

\section{Application to Neutron-Proton Scattering}
\label{sec:np-scatt}

The previous results can be applied to the case neutron-proton (np)
scattering in nuclear effective field theory, where the resulting
potentials are in many cases singular and attractive.
In principle we can relate any two partial waves for which the
potential diverges in the same way, i.e. the {\it same} power and the
{\it same} coefficient. This means that correlations will emerge only
between channels with the same spin and isospin values.
For the spin triplet channels it is also necessary to consider whether the
channels are coupled ($l = j\pm1$) or uncoupled ($l = j$).
Therefore, we obtain a total of six sets of correlated waves, namely 
(i) singlet isovectors, 
(ii) singlet isoscalars, 
(iii) uncoupled triplet isovectors, 
(iv) uncoupled triplet isoscalars, 
(v) coupled triplet isovectors, 
and (vi) coupled triplet isoscalars
.
We limit ourselves to the $j \leq 5$ partial waves.
The $^3P_0$ wave remains uncorrelated in our current scheme and it is not
further considered in this work.

\subsection{The Chiral Neutron-Proton Potential}

The finite range piece of the nucleon-nucleon (NN) potential in 
chiral perturbation theory is expressed as an expansion
in powers of Q
\begin{eqnarray}
V_{\rm NN}(\vec r) = V_{\chi}^{(0)}(\vec r) + V_{\chi}^{(2)}(\vec r) +
V_{\chi}^{(3)}(\vec r) + {\mathcal O}(Q^4) \, ,
\end{eqnarray}
where $Q$ represents either the pion mass, the nucleon-$\Delta$ splitting or
the momentum of the nucleons.
We only consider here chiral potentials in which the $\Delta$ isobar has been
explicitly included~\cite{Kaiser:1998wa,Krebs:2007rh}.
The reasons for this decision are that (i)
they have better convergence properties than their $\Delta$-less counterparts,
and (ii) they are attractive-attractive singular interactions in all coupled
channels at orders $Q^2$ (${\rm NLO}$) and $Q^3$ (${\rm N}^2{\rm LO}$),
leading to a simpler analysis in general.
For the finite range piece of the interaction
we adopt the original Weinberg counting~\cite{Weinberg:1990rz},
in which $1/M_N$ corrections are treated
as higher order as it is done
in Refs.~\cite{Ordonez:1995rz,Epelbaum:2003gr,Epelbaum:2003xx}.
At the orders considered in this work,
the potential can be decomposed as a central, spin-spin
and a tensor component which in coordinate space reads
\begin{eqnarray}
V_{\rm NN}(\vec r) &=& V_C(r) + \tau W_C(r) + \sigma\,(V_S(r) + \tau W_S(r))
\nonumber \\
&+& {S}_{12}(\hat{r})\,(V_T(r) + \tau W_T(r)) \, ,
\end{eqnarray}
where spin-orbit and quadratic spin-orbit terms have been ignored as
they do not appear up to higher orders.
The use of previous counting rule for the $1/M_N$ corrections is necessary
if we plan to correlate at short distance the behaviour of the different waves,
as it generates a spin-orbit term which is less singular
than the other components of the interaction.
The operators $\tau$, $\sigma$ and $S_{12}$ are given by
\begin{eqnarray}
\tau &=& \vec \tau_1 \cdot
\vec \tau_2 = 2 t(t+1) -3 \, , \nonumber \\ 
\sigma &=& \vec \sigma_1 \cdot \vec \sigma_2 = 2 s(s+1) -3 \, ,
\nonumber \\ 
{S}_{12}(\hat{r}) &=&
3\,\vec \sigma_1 \cdot \hat r \vec \sigma_2 \cdot \hat
r - \vec\sigma_1 \cdot \vec \sigma_2 \, ,
\label{eq:S12-def}
\end{eqnarray} 
where $\vec{\tau}_{1(2)}$ and $\vec{\sigma}_{1(2)}$ are the
proton(neutron) isospin and spin operators;
$t$ and $s$ represent the total isospin $t=0,1$ and total spin $s=0,1$ of
the np system.
The precise form of the chiral $\Delta$ potential is taken from
Ref.~\cite{Krebs:2007rh}.

Note that in the singlet channel cases ($s = 0$) the tensor force
operator does not contribute.
For symmetry reasons (Fermi-Dirac statistics) we have $(-)^{l+s+t}=-1$,
where $l$ is the orbital angular momentum.
This means in particular that even partial waves are isovectors ($t = 1$)
and odd partial waves are isoscalars ($t = 0$).
The NN potential reads for the singlet channels
\begin{eqnarray}
V_{^1S_0}(r) &=& V_{^1D_2}(r) = V_{^1G_4}(r) \nonumber \\ &=& 
V_C(r) + W_C(r) - 3 V_S(r) -3 W_S(r) \, ,  \nonumber \\
V_{^1P_1}(r) &=& V_{^1F_3}(r) = V_{^1H_5}(r) \nonumber \\ &=& 
V_C(r) - 3 W_C(r) - 3 V_S(r) + 9 W_S(r) \, ,  \nonumber \\
\end{eqnarray}
that is, all the singlet channels can be described with two different
potentials depending on whether we are in the isoscalar or isovector case
(or equivalently, on whether even or odd partial waves are considered).

In the spin triplet channels ($s=1$) we must distinguish between uncoupled
($l=j$) and coupled waves ($l = j \pm 1$).
In the uncoupled waves, we can again distinguish between the potential
in the isoscalar ($^3D_2$, $^3G_4$) and isovector
($^3P_1$, $^3D_2$) waves
\begin{eqnarray}
V_{^3D_2}(r) &=& V_{^3G_4}(r) \, , \nonumber \\
V_{^3P_1}(r) &=& V_{^3F_3}(r) = V_{^3H_5}(r) \, , \nonumber \\
\end{eqnarray}
where the explicit expressions of the previous potentials in terms of the
central, spin-spin and spin-tensor components is given by
\begin{eqnarray}
V_{^3D_2}(r) &=& V_C(r) - 3 W_C(r) + V_S(r) - 3 W_S(r) \nonumber \\ 
&&+ 
2 V_T(r) - 6 W_T(r)  , \\
V_{^3P_1}(r) &=& V_C(r) + W_C(r) +V_S(r) + W_S(r) 
\nonumber \\ 
&& +
2 V_T(r) + 2 W_T(r)\, . \nonumber \\
\end{eqnarray}
Equivalently, for the coupled waves we have
\begin{eqnarray}
{\bf R}_1\,{\bf V}_{^3C_1}(r)\,{\bf R}_1^{\rm T} &=& 
{\bf R}_3\,{\bf V}_{^3C_3}(r)\,{\bf R}_3^{\rm T} = 
{\bf R}_5\,{\bf V}_{^3C_5}(r)\,{\bf R}_5^{\rm T} \, , \nonumber  \\
{\bf R}_2\,{\bf V}_{^3C_2}(r)\,{\bf R}_2^{\rm T} &=&  
{\bf R}_4\,{\bf V}_{^3C_4}(r)\,{\bf R}_4^{\rm T} \, , \nonumber \\
\end{eqnarray}
for the isoscalar ($^3C_1$, $^3C_3$, $^3C_5$) and isovector 
($^3C_2$, $^3C_4$) waves, and where the expressions are
more cumbersome as they involve matrices.
In the expressions above, ${\bf R}_j$ are the rotation matrices defined
in Eq.~(\ref{eq:tensor-rotation}).
The notation $^3C_1$, $^3C_2$, etc, is a short hand for
$^3S_1-{}^3D_1$, $^3P_2-{}^3F_2$, etc,
and the explicit form of the rotated potentials is given by
\begin{eqnarray}
{\bf R}_1\,{\bf V}_{^3C_1}(r)\,{\bf R}_1^{\rm T} &=& 
{\bf 1}\,\left(V_C(r) - 3 W_C(r) + V_S(r) - 3 W_S(r) \right) \nonumber \\
&+& {\bf S}^j_{12,\rm D}\,(V_T(r) - 3 W_T(r)) \, , \nonumber \\
{\bf R}_2\,{\bf V}_{^3C_2}(r)\,{\bf R}_2^{\rm T} &=&  
{\bf 1}\,\left(V_C(r) + W_C(r) + V_S(r) + W_S(r) \right) \nonumber \\ &+&
{\bf S}^j_{12,\rm D}\,(V_T(r) + W_T(r)) \, , \nonumber \\
\end{eqnarray}
with ${\bf 1}$ the 2x2 identity matrix, and ${\bf S}^j_{12,\rm D}$
the diagonalized tensor matrix represented by Eq.~(\ref{eq:S12-rotated}).

\subsection{Van der Waals Behaviour of the Chiral Potentials}

At distances below the pion Compton wavelength, $m_{\pi} r \ll 1$,
the chiral potentials exhibit at orders $Q^2$ and $Q^3$
a power-law behaviour of the type
\begin{eqnarray}
V^{(\nu)}(r) &\to& \frac{C_6^{(\nu)}}{r^{6}} \, ,
\end{eqnarray}
with $\nu = 2,3$ and where the value of $C_6^{(\nu)}$ depends on the particular
component of the potential considered.
These coefficients were computed in Ref.~\cite{Valderrama:2008kj} based
on the spectral function representation of the potentials of
Krebs, Epelbaum and Mei{\ss}ner~\cite{Krebs:2007rh}.
It should be noted though that the exact behaviour of the potential at short
distances is inessential for the angular momentum correlations.
What really matters is
(i) that the potential is a singular attractive interaction and
(ii) that it is much stronger than the centrifugal barrier
at the chosen cut-off radius $r_c$.
In the case of the order $Q^2$ and $Q^3$ chiral $\Delta$-full potentials these
conditions are fulfilled in all partial waves with $j \leq 5$
for cut-off radii as big as $1.0\,{\rm fm}$.

\subsection{Correlated Renormalization of the Uncoupled Waves}

We describe the scattering states in the uncoupled waves by solving
the following reduced Schr\"odinger equation for $r > r_c$
\begin{eqnarray}
\label{eq:schroe-NN}
- u_{k,l}'' + \left[ M_N\,V_{\rm NN}(r) +
\frac{l (l+1)}{r^2} \right]\,u_{k,l}(r) &=& k^2 u_{k,l}(r) \, , \nonumber \\ 
\end{eqnarray}
where $u_{k,l}$ is the reduced wave function, $V_{NN}$ the corresponding
chiral potential for the particular partial wave considered,
$k$ the center-of-mass momentum, $l$ the angular momentum,
and $M_N$ is twice the neutron-proton reduced mass,
i.e. $M_N = 2 M_p M_n / (M_p + M_n)$.
The reduced wave function is asymptotically normalized to
\begin{eqnarray}
u_{k,l}(r) \to k^l\,
\left( \cot{\delta_l}\,\hat{j}_l(k r) - \hat{y}_l(k r) \right) \, ,
\end{eqnarray}
for $r \to \infty$, with $\delta_l$ the phase shift, and
$\hat{j}_l(x) = x\,j_l(x)$ and $\hat{y}_l(x) = x\,y_l(x)$
the reduced spherical Bessel functions.
The normalization factor $k^l$ is added in order to have a well-defined
normalization of the wave function in the $k \to 0$ limit.
At $r = r_c$ the wave function can be determined by several means.
One is by solving Eq.~(\ref{eq:C0-singular-running}) for some value of
the counterterm $C_0$, which can be later fitted to reproduce some
observable, like for example, the $^1S_0$ (or $^1P_1$/$^3P_1$/$^3D_2$)
scattering length.
A different way is to construct an asymptotic wave function ($r \to \infty$)
reproducing the desired scattering length.
In the case of the $^1S_0$ channel, this wave function is given by
\begin{eqnarray}
\label{eq:uk-asymp}
u_{0,{{}^1S_0}}(r) \to 1 - \frac{r}{a_0} \, ,
\end{eqnarray}
with $a_0$ the $^1S_0$ scattering length, and then integrate the 
reduced Schr\"odinger equation, Eq.~(\ref{eq:schroe-NN}), downwards
from $r \to \infty$ to $r = r_c$.
Then we use the different relations derived previously to obtain
the logarithmic boundary condition at $r = r_c$ for the different
energies and partial waves considered.
For the particular case of the $^1S_0$ channel and its correlated channels
$^1D_2$ and $^1G_4$, the relation takes the form
\begin{eqnarray}
\label{eq:1S0-1D2-relation}
\left. \frac{u_{0,{{}^1S_0}}'}{u_{0,{{}^1S_0}}} \right|_{r=r_c} &=& 
\left. \frac{u_{k,{{}^1S_0}}'}{u_{k,{{}^1S_0}}} \right|_{r=r_c} \, , \\
\left. \frac{u_{k,{{}^1S_0}}'}{u_{k,{{}^1S_0}}} \right|_{r=r_c} &=& 
\left. \frac{u_{k,{{}^1D_2}}'}{u_{k,{{}^1D_2}}} \right|_{r=r_c} =
\left. \frac{u_{k,{{}^1G_4}}'}{u_{k,{{}^1G_4}}} \right|_{r=r_c} \, ,
\end{eqnarray}
where the first equation relates the zero and finite energy states of
the $^1S_0$ wave, and the second one represents the partial wave
correlations.
For the other correlated channels, we have the correlation conditions
\begin{eqnarray}
\left. \frac{u_{k,{{}^1P_1}}'}{u_{k,{{}^1P_1}}} \right|_{r=r_c} &=& 
\left. \frac{u_{k,{{}^1F_3}}'}{u_{k,{{}^1F_3}}} \right|_{r=r_c} =
\left. \frac{u_{k,{{}^1H_5}}'}{u_{k,{{}^1H_5}}} \right|_{r=r_c} \, , 
\nonumber \\ \label{eq:1P1-1F3-relation} \\
\left. \frac{u_{k,{{}^3P_1}}'}{u_{k,{{}^3P_1}}} \right|_{r=r_c} &=& 
\left. \frac{u_{k,{{}^3F_3}}'}{u_{k,{{}^3F_3}}} \right|_{r=r_c} =
\left. \frac{u_{k,{{}^3H_5}}'}{u_{k,{{}^3H_5}}} \right|_{r=r_c} \, , 
\nonumber \\ \label{eq:3P1-3F3-relation}  \\
\left. \frac{u_{k,{{}^3D_2}}'}{u_{k,{{}^3D_2}}} \right|_{r=r_c} &=& 
\left. \frac{u_{k,{{}^3G_4}}'}{u_{k,{{}^3G_4}}} \right|_{r=r_c} \, ,
\label{eq:3D2-3G4-relation}
\end{eqnarray}
which are to be supplemented with the regularization conditions
for the {\it base} waves
\begin{eqnarray}
\left. \frac{u_{0,{{}^1P_1}}'}{u_{0,{{}^1P_1}}} \right|_{r=r_c} &=& 
\left. \frac{u_{k,{{}^1P_1}}'}{u_{k,{{}^1P_1}}} \right|_{r=r_c} \, , \\
\left. \frac{u_{0,{{}^3P_1}}'}{u_{0,{{}^3P_1}}} \right|_{r=r_c} &=& 
\left. \frac{u_{k,{{}^3P_1}}'}{u_{k,{{}^3P_1}}} \right|_{r=r_c} \, , \\
\left. \frac{u_{0,{{}^3D_2}}'}{u_{0,{{}^3D_2}}} \right|_{r=r_c} &=& 
\left. \frac{u_{k,{{}^3D_2}}'}{u_{k,{{}^3D_2}}} \right|_{r=r_c} \, .
\end{eqnarray}
These boundary conditions can be used as initial integration conditions
for the corresponding Schr\"odinger equation Eq.~(\ref{eq:schroe-NN}).
After integrating upwards from $r = r_c$ to $r \to \infty$, we match
to the asymptotic behaviour of the wave functions, Eq.~(\ref{eq:uk-asymp}),
in order to obtain the phase shifts.
The equivalent value for the counterterm coupling $C_0(r_c)$ can be obtained
from Eq.~(\ref{eq:C0-singular-running}), giving in the $r_c \to 0$ limit
\begin{eqnarray}
\frac{M_N\,C_{^1S_0}(r_c)}{4 \pi r_c^2} \simeq 
\frac{u_{0,{{}^1S_0}}' (r_c)}{u_{0,{{}^1S_0}}(r_c)} \, ,
\end{eqnarray}
plus the corresponding expressions for the other base waves. 

\subsection{Correlated Renormalization of the Coupled Waves}

For the coupled channels we solve the coupled reduced Schr\"odinger equation
in its matrix form
\begin{eqnarray}
\label{eq:schroe-NN-coupled}
-{{\bf u}_{k,j}}'' + 
\left[ 2\mu\,{\bf V}_{\rm NN}(r) +
\frac{{\bf L}^2}{r^2} \right]\,{\bf u}_{k,j}(r)
= k^2\,{\bf u}_{k,j}(r)
\end{eqnarray}
where we now use the notation of Sect.~(\ref{sec:tensor}) in which
${\bf u}_{k,j}$, ${\bf V}_{\rm NN}$ and ${\bf L}^2$ are matrices.
The reduced wave function is asymptotically ($r \to \infty$) normalized to
\begin{eqnarray}
{\bf u}_{k,j} \to
\left( {\bf j}_{j}(k r)\,{\bf M}_j(k) - {\bf y}_{j}(k r) \right)
\,{\bf F}_j (k)\, ,
\end{eqnarray}
where ${\bf j}_j(k r)$, ${\bf y}_j(k r)$ and ${\bf F}_j (k)$
are diagonal matrices defined as
\begin{eqnarray}
{\bf j}_{j}(k r) &=&
\begin{pmatrix}
\hat{j}_{j-1}(k r) & 0 \\
0 & \hat{j}_{j+1}(k r)
\end{pmatrix} \, , \\
{\bf y}_{j}(k r) &=&
\begin{pmatrix}
\hat{y}_{j-1}(k r) & 0 \\
0 & \hat{y}_{j+1}(k r)
\end{pmatrix} \, , \\
{\bf F}_j(k) &=&
\begin{pmatrix}
k^{j-1} & 0 \\
0 & k^{j+1}
\end{pmatrix} \, ,
\end{eqnarray}
with $\hat{j}_l(x) = x\,j_l(x)$ and $\hat{y}_l(x) = x\,y_l(x)$ the reduced
spherical Bessel functions.
The normalization factor ${\bf F}_j(k)$ is included in order to have a
well-defined asymptotic ($r \to \infty$) wave function at $k = 0$.
The ${\bf M}_j(k)$ matrix is the coupled-channel equivalent
of $\cot{\delta_l(k)}$ and is related to the S-matrix ${\bf S}_j(k)$
by ${\bf M}_j(k) = i ({\bf S}_j(k) + {\bf 1}) / ({\bf S}_j(k) - {\bf 1})$
with ${\bf 1}$ the 2x2 identity matrix.

For the chiral $\Delta$-full potential of Ref.~\cite{Krebs:2007rh}
all the coupled channels are attractive-attractive singular potentials
at distances below the pion Compton wavelength.
Thus three renormalization conditions or counterterms are needed
in order to obtain well-defined results.
The usual procedure is to fix the asymptotic ($r\to\infty$) behaviour of
the wave function at $k = 0$. That is, we fix the three scattering
lengths of the coupled system.
Then we integrate the Schr\"odinger equation, Eq.~(\ref{eq:schroe-NN-coupled}),
downwards from $r \to \infty$ to $r = r_c$.
If we define ${\bf L}_{k,j}(r)$ as
\begin{eqnarray}
{\bf L}_{k,j} (r) = {\bf u}_{k,j}'(r)\,{{\bf u}_{k,j}}^{-1}(r) \, ,
\end{eqnarray}
then, the finite energy solution is constructed from 
the following boundary condition at $r = r_c$
\begin{eqnarray}
{\bf L}_{k,j} (r_c) = {\bf L}_{0,j} (r_c) \, .
\end{eqnarray}

\begin{figure*}[htb]
\begin{center}
\epsfig{figure=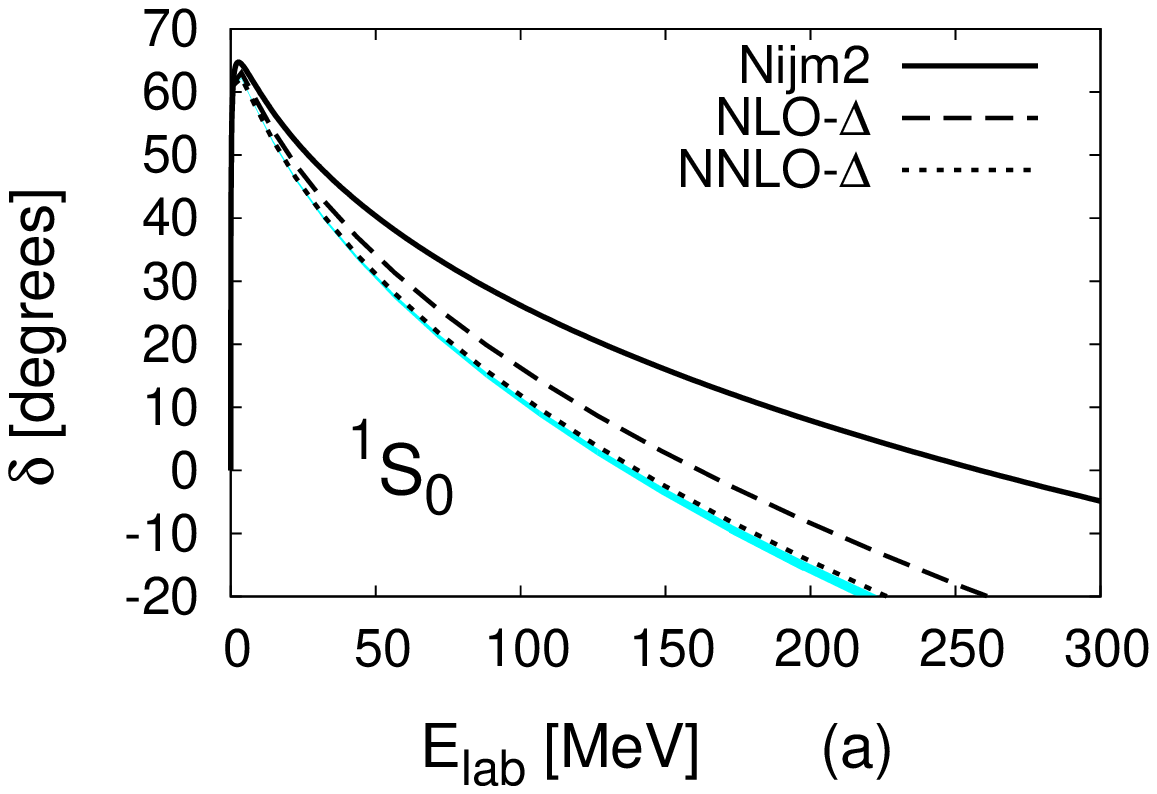, 
	height=5.0cm, width=5.5cm}
\epsfig{figure=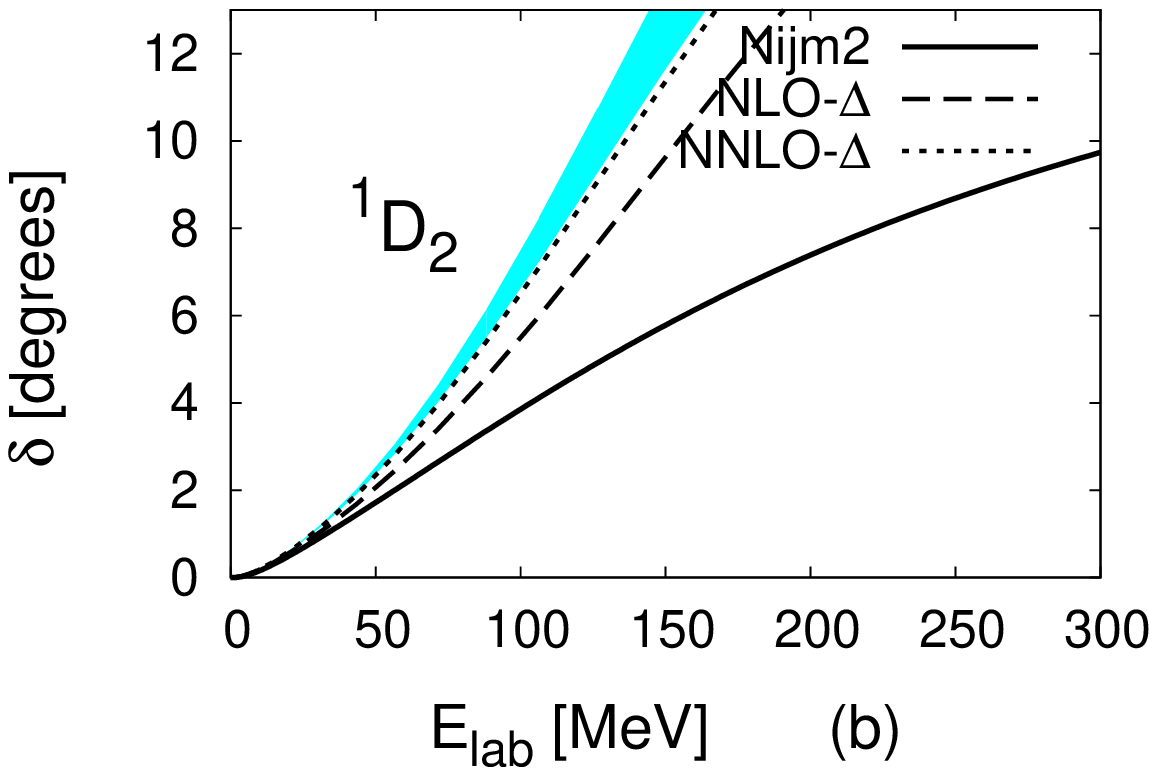, 
	height=5.0cm, width=5.5cm}
\epsfig{figure=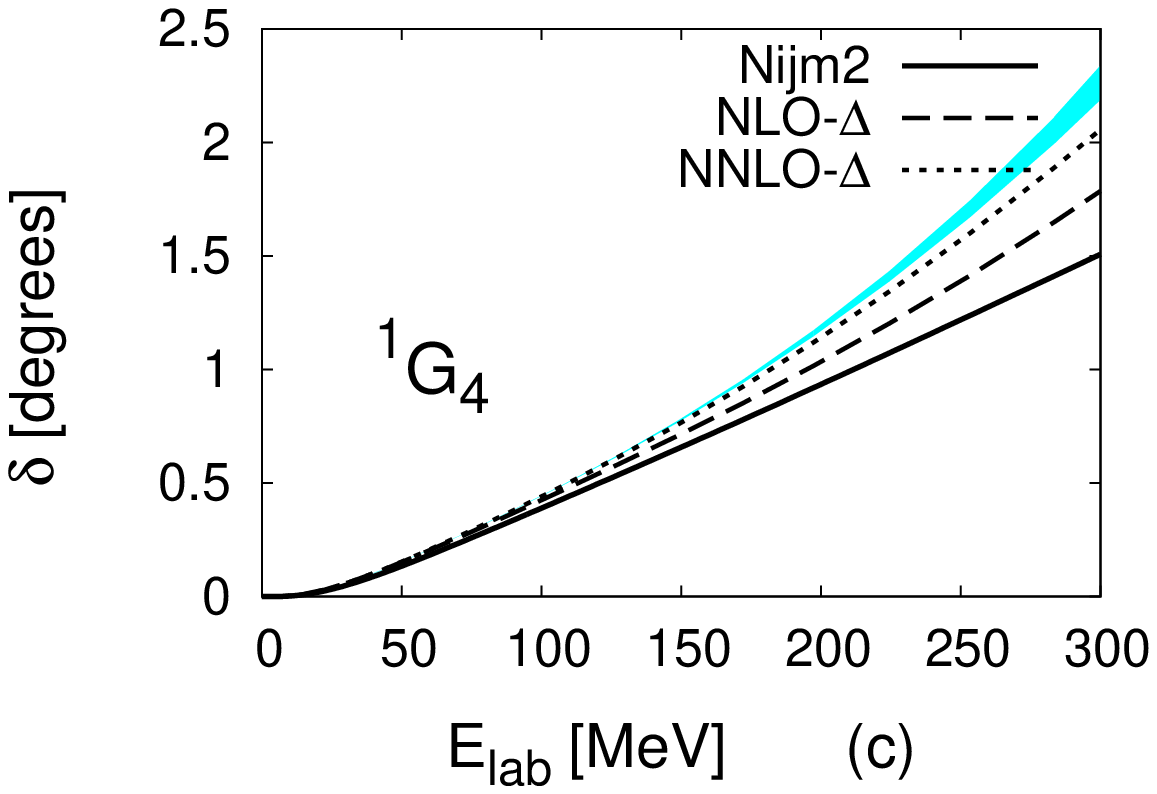, 
	height=5.0cm, width=5.5cm}
\epsfig{figure=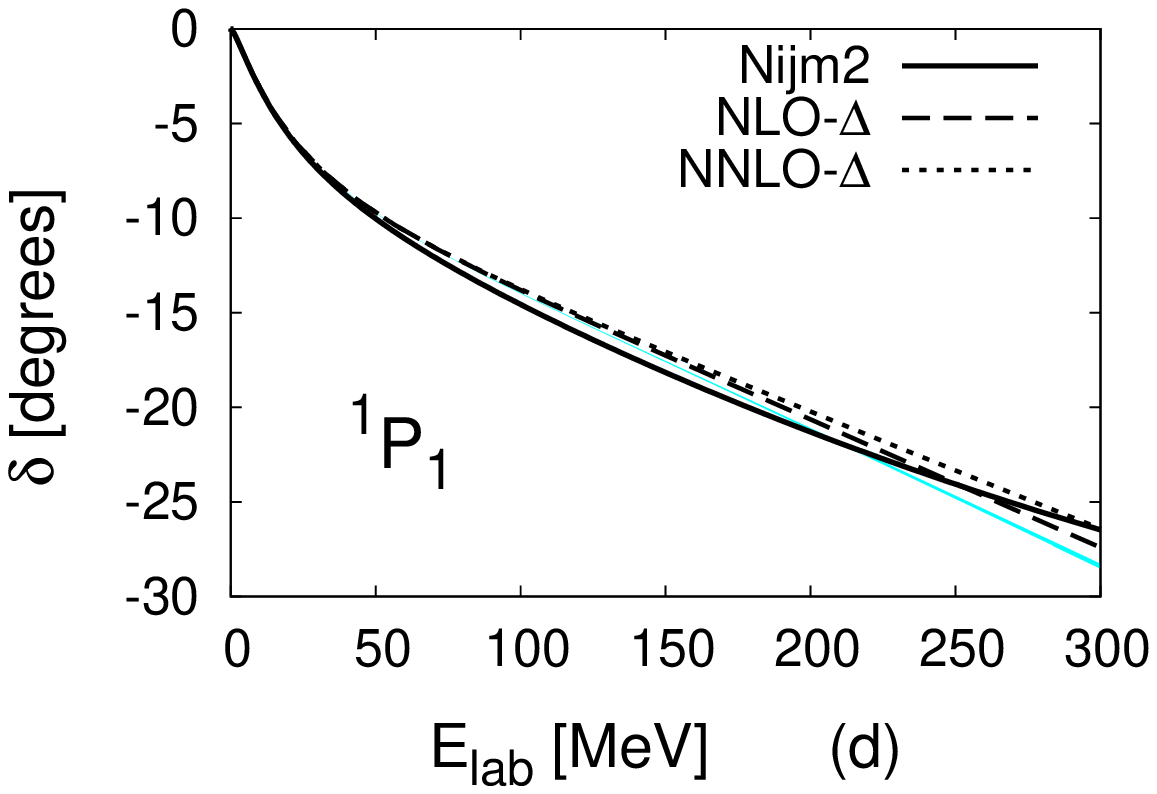, 
	height=5.0cm, width=5.5cm}
\epsfig{figure=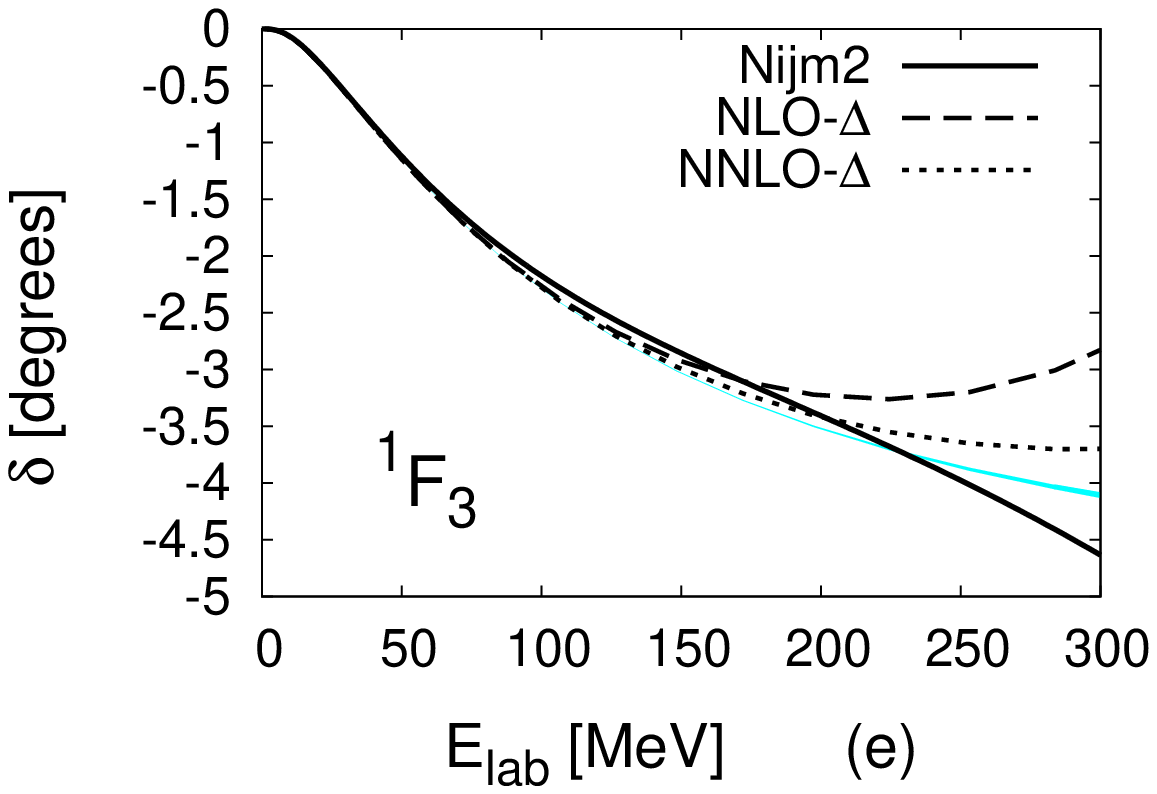, 
	height=5.0cm, width=5.5cm}
\epsfig{figure=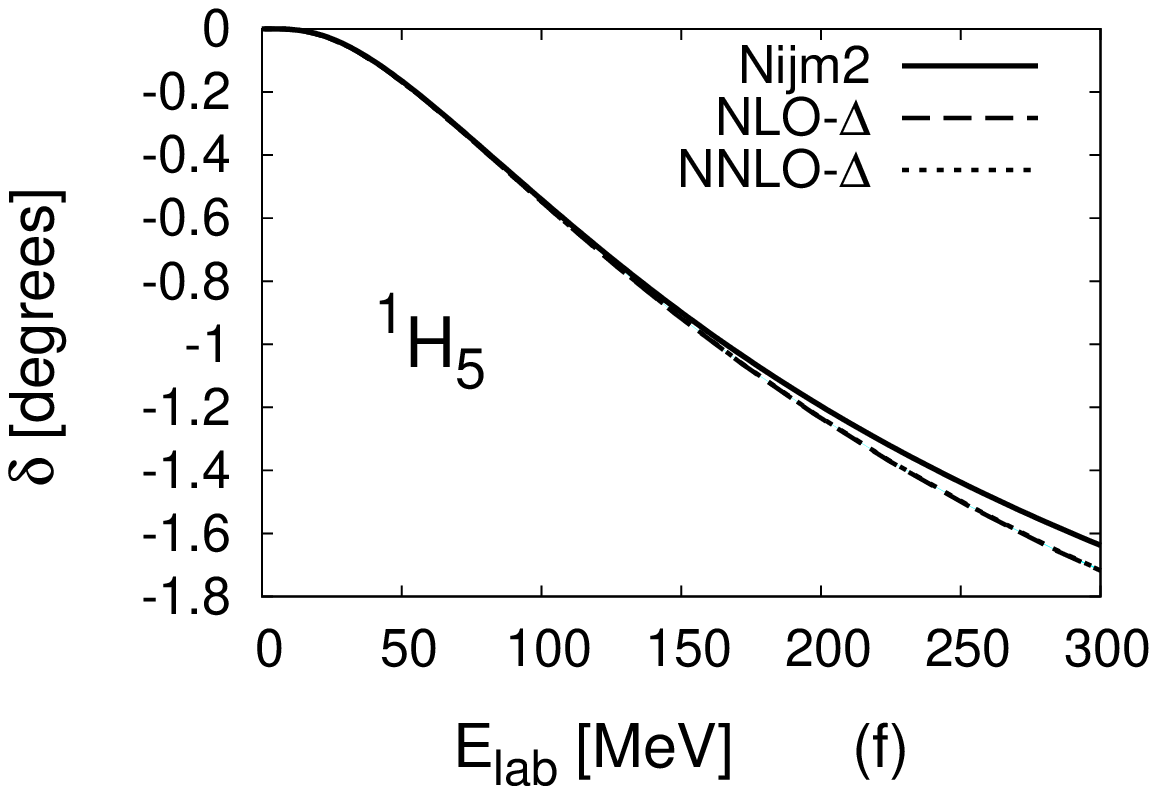, 
	height=5.0cm, width=5.5cm}
\end{center}
\caption{(Color online) (Upper panel) $^1S_0$, $^1D_2$ and $^1G_4$
phase shifts computed from
Eq.~(\ref{eq:1S0-1D2-relation}), using the $^1S_0$ scattering length as
an input parameter with a coordinate space cut-off
$r_c = 0.3\,{\rm fm}$.
(Lower panel) $^1P_1$, $^1F_3$ and $^1H_5$ phase shifts computed from
Eq.~(\ref{eq:1P1-1F3-relation}), using the $^1P_1$ scattering length as
an input parameter with a coordinate space cut-off
$r_c = 0.3\,{\rm fm}$.
The light blue band is generated by varying the cut-off radius within
the range $r_c = 0.6-0.8\,{\rm fm}$.
}
\label{fig:singlet-NNLO-DD}
\end{figure*}

The procedure for correlating the different partial waves considered is similar
to the one employed in constructing the finite energy solutions, the only
difference being the rotation to the basis in which the tensor force
is diagonal.
For the two sets of correlated coupled channels,
those with $j=1,3,5$ and those with $j=2,4$,
we have the boundary conditions
\begin{eqnarray} 
\label{eq:3C1-3C3-3C5-relation}
{\bf R_1}\,{\bf L}_{k,1}(r_c)\,{\bf R_1}^{\rm T} &=&
{\bf R_3}\,{\bf L}_{k,3}(r_c)\,{\bf R_3}^{\rm T} =
{\bf R_5}\,{\bf L}_{k,5}(r_c)\,{\bf R_5}^{\rm T} \, , \nonumber \\ \\
\label{eq:3C2-3C4-relation}
{\bf R_2}\,{\bf L}_{k,2}(r_c)\,{\bf R_2}^{\rm T} &=&
{\bf R_4}\,{\bf L}_{k,4}(r_c)\,{\bf R_4}^{\rm T} \, ,
\end{eqnarray}
from which the ${\bf M}_{k,j}(k)$ matrix (and the corresponding phase shifts)
can be obtained.

\subsection{Cut-off Dependence of the Phase Shifts}
\label{subsec:cutoff-dependence}

The cut-off dependence of the phase shifts in the correlated renormalization
procedure can be easily estimated by making use of the renormalization
group analysis of boundary condition regularization
of Ref.~\cite{PavonValderrama:2007nu}.
For simplicity, we only consider in detail the uncoupled channel case.
According to Ref.~\cite{PavonValderrama:2007nu}, the cut-off dependence
of the phase shift for an uncoupled channel is given by
\begin{eqnarray}
\frac{d\,\delta_l(k; r_c)}{d r_c} &=& \Big[ M_N\, V_{\rm NN}(r_c) - k^2 + 
\frac{l(l+1)}{r_c^2} \nonumber\\
&+& L_{k,l}'(r_c) + L_{k,l}^2(r_c) \Big]\,u_{k,l}^2(r_c) \, ,
\end{eqnarray}
where $\delta_l(k)$ is the phase shift, $r_c$ the cut-off radius,
and with $u_{k,l}$, $V_{\rm NN}$, $k$, $l$ and $M_N$ as defined
in Eq.~(\ref{eq:schroe-NN}).
In the previous formula $L_{k,l}(r_c)$ is the logarithmic derivative of the
$u_{k,l}$ reduced wave function at the cut-off radius, i.e.
\begin{eqnarray}
L_{k,l}(r_c) = \frac{u_{k,l}'(r_c)}{u_{k,l}(r_c)} \, .
\end{eqnarray}
If we are correlating the $l_0$- and $l$-waves, we have for the logarithmic
derivatives at the cut-off radius $r_c$ that
\begin{eqnarray}
L_{k_0,l_0}(r_c) = L_{k,l}(r_c) \, ,
\end{eqnarray}
where the partial wave $l_0$ is taken to be the {\it base} wave,
i.e. the wave for which we have fixed the value of
the phase shift at $k = k_0$ (or the scattering length if $k_0 = 0$).
By taking into account that the reduced wave function $u_{k_0,l_0}(r)$
obeys the following Schr\"odinger equation
\begin{eqnarray}
-u_{k_0,l_0}''(r) &+& \left[ M_N V_{\rm NN}(r) + \frac{l_0 (l_0+1)}{r^2} \right]
\, u_{k_0,l_0}(r) 
\nonumber \\
&=& k_0^2\,u_{k_0,l_0}(r) \, , 
\end{eqnarray}
it is trivial to check that the logarithmic boundary condition
for $k=k_0$, $l=l_0$ fulfills the differential equation
\begin{eqnarray}
M_N\,V_{\rm NN}(r_c) &+& \frac{l_0(l_0+1)}{r_c^2} - k_0^2 \nonumber \\ 
&+& L_{k_0,l_0}'(r_c) + L_{k_0,l_0}^2(r_c) = 0 \, ,
\end{eqnarray}
which is also the differential equation obeyed by $L_{k,l}(r_c)$.
In particular, the previous means that the cut-off dependence of the
phase shift simplifies to
\begin{eqnarray}
\label{eq:cut-off-dependence}
\frac{d\,\delta_l(k; r_c)}{d r_c} &=& \Big[ 
\frac{l(l+1)-l_0(l_0+1)}{r_c^2} \nonumber\\
&-& \left( k^2 - k_0^2 \right) \Big]\,u_{k,l}^2(r_c) \, .
\end{eqnarray}
For cut-off radii such that $2 m_{\pi} r_c \ll 1$, the behaviour of the wave
functions will be determined by the van der Waals piece of the interaction,
i.e. $u_{k,l}^2(r_c) \sim r_c^3$, up to oscillations (see Eq.~(\ref{eq:uk-uv}))
for the chiral ${\rm NLO}$- and ${\rm N^2LO}$-$\Delta$ potentials
of Ref.~\cite{Krebs:2007rh}.
This implies that the cut-off dependence of the phase shifts
can be approximated by
\begin{eqnarray}
\label{eq:cutoff-base}
\delta_{l_0}(k,r_c) - \delta_{l_0}(k,0) &\propto& - (k^2 - k_0^2)\,r_c^4 \, ,
\end{eqnarray}
for $l = l_0$ (that is, the {\it base} wave), and
\begin{eqnarray}
\label{eq:cutoff-derived}
\delta_{l}(k,r_c) - \delta_{l}(k,0) &\propto& \left[ l(l+1)-l_0(l_0+1) \right]
\,r_c^2 \, ,
\end{eqnarray}
for $l \neq l_0$.
At the end of Section~\ref{sec:num-res}, we illustrate
these expectations for the chiral $\Delta$-potentials
of Ref.~\cite{Krebs:2007rh}.
It should be noted that Eq.~(\ref{eq:cut-off-dependence}) implies that
the correlated renormalization procedure only generates converging
phase shifts if the potential $V_{\rm NN}$ is singular,
as expected from the discussion in Section~\ref{sec:central}.
The extension of the previous results to coupled channels is
straightforward and leads to the same conclusion and
cut-off dependence as the uncoupled channel case.

\section{Numerical Results}
\label{sec:num-res}

As we have shown we can relate the phase shifts in different
partial waves using the short range relation described by
Eqs.~(\ref{eq:1S0-1D2-relation}-\ref{eq:3D2-3G4-relation}).
We take in our numerical computations 
$f_\pi=92.4 {\rm MeV}$, $m_\pi=138.03 {\rm MeV}$,  
$2 \mu_{np}= M_N = 2M_p M_n /(M_p+M_n) = 938.918 {\rm MeV}$, 
$g_A =1.29$ in the OPE piece to account for the Goldberger-Treimann
discrepancy and $g_A=1.26$ in the TPE piece of the potential.
The discussion of the standard OPE potential corresponds to the
attractive-repulsive case and is relegated to Appendix~\ref{app:OPE}.
We discuss here the TPE chiral potential with $\Delta$ excitations as
obtained from Ref.~\cite{Krebs:2007rh} (however with the spectral
cut-off removed). For $h_A$, the chiral couplings $c_1$, $c_3$ and
$c_4$ and $\tilde{b} = b_3 + b_8$ we take the values corresponding to
``Fit 1'' of Ref.~\cite{Krebs:2007rh} (see table I inside the previous
reference).

All the partial waves are renormalized at a cut-off radius
$r_c = 0.3\,{\rm fm}$.
This cut-off is small enough by far: in most partial waves
the phase shifts have already converged in the range
$r_c = 0.6-0.8\,{\rm fm}$.
Smaller cut-off radii are in principle possible, but require too much
computing time for the higher partial waves,
while cut-off radii larger than $0.8-1.0\,{\rm fm}$ yield amplitudes which
depend linearly on the cut-off for low partial waves.
The appearance of the first deeply bound state usually happens
in the $0.5-1.0\,{\rm fm}$ region,
the exact location depending on the particular partial wave considered.
Even in the case of G- and H-waves there are usually between two and three
deeply bound states at $r_c = 0.3\,{\rm fm}$.
At these distances the wave functions are dominated
by the van der Waals behaviour of the ${\rm NLO}-{\Delta}$ and
${\rm N^2LO}-{\Delta}$ potentials, meaning that the correlated
renormalization procedure is guaranteed to work.
This does not imply however that the low energy phase shifts are
dominated by the singular structure of the chiral potentials
at distances below the pion Compton wavelength.
In fact, as will be commented in the following paragraphs,
the results for peripheral waves do not significantly differ
from those computed in first order perturbation theory~\cite{Krebs:2007rh}.
The explicit cut-off dependence of the phase shifts is discussed
in more detail for selected partial waves at the end of
this section.

In Fig.~(\ref{fig:singlet-NNLO-DD}), we show the results for the singlet
waves.
For the $^1S_0$-$^1D_2$-$^1G_4$ ($^1P_1$-$^1F_3$-$^1H_5$) correlation
we have taken as input parameter the $^1S_0$ ($^1P_1$) scattering length
from the Nijmegen II potential~\cite{Stoks:1994wp},
which was computed in Ref.~\cite{PavonValderrama:2005ku}
yielding the result
$a_{^1S_0} = -23.727\,{\rm fm}$ ($a_{^1P_1} = 2.797\,{\rm fm}^3$).
For the $^1G_4$ phase in the isovector channels and the $^1F_3$ 
wave in the isosinglet, the phase shifts do not differ much
from those obtained in Ref.~\cite{Krebs:2007rh}
in the Born approximation (in the previous reference
only waves with $l=2,3,4$ were considered).
The $^1H_5$ is also very similar to the phases obtained in
Refs.~\cite{Kaiser:1997mw,Kaiser:1998wa}
for the ${\rm NLO}$-${\Delta}$ potential.
These waves are also quite similar to those obtained
in Ref.~\cite{PavonValderrama:2005uj}
by renormalizing the ${\rm N^2LO}$ potential for the $\Delta$-less theory
in a wave-by-wave basis, that is, by fixing the scattering lengths separately
in each of the channels to their Nijmegen II values.
In general, peripheral partial waves will not notice too much
the inclusion of the two pion exchange interaction or the $\Delta$ excitation
and will behave very similarly as in first order perturbation theory.
In this regard, the partial wave correlation is useful mainly as a way
to renormalize all the peripheral waves with a minimum number of
counterterms, but not necessarily as a real correlation.
The only wave in which it can be effectively noticed is in the $^1D_2$ one,
in which the ${}^1S_0-{}^1D_2$ correlation predicts a scattering length
of $a_{^1D_2} = -1.728\,{\rm fm}^5$ for the $^1D_2$ wave~\footnote{
The quoted scattering lengths have been computed for the fixed cut-off radius
$r_c = 0.3\,{\rm fm}$ for the ${\rm N^2LO}$-$\Delta$ case
and are accurate within the numerical error.
The systematic uncertainty of taking a cut-off between $0.3$ and $0.6\,{\rm fm}$
typically only influences the last digit (in the particular case
of the $^1D_2$ scattering length we have a $0.005\,{\rm fm^3}$
change in the previous cut-off window). 
Note that a factor of two in the cut-off corresponds to doubling
the momentum space cut-off. 
Actually, using the ``equivalence'' $\Lambda = \pi /2 r_c$~\cite{Entem:2007jg,Valderrama:2007ja},
we are testing the $\Lambda = 0.5-1\,{\rm GeV}$ region which seems reasonable.
},
to be compared with an optimal scattering length of
$a_{^1D_2} = -1.686\,{\rm fm}^5$
for which the ${\rm N^2LO}$-$\Delta$ potential effectively reproduces
the Nijmegen II results for $E_{\rm LAB} \leq 150\,{\rm MeV}$.
The previous observation indicates the necessity of the specific $^1D_2$ wave
${\rm N^3LO}$ counterterm in order to reproduce the results
in this partial wave.
The predicted value greatly differs from the one corresponding to the
Nijmegen II or Reid93 potentials~\cite{Stoks:1994wp},
namely $a_{^1D_2,\rm Nijm2} = -1.389\,{\rm fm}^5$
and $a_{^1D_2,\rm Reid93} = -1.377\,{\rm fm}^5$,
which were computed in Ref.~\cite{PavonValderrama:2005ku}.
This discrepancy is however common in most effective field theory computations
in which the scattering length is fixed,
see for example Ref.~\cite{PavonValderrama:2005uj},
or the related comments in Refs.~\cite{Yang:2009kx,Yang:2009pn,Yang:2009fm},
where a subtractive regularization approach is employed.
This inconsistency between the Nijmegen low energy parameters and the chiral
$\Delta$ potentials is explained by the fact that the Nijmegen phenomenological
potentials do not contain either two pion exchange contributions or $\Delta$
excitations, and was briefly commented in the previous references.

\begin{figure*}[htb]
\begin{center}
\epsfig{figure=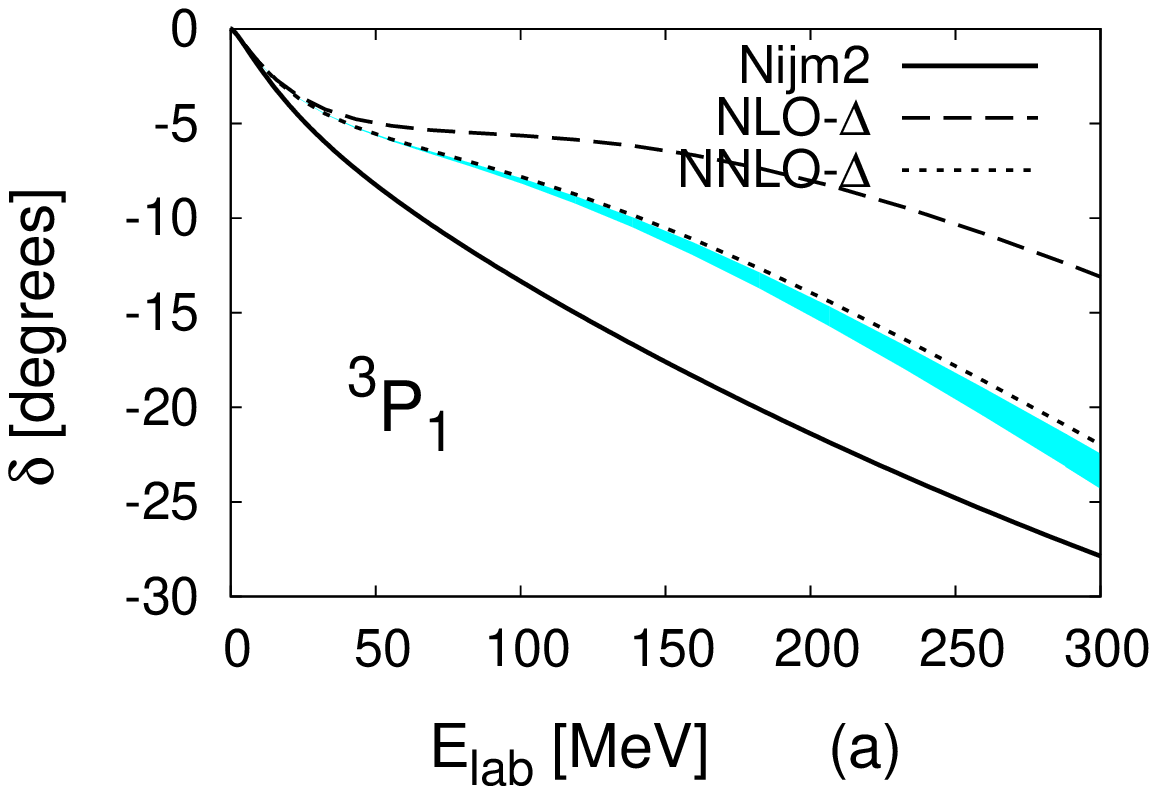, 
	height=5.0cm, width=5.5cm}
\epsfig{figure=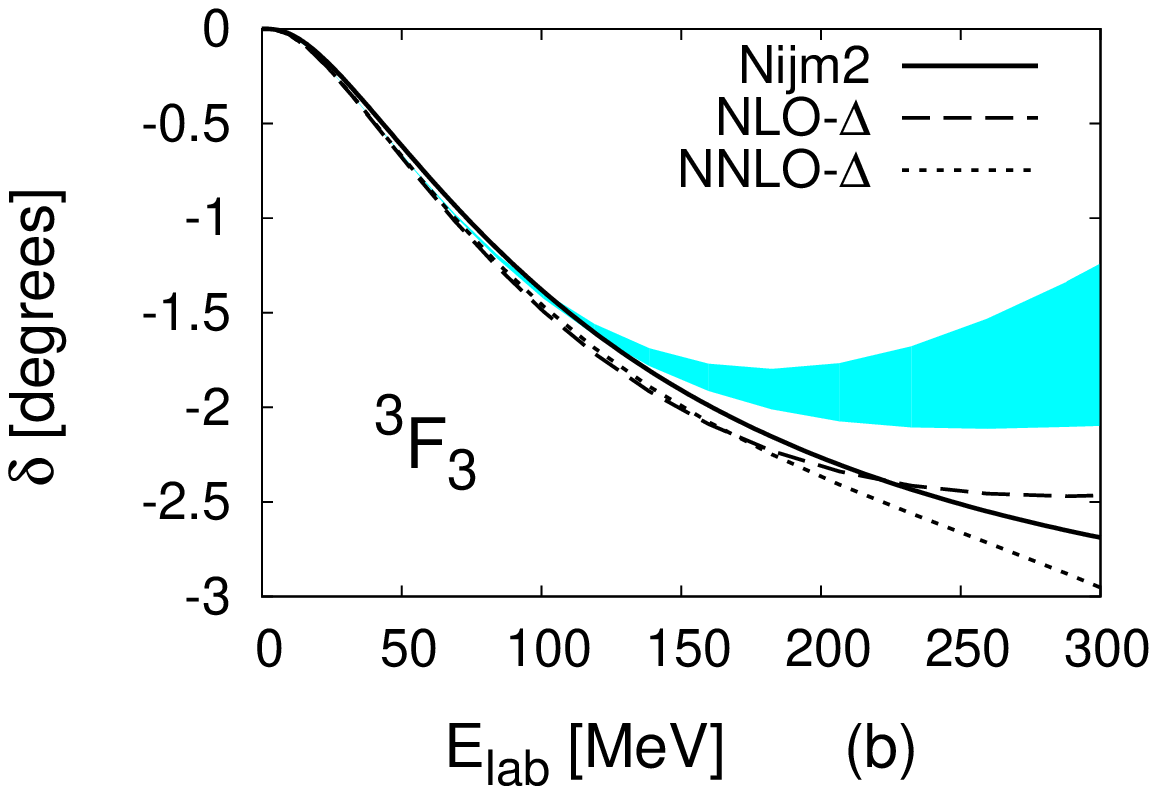, 
	height=5.0cm, width=5.5cm}
\epsfig{figure=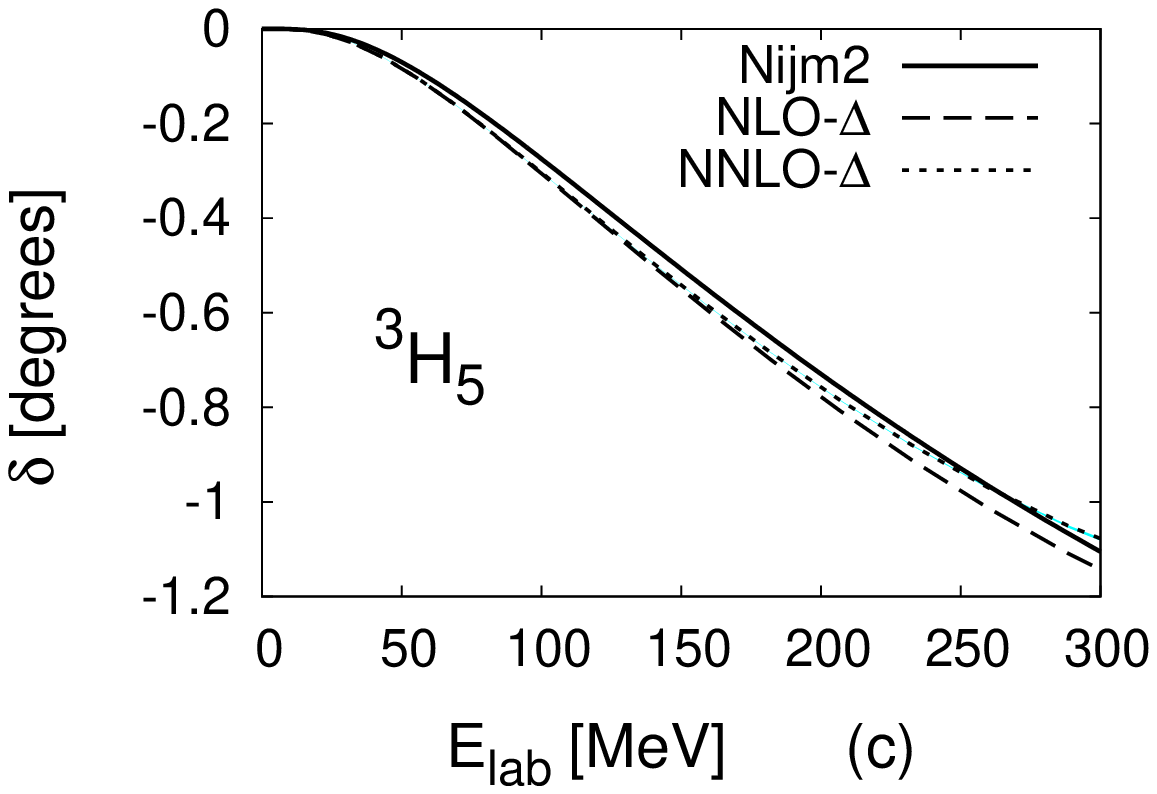, 
	height=5.0cm, width=5.5cm}
\epsfig{figure=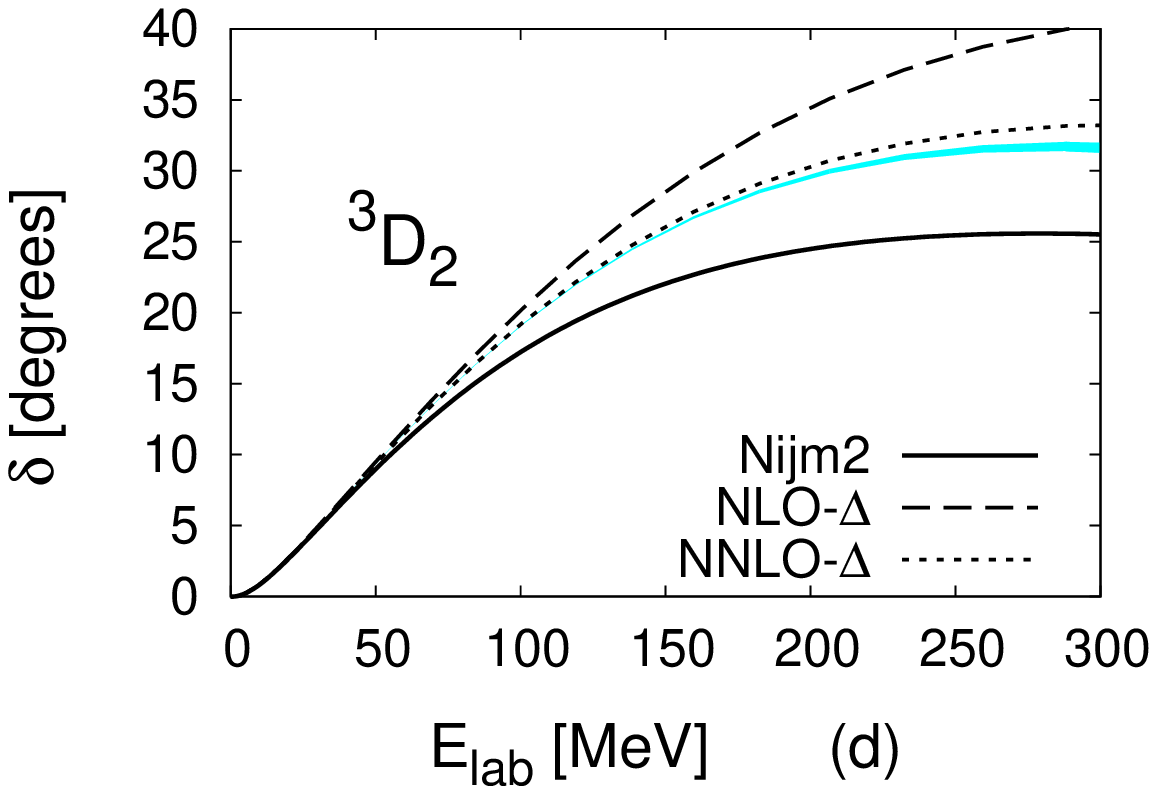, 
	height=5.0cm, width=5.5cm}
\epsfig{figure=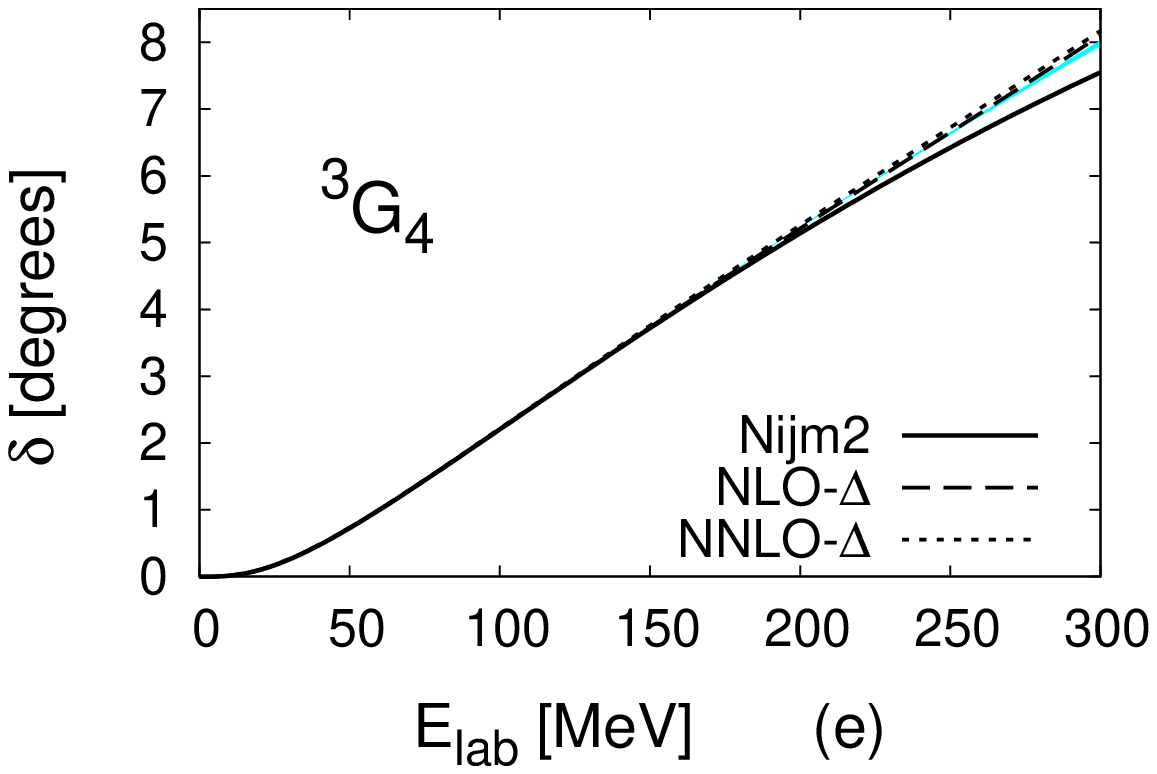, 
	height=5.0cm, width=5.5cm}
\end{center}
\caption{(Color online)
(Upper panel) $^3P_1$, $^3F_3$ and $^3H_5$ phase shifts computed from
Eq.~(\ref{eq:3P1-3F3-relation}), using the $^3P_1$ scattering length as
an input parameter with a coordinate space cut-off
$r_c = 0.3\,{\rm fm}$.
(Lower panel) $^3D_2$ and $^3G_4$ phase shifts computed from
Eq.~(\ref{eq:3D2-3G4-relation}), using the $^3D_2$ scattering length as
an input parameter with a coordinate space cut-off
$r_c = 0.3\,{\rm fm}$.
The light blue band is generated by varying the cut-off radius within
the range $r_c = 0.6-0.8\,{\rm fm}$.
}
\label{fig:uncoupled-triplet-NNLO-DD}
\end{figure*}

In Fig.~(\ref{fig:uncoupled-triplet-NNLO-DD}), we show the results for the 
uncoupled triplet waves.
We have taken as input parameter for the $^3P_1$ ($^3D_2$) correlation
the Nijmegen II scattering length~\cite{PavonValderrama:2005ku},
namely $a_{^3P_1} = 1.529\,{\rm fm}^3$
($a_{^3D_2} = -7.405\,{\rm fm}^5$).
Taking these values does not yield the better possible results for the
$^3P_1$ ($^3D_1$) wave, but generates renormalized results for the
$^3F_3-{}^3H_5$ ($^3G_4$) waves.
As it happened in the singlet case, the phase shifts for the higher
partial waves are very similar to the values obtained in first order
perturbation theory either in the $\Delta$-less~\cite{Kaiser:1997mw}
and $\Delta$-full~\cite{Kaiser:1998wa,Krebs:2007rh} cases.

\begin{figure*}[htb]
\begin{center}
\epsfig{figure=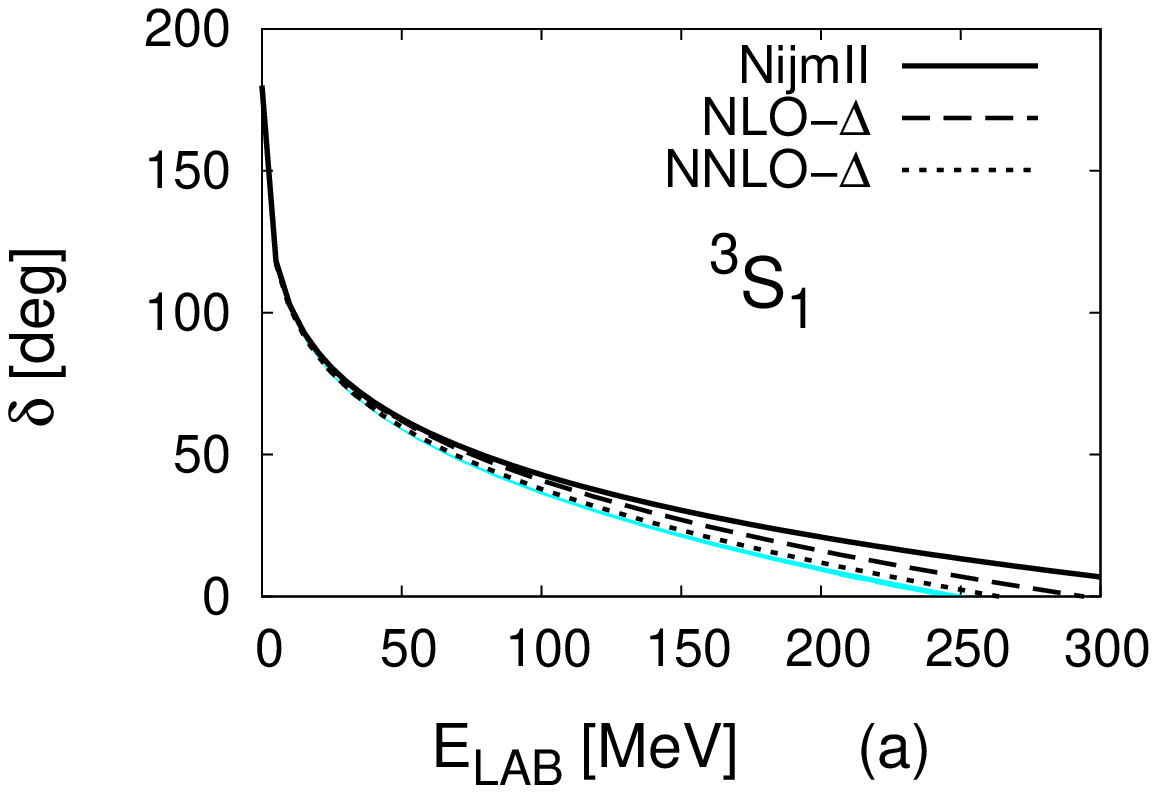, 
	height=5.0cm, width=5.5cm}
\epsfig{figure=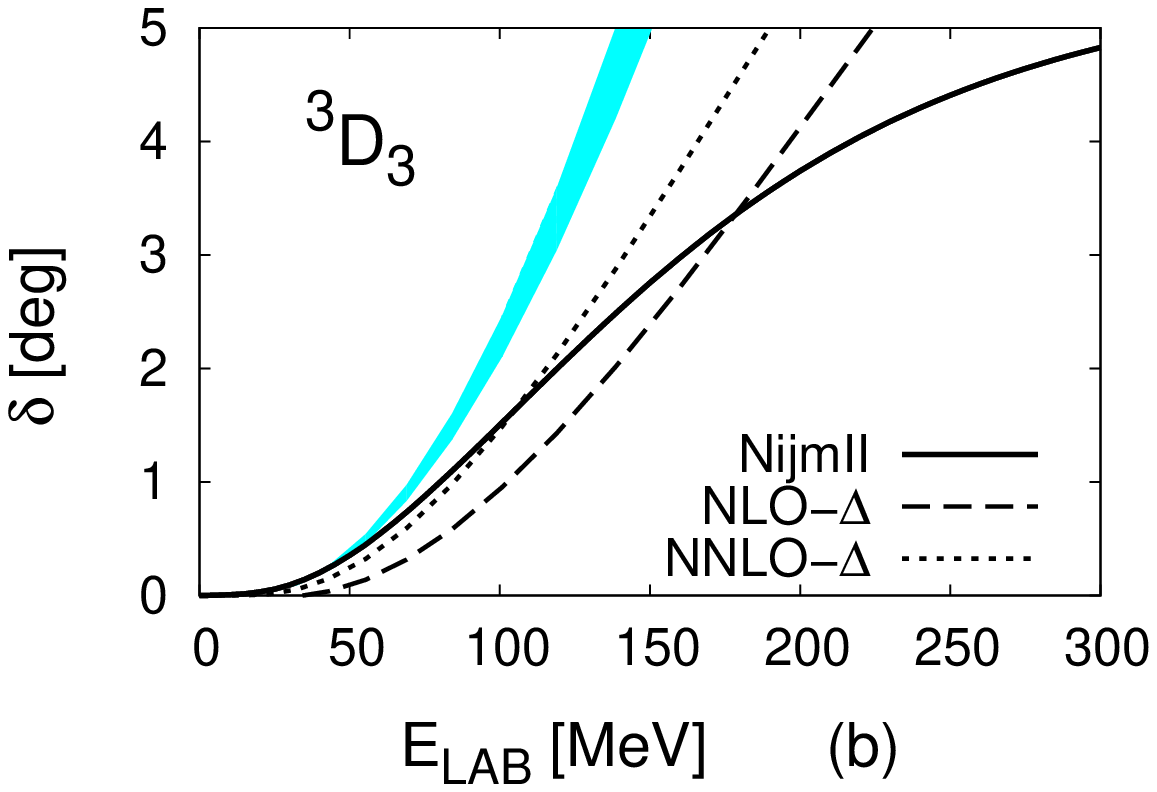, 
	height=5.0cm, width=5.5cm}
\epsfig{figure=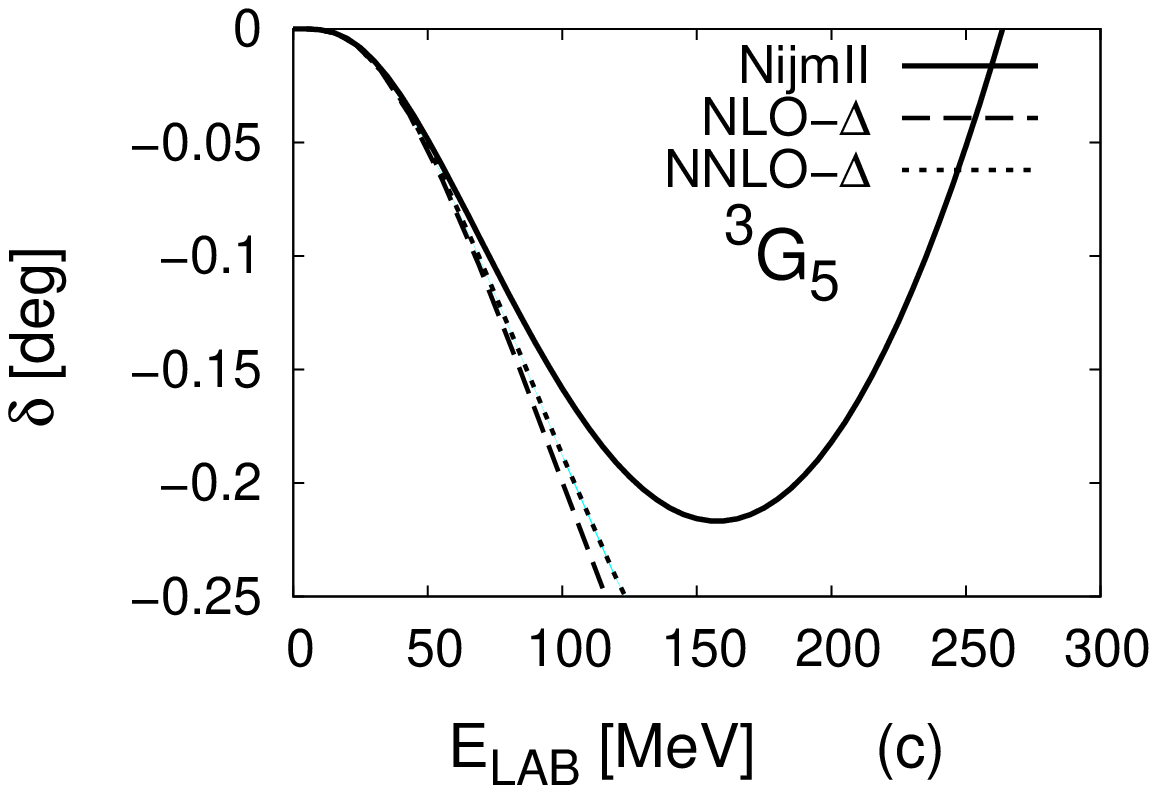, 
	height=5.0cm, width=5.5cm}
\epsfig{figure=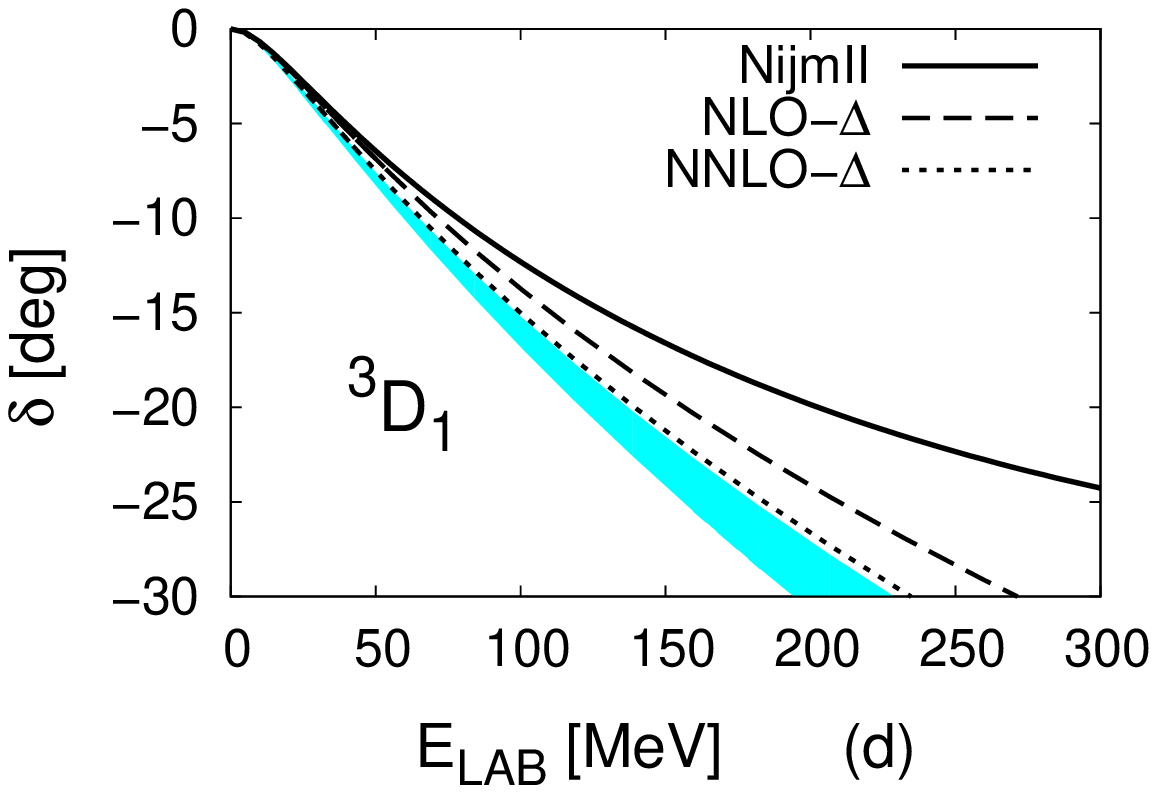, 
	height=5.0cm, width=5.5cm}
\epsfig{figure=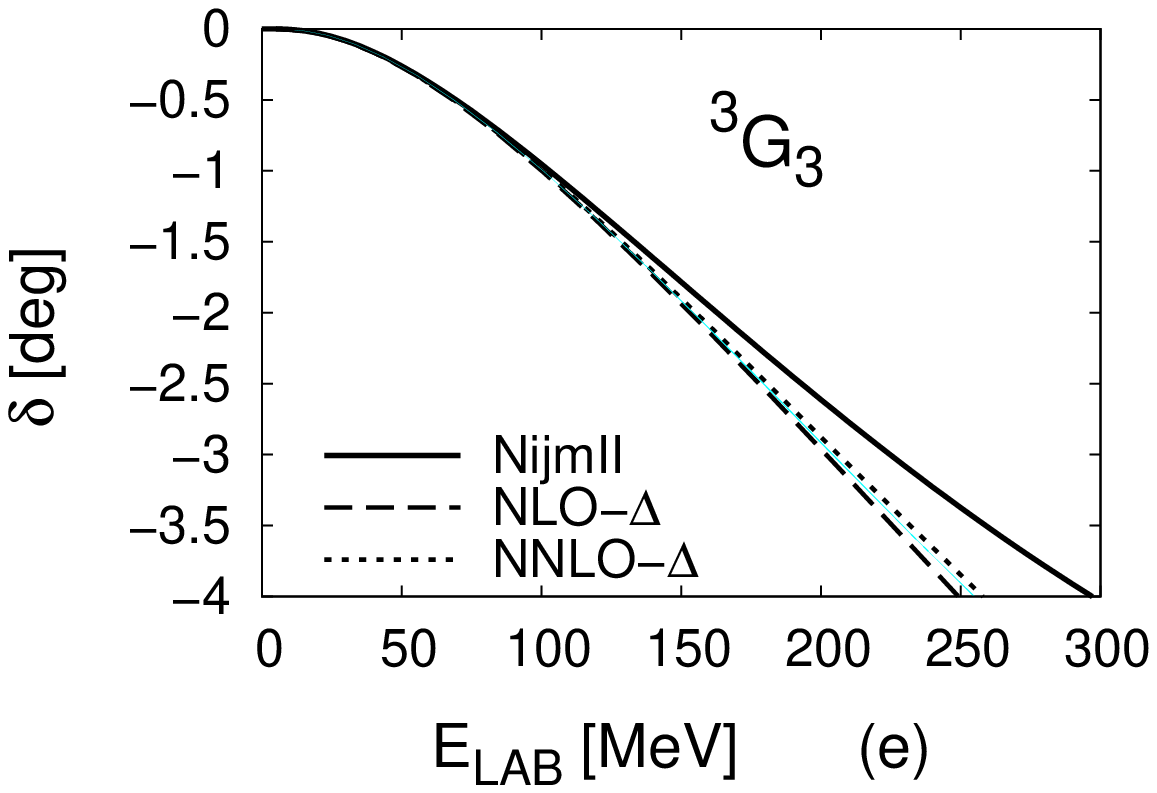, 
	height=5.0cm, width=5.5cm}
\epsfig{figure=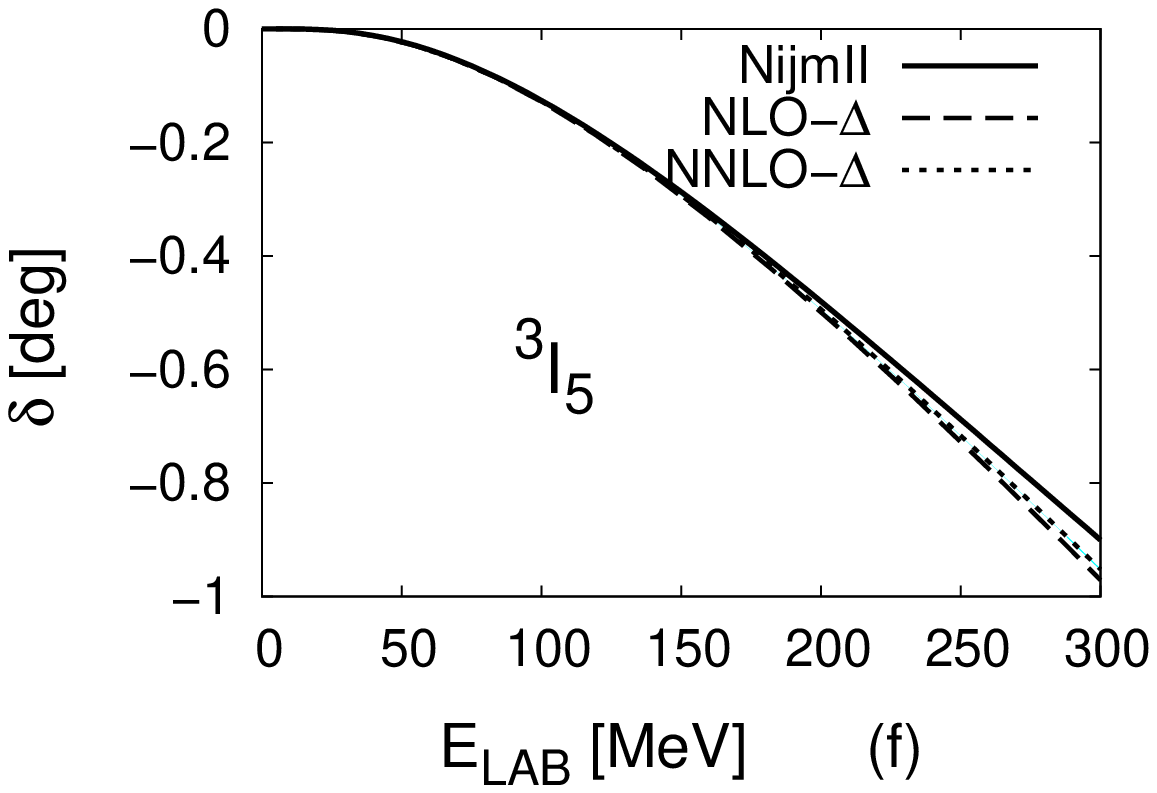, 
	height=5.0cm, width=5.5cm}
\epsfig{figure=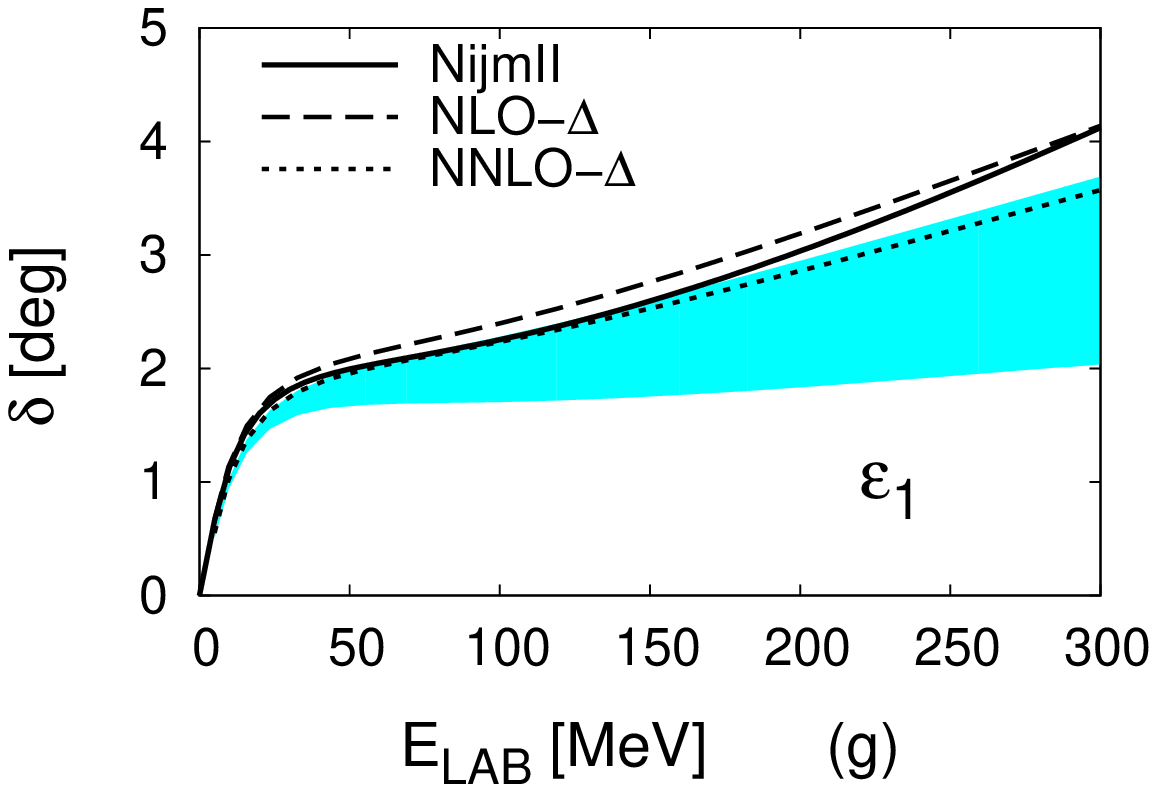, 
	height=5.0cm, width=5.5cm}
\epsfig{figure=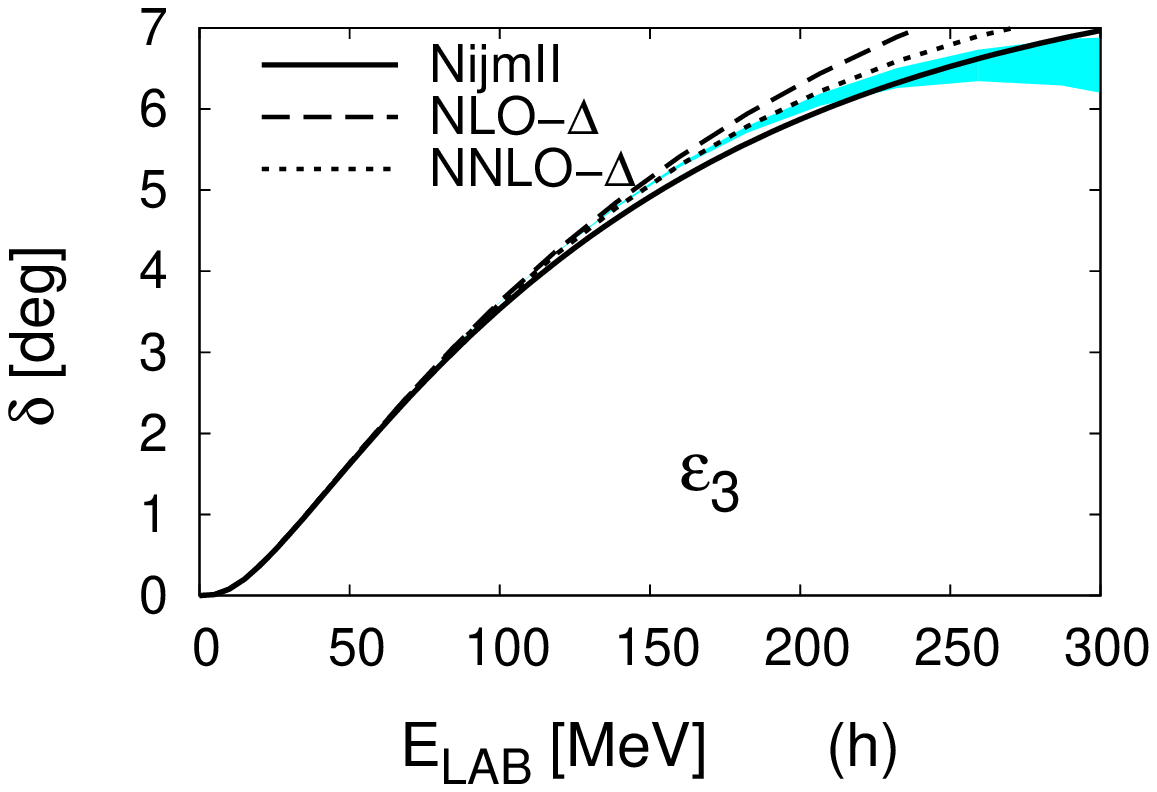, 
	height=5.0cm, width=5.5cm}
\epsfig{figure=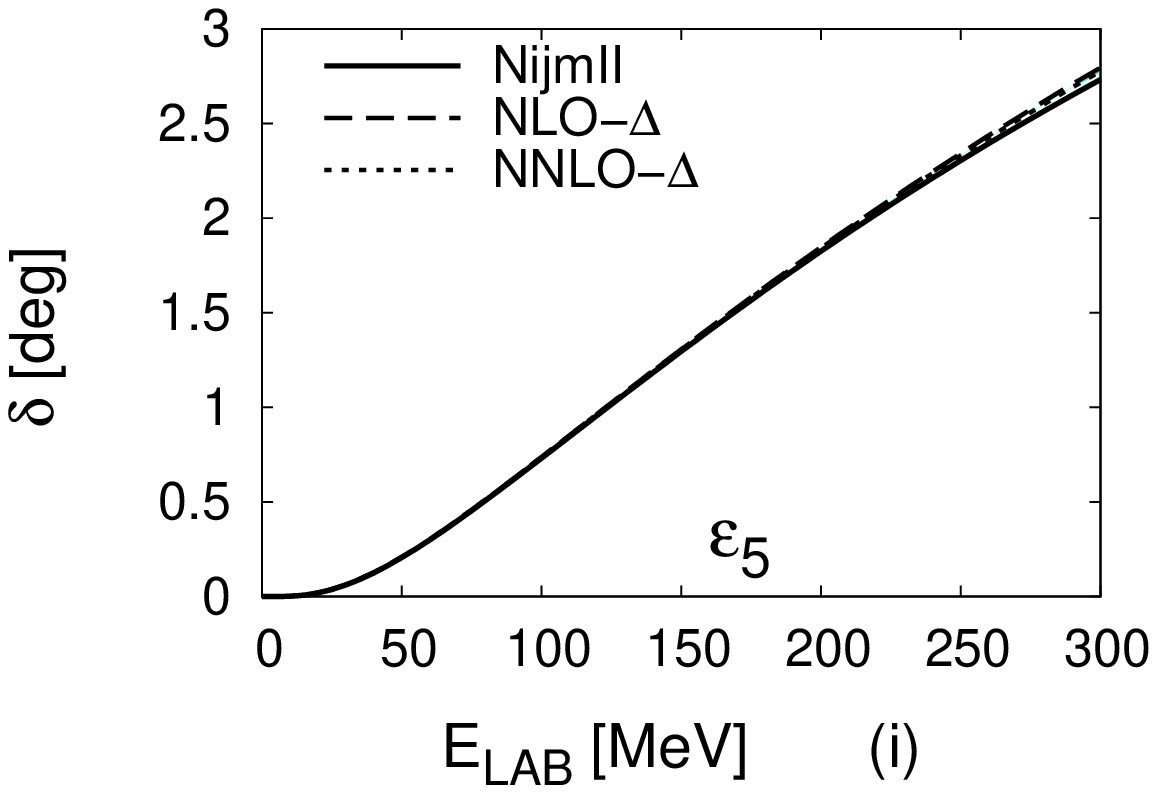, 
	height=5.0cm, width=5.5cm}
\end{center}
\caption{(Color online) $^3S_1-{}^3D_1$, $^3D_3-{}^3G_3$ and $^3G_5-{}^3I_5$
coupled channel phase shifts.
The $^3S_1-{}^3D_1$ wave is computed from orthogonality to the deuteron
bound state and from the triplet scattering length
$a_{0,t} = 5.419\,{\rm fm}$.
The $^3D_3-{}^3G_3$ and $^3G_5-{}^3I_5$ coupled channels are
computed from the partial wave correlation given by
Eq.~(\ref{eq:3C1-3C3-3C5-relation})
{\it without} introducing new counterterms.
We use the same cut-off values as in Figs.~(\ref{fig:singlet-NNLO-DD})
and (\ref{fig:uncoupled-triplet-NNLO-DD}).
}
\label{fig:coupled-triplet-deuteron-NNLO-DD}
\end{figure*}

\begin{figure*}[htb]
\begin{center}
\epsfig{figure=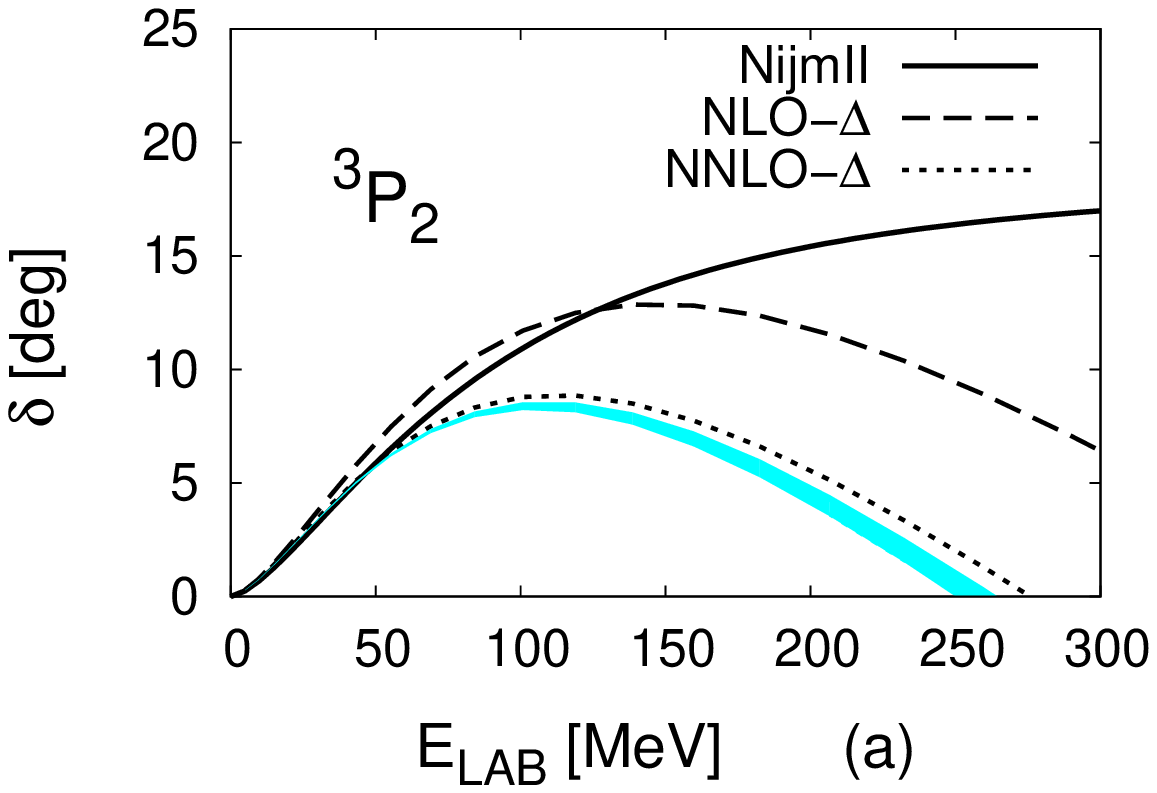, 
	height=5.0cm, width=5.5cm}
\epsfig{figure=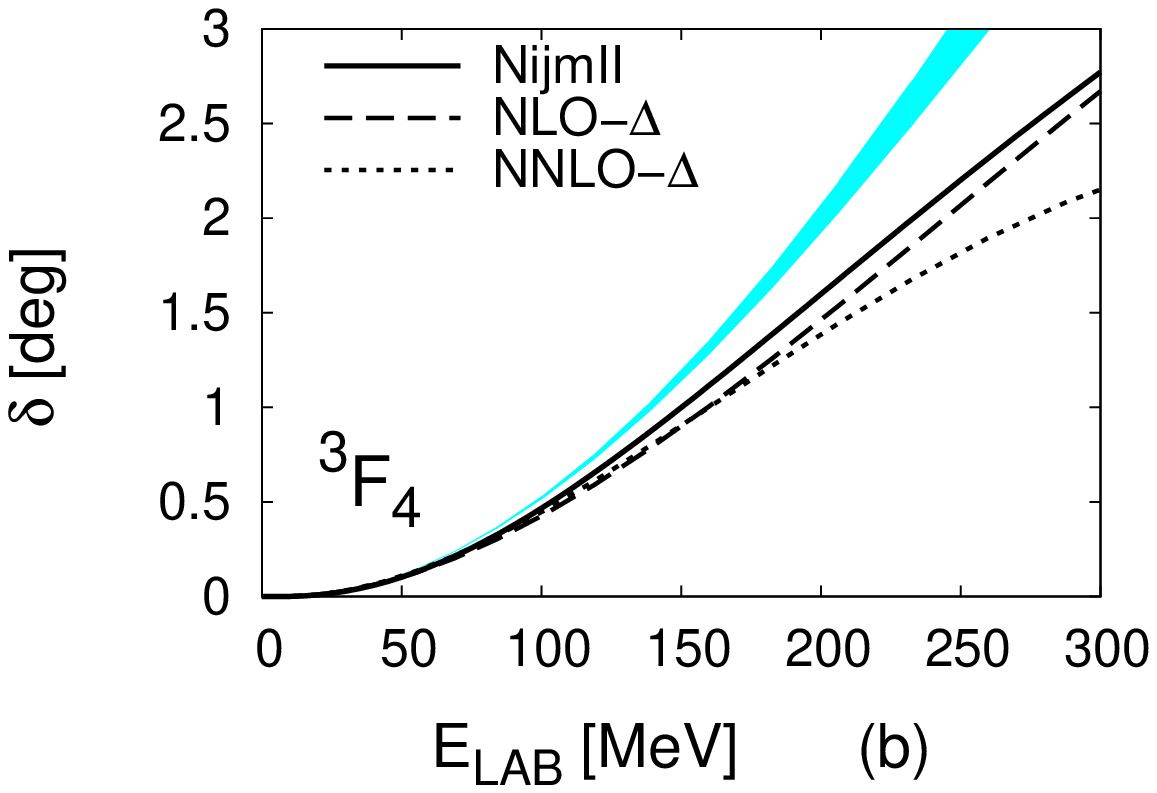, 
	height=5.0cm, width=5.5cm} \\
\epsfig{figure=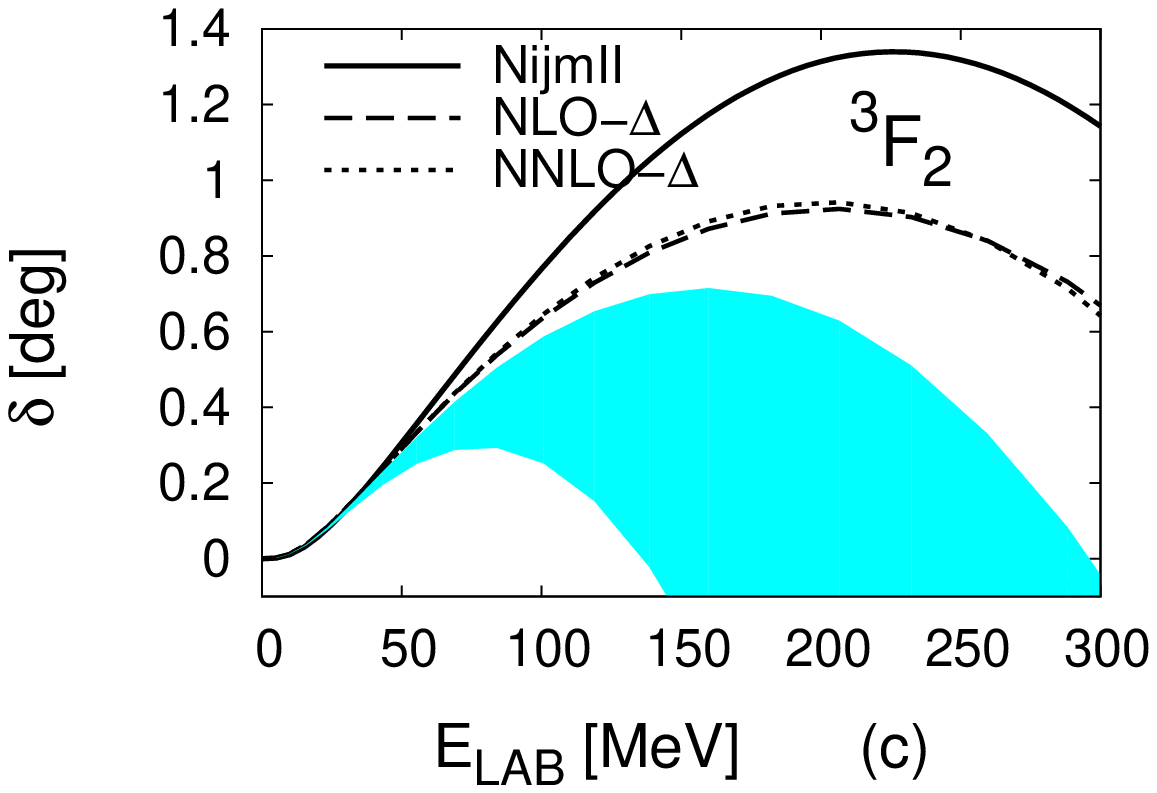, 
	height=5.0cm, width=5.5cm}
\epsfig{figure=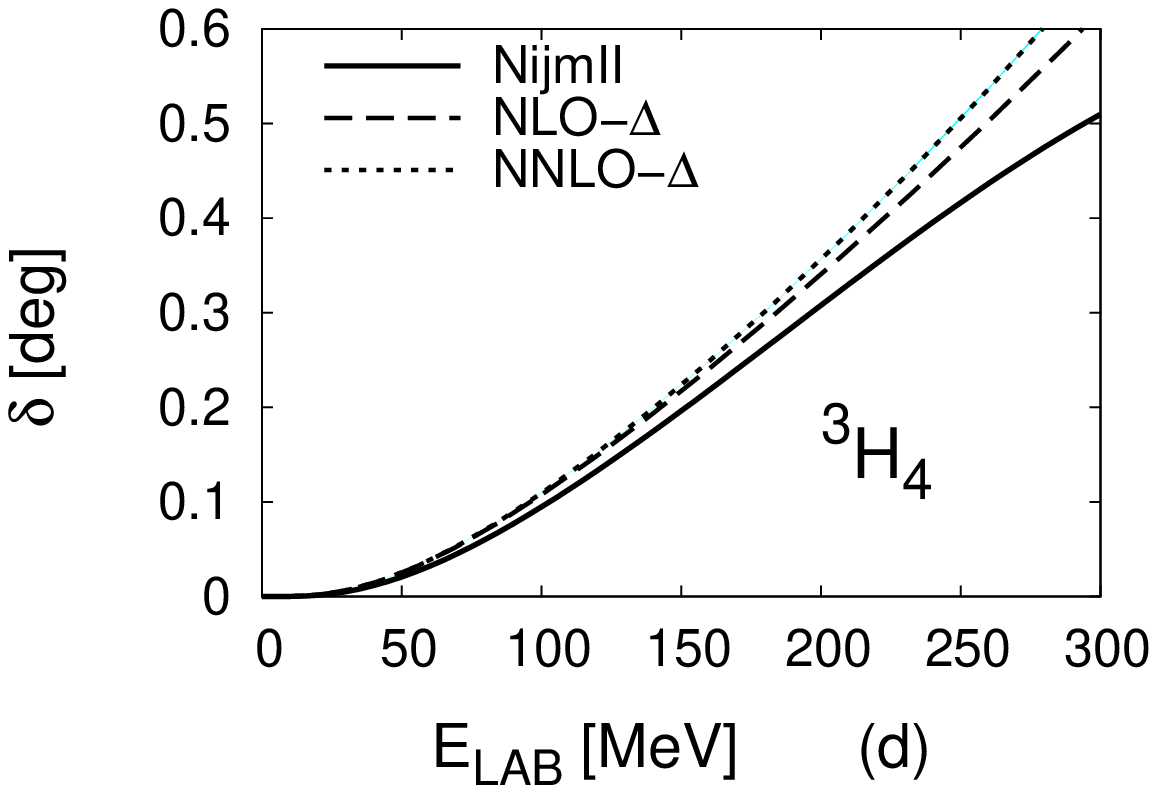, 
	height=5.0cm, width=5.5cm} \\
\epsfig{figure=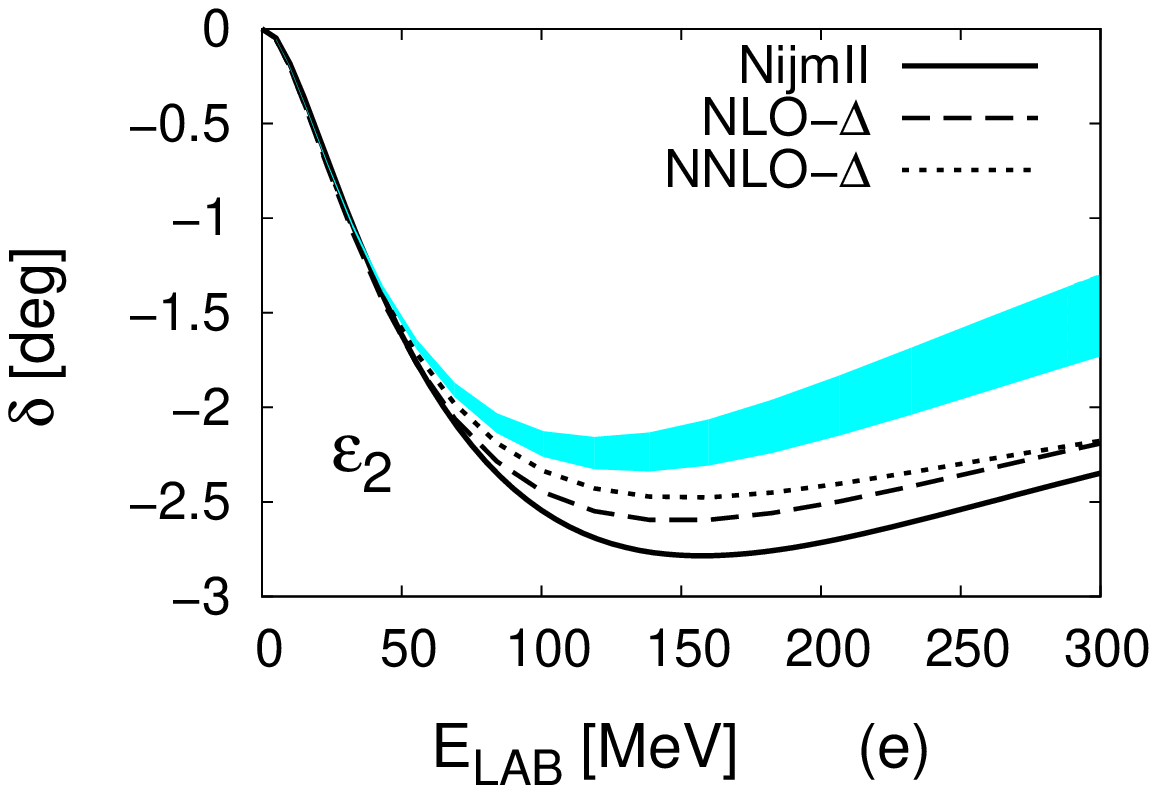, 
	height=5.0cm, width=5.5cm}
\epsfig{figure=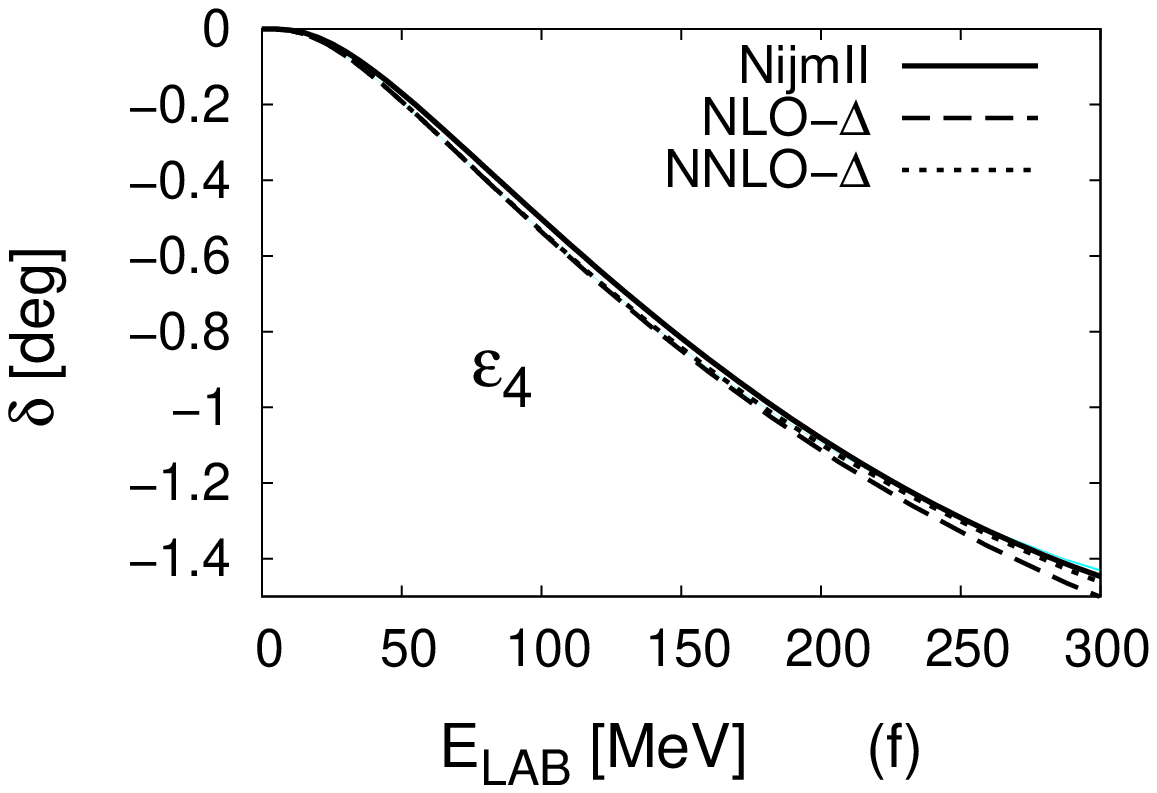, 
	height=5.0cm, width=5.5cm}
\end{center}
\caption{(Color online) $^3P_2-{}^3F_2$ and $^3F_4-{}^3H_4$ phase shifts.
The $^3P_2-{}^3F_2$ wave is constructed to reproduce the Nijmegen II scattering
lengths for this coupled channel, while the $^3F_4-{}^3H_4$ wave is obtained
from the partial wave correlation described in Eq.~(\ref{eq:3C2-3C4-relation}) {\it without} introducing new counterterms.
The values of the coordinate space cut-offs are the same
as in Figs.~(\ref{fig:singlet-NNLO-DD})
and (\ref{fig:uncoupled-triplet-NNLO-DD}).
}
\label{fig:coupled-triplet-NNLO-DD}
\end{figure*}

In Figs.~(\ref{fig:coupled-triplet-deuteron-NNLO-DD}) and 
(\ref{fig:coupled-triplet-NNLO-DD})
we show the results for the coupled triplet waves.
For the $^3S_1-{}^3D_1$ correlation, we have taken as input parameters the
deuteron binding energy $B_d = 2.224575\,{\rm MeV}$ and D/S asymptotic ratio
$\eta = 0.0256$, and the $^3S_1$ scattering length
$a_{^3S_1} = 5.419\,{\rm fm}$.
The scattering solutions are then constructed by orthogonality with respect
to the deuteron wave function and the $^3S_1$ scattering state.
The procedure is described in detail in Ref.~\cite{Valderrama:2005wv}, and
was already used in Ref.~\cite{Valderrama:2008kj} to construct the
scattering solutions in the $^3S_1-{}^3D_1$ channel for the $\Delta$
potentials of Ref.~\cite{Krebs:2007rh}.
For $r_c = 0.3\,{\rm fm}$, we obtain the values
$a_{E_1} = 1.953\,{\rm fm}^3$ and $a_{^3D_1} = 5.034\,{\rm fm}^5$
for the scattering lengths.
The previous low energy information yields much better results than the
use of the Nijmegen scattering length for this channels
($a_{E_1} = 1.647\,{\rm fm}^3$ and $a_{^3D_1} = 6.505\,{\rm fm}^5$), 
which induce a spurious resonance at $k_{\rm cm} \simeq 100\,{\rm MeV}$
when the ${\rm N^2LO}$-$\Delta$ potentials are employed.
This behaviour can also happen when using the standard chiral potentials
without explicit $\Delta$ degrees of freedom, as was discussed
in Refs.~\cite{Yang:2009kx,Yang:2009pn,Yang:2009fm}.
The corresponding phase shift for the $^3D_3$ is slightly better than the
one obtained in Ref.~\cite{Krebs:2007rh}, while the ${}^3G_3$ phase
and the $\epsilon_3$ mixing parameter are quite similar to
the ones obtained in the previous reference.
The results for the $^3G_5-{}^3I_5$ coupled channel are good in general
with the exception of the $^3G_5$ phase in which only the threshold
behaviour is correctly reproduced.

In the case of the $^3P_2-{}^3F_2$ waves, we renormalize these waves by fixing
the scattering lengths to the values
$a_{^3P_2} = -0.320\,{\rm fm}^3$, $a_{E_2} = 1.936\,{\rm fm}^5$ and
$a_{^3F_2} = -1.289\,{\rm fm}^7$,
which provide an acceptable description of the phase shifts for this channel,
see Fig.~(\ref{fig:coupled-triplet-NNLO-DD}).
As happened in the $^3S_1-{}^3D_1$ channel, the Nijmegen II values for the
scattering lengths~\cite{PavonValderrama:2005ku}
do not yield good results with the ${\rm NLO}-{\Delta}$
and ${\rm N^2LO}-{\Delta}$ potentials.
However, in the ${\Delta}$-less theory
the Nijmegen II scattering lengths generated good
results at ${\rm N^2LO}$~\cite{PavonValderrama:2005uj},
meaning that the discrepancy is due to the long range physics
introduced by the $\Delta$ excitations.
The $^3F_2$ phases show a strong relative cut-off dependence in the range
$r_c = 0.6-0.8\,{\rm fm}$, although this is partly due to the small value
of this phase.
As in the previous cases, the $^3F_4-{}^3H_4$ waves are very similar to those
of Ref.~\cite{Krebs:2007rh}.

\begin{figure*}[htb]
\begin{center}
\epsfig{figure=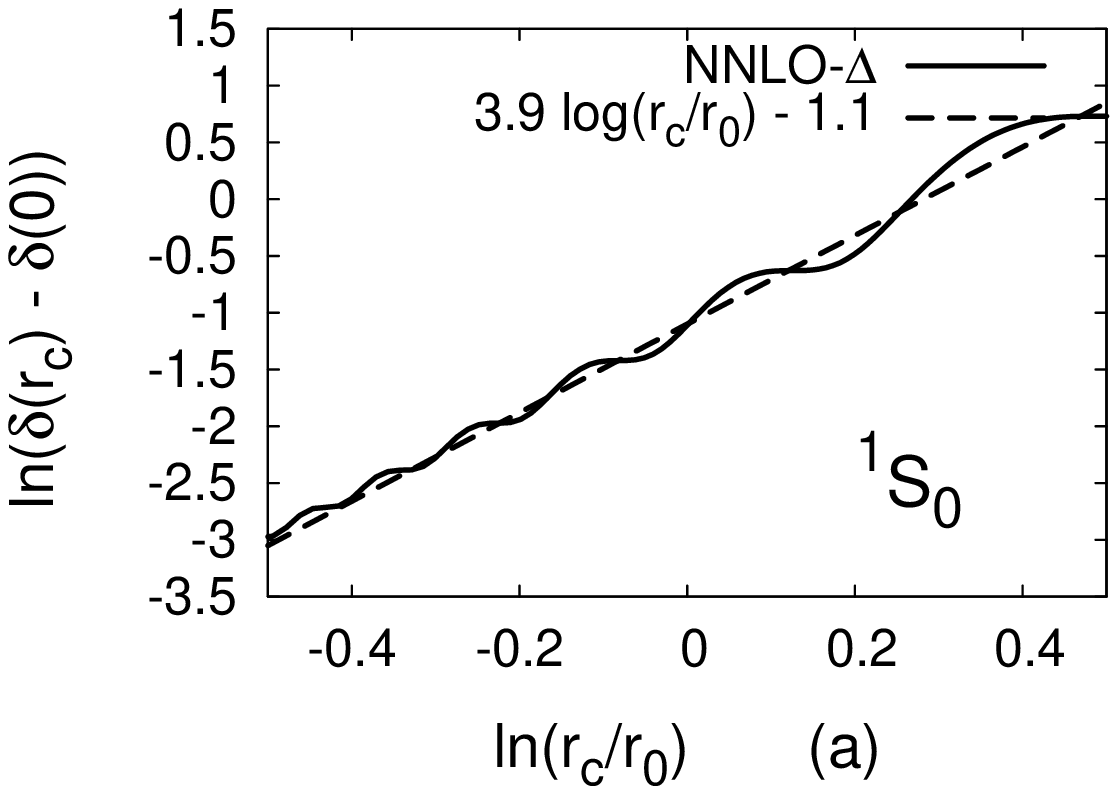, 
	height=5.0cm, width=5.5cm}
\epsfig{figure=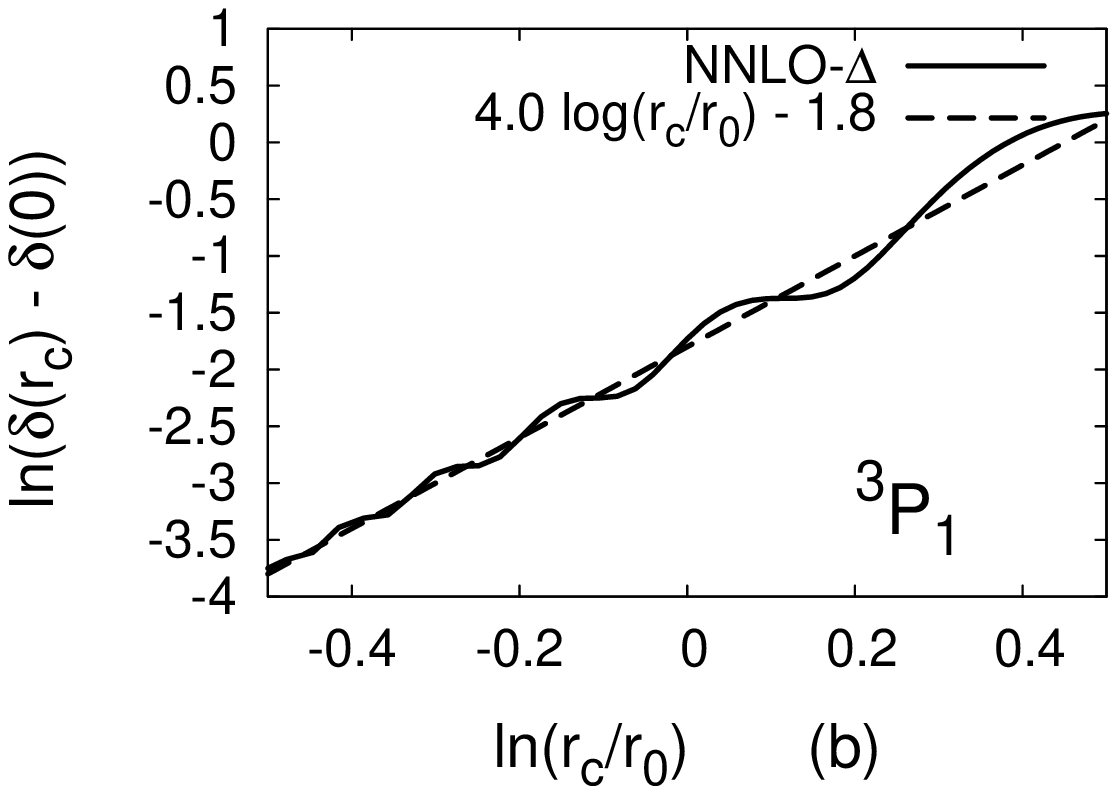, 
	height=5.0cm, width=5.5cm} \\
\epsfig{figure=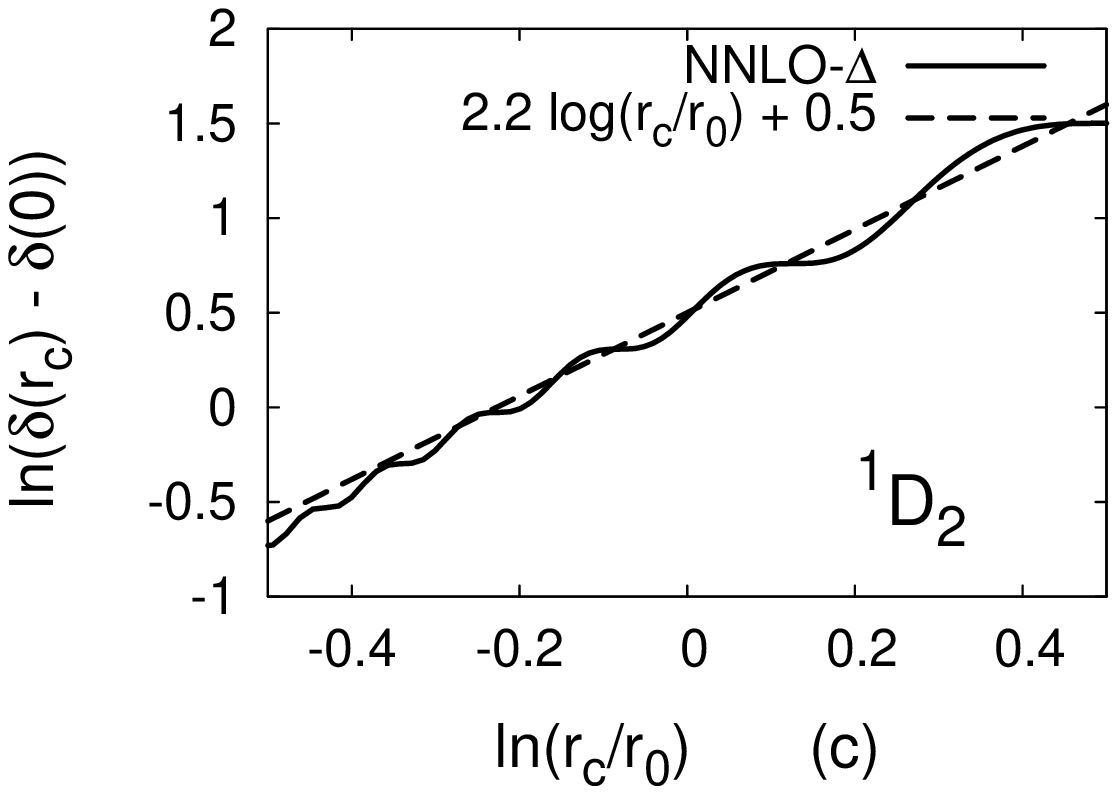,  
	height=5.0cm, width=5.5cm}
\epsfig{figure=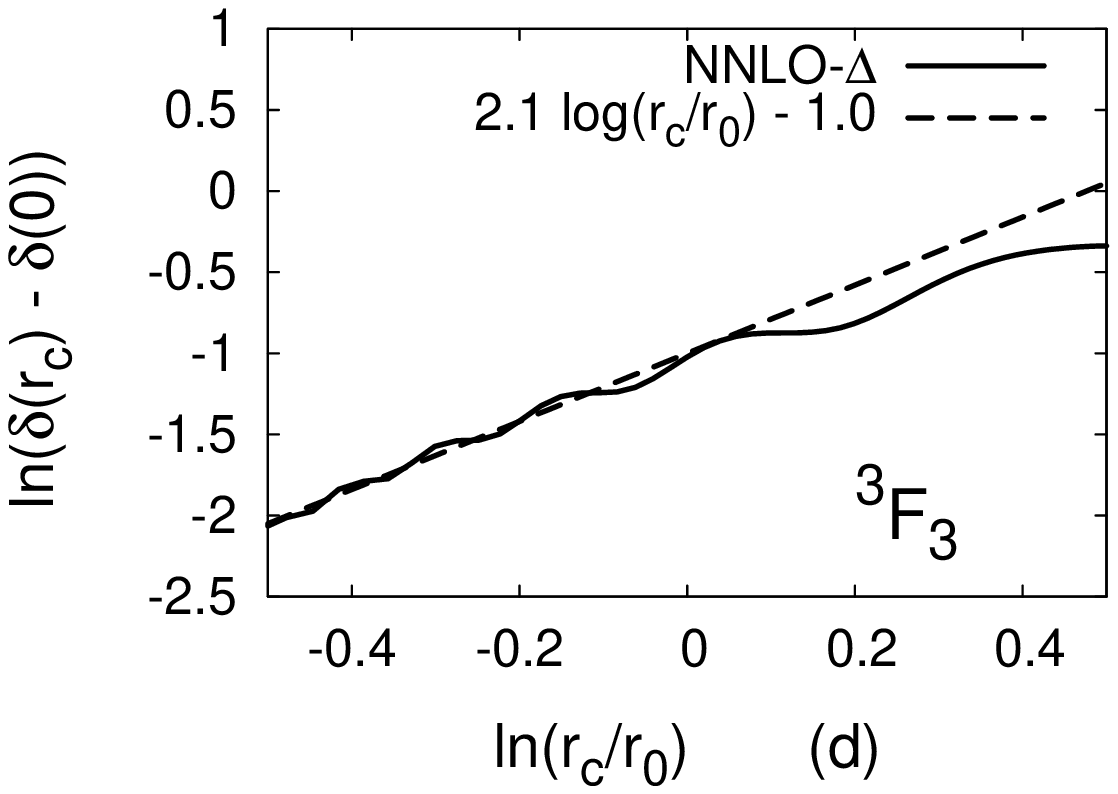, 
	height=5.0cm, width=5.5cm} \\
\epsfig{figure=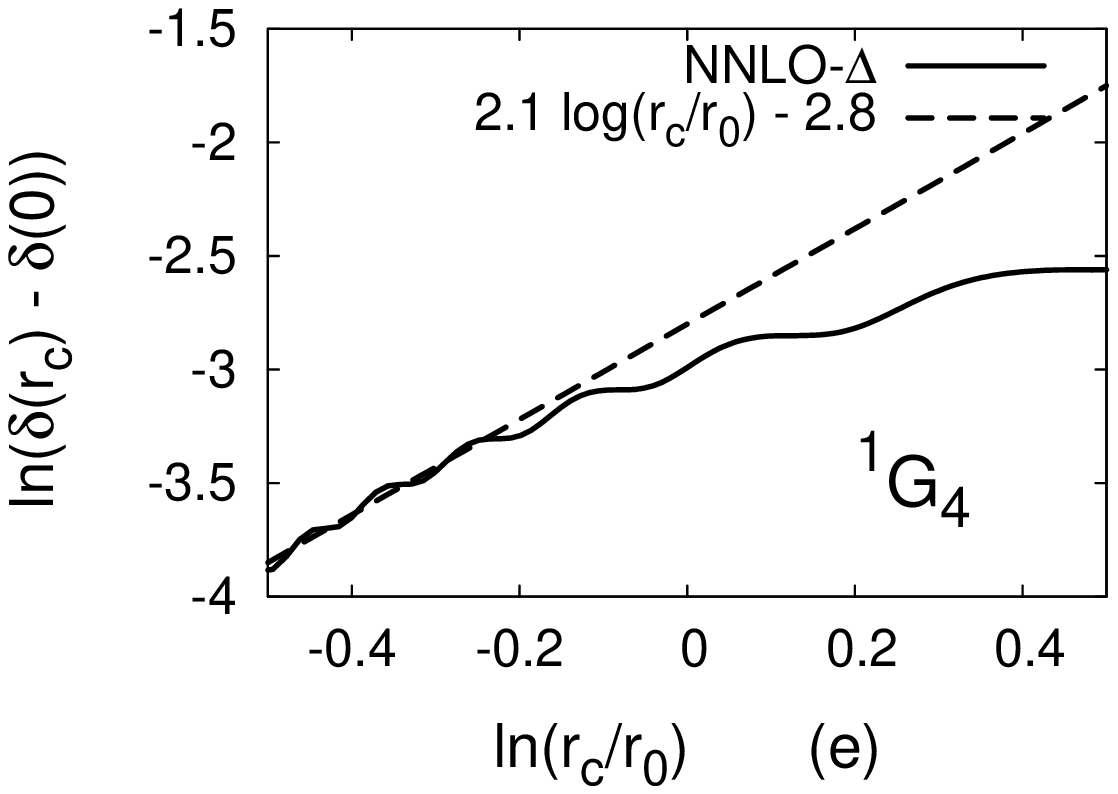, 
	height=5.0cm, width=5.5cm}
\epsfig{figure=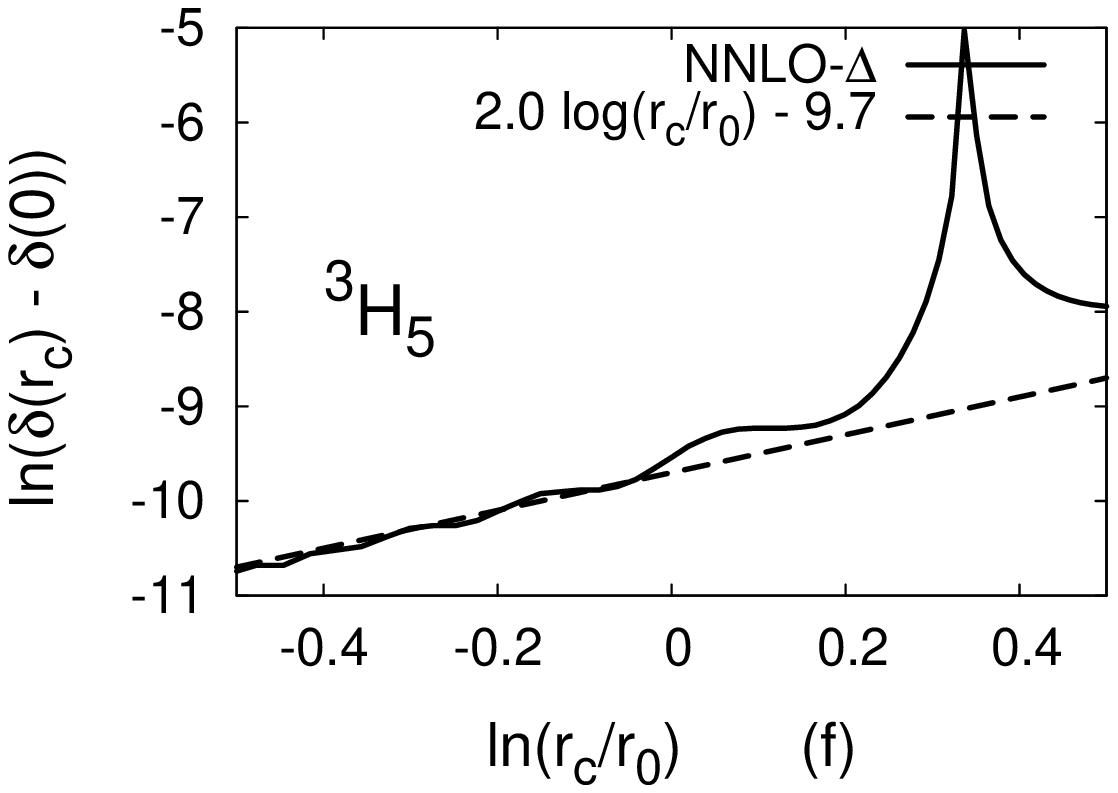, 
	height=5.0cm, width=5.5cm}
\end{center}
\caption{(Color online) Dependence of the phase shifts with
respect to the cut-off radius  in the $^1S_0-{}^1D_2-{}^1G_4$ (left panel)
and $^3P_1-{}^3F_3-{}^3H_5$ waves (right panel) at ${\rm N^2LO}$-$\Delta$
for $E_{\rm LAB} = 200\,{\rm MeV}$.
In the x-axis we plot the natural logarithm of the ratio of the cut-off radius
with respect to a scaling radius of $r_0 = 0.5\,{\rm fm}$,
while the y-axis displays the natural logarithm of the difference
between the phase shift at the cut-off radius $r_c$ and
the phase shifts in the $r_c \to 0$ limit.
The value of the phase shift in this limit is deduced by extrapolating
from the scaling described in Eqs.~(\ref{eq:cutoff-base})
and (\ref{eq:cutoff-derived}).
The figures also show a power law approximation to the cut-off
dependence of the phase shifts, which confirm the expected scaling given
by Eqs.~(\ref{eq:plot-dependence-base}) and (\ref{eq:plot-dependence-derived}).
The approximation confirms the dominance of the chiral van der Waals component
($\sim 1/r^6$) of the interaction for cut-off radii below
$0.8-0.9\,{\rm fm}$ ($\ln{(r_c/r_0)} \simeq 0.5$)
for the lower partial waves ($l \leq 2$) and
$0.5\,{\rm fm}$ ($\ln{(r_c/r_0)} \simeq 0.0$)
for the higher partial waves ($l \geq 2$).
}
\label{fig:Lepage-plot-NNLO-DD}
\end{figure*}

Finally, in Fig.~(\ref{fig:Lepage-plot-NNLO-DD})
we consider the cut-off dependence of the phase shifts
in the form of a logarithmic (or Lepage's~\cite{Lepage:1997cs}) plot.
For simplicity, we only consider two sets of correlated partial waves in detail,
${}^1 S_0$-${}^1 D_2$-${}^1G_4$ and ${}^3 P_1$-${}^3 F_3$-${}^3 H_5$,
a singlet and a triplet.
For the other partial waves the cut-off dependence follows a similar pattern. 
According to Sect.~\ref{subsec:cutoff-dependence},
for small enough cut-offs ($2 m_{\pi} r_c \ll 1$)
the convergence of the phase shift in the lower partial wave
of the correlation (i.e. ${{}^1S_0}$/${}^3P_1$ in this case) is given by
\begin{eqnarray}
\label{eq:plot-dependence-base}
\log{\left| \delta_{A}(k,r_c) - \delta_{A}(k,0) \right|}
\simeq 4\,\log{r_c} + C_{A} + f_{A}(r_c) \, , \nonumber \\
\end{eqnarray}
where $A = {{}^1S_0} ({}^3P_1)$, $\delta_{A} (k, r_c)$ is the phase shift
computed at the cut-off radius $r_c$, $\delta_{A} (k, 0)$ the phase shift
in the $r_c \to 0$ limit, $C_{A}$ a constant,
and $f_{A}(x)$ a small oscillatory contribution which takes into account
the sine factor of the reduced wave function at short distances,
see Eq.~(\ref{eq:uk-uv}).
As can be seen in Fig.~(\ref{fig:Lepage-plot-NNLO-DD}),
this behaviour is indeed fulfilled up to $r_c \sim 0.8-0.9\,{\rm fm}$.
In particular, the numerical factors multiplying the logarithms
in the fits of Fig.~(\ref{fig:Lepage-plot-NNLO-DD})
are very close to $4$, indicating that the van der Waals contribution
to the chiral potential dominates the behaviour of the wave functions
at short distances.
For the higher partial waves in the correlation, the expected scaling is
\begin{eqnarray}
\label{eq:plot-dependence-derived}
\log{\left| \delta_{B}(k,r_c) - \delta_{B}(k,0) \right|}
\simeq 2\,\log{r_c} + C_{B} + f_B(r_c) \, , \nonumber \\
\end{eqnarray}
with $B = {}^1D_2 / {}^1G_4 ({}^3F_3/{}^3H_5)$.
In these waves the van der Waals dominance is apparent
for cut-off radii below $r_c \simeq 0.5-0.8\,{\rm fm}$,
with the lower bound corresponding to the most peripheral partial waves.
It should be noted however that the appearance of van der Waals scaling
in the renormalization group (RG) flow of the phase shifts
for the higher partial waves does not imply
that the phase shifts themselves are dominated by the $1/r^6$ piece
of the interaction.
The region in which the RG flow is driven by the chiral van der Waals force
only amounts for a tiny contribution to the total phase shifts
of the peripheral waves, as can be deduced from the large negative values
of $\log{\left| \delta(k,r_c) - \delta(k,0) \right|}$ 
in the case of the $^1G_4$ and $^3H_5$ waves,
see Fig.~(\ref{fig:Lepage-plot-NNLO-DD}).
This feature fully agrees with the expectations of the renormalization
approach.

\section{Conclusions}
\label{sec:concl}

In the present paper we have considered the relation between
the renormalization of attractive singular potentials
and the partial wave expansion.
Given that attractive singular interactions can be renormalized by
including one counterterm per partial wave, each counterterm
stabilizes the cut-off dependence in each one of the channels
separately. While this is a sufficient condition for renormalizability
it is actually not necessary.
We have shown that if the finite range (attractive singular) interaction
is central, then it can be renormalized by means of a single
delta-shell central potential in coordinate space,
in contrast with the previous situation
in which the predictive power is lost as there are
an infinite number of partial waves.
Of course, this result depends on the assumption that the unknown short
range potential which is represented by a single delta-shell
counterterm is central. Phenomenological potentials do depend on the
orbital angular momentum at short
distances~\cite{Stoks:1993tb,Stoks:1994wp,Wiringa:1994wb,Machleidt:2000ge}.
For this more general situation in which nothing can be assumed about
the short range interaction, the usual result of one counterterm per
channel will be recovered.

Our analysis has been carried out in coordinate space, which on the other hand
has been proven to be equivalent to momentum space
calculations~\cite{Entem:2007jg,Valderrama:2007ja}.
The particularly interesting issue of extending the correlated
renormalization method to momentum space, not addressed
in the present work, is left for future research.
A possible clue might be provided by the observation that the high momentum
behaviour of the chiral potentials ought to reflect the partial wave
independence observed and exploited in the present paper
at short distances, suggesting a common subtraction
perhaps along the lines of Refs.~\cite{Yang:2007hb,Yang:2009kx,Yang:2009pn}.

We have extended the previous result to the case of a finite range potential
containing a tensor piece, which is of great interest for the renormalization
of nuclear forces in the effective field theory approach.
In that case, the number of counterterms depends on the sign of
the eigenvalues of the coupled channel potential.
The application to the chiral NN potentials with $\Delta$ excitations
is possible and straightforward, and only requires to take into account
the additional spin and isospin structure of the NN interaction. We
stress that this is based on taking a counterterm structure based on
the longer range OPE and TPE components of the interaction.
For the order $Q^2$ and $Q^3$ $\Delta$-potentials of Ref.~\cite{Krebs:2007rh}
a total of eleven counterterm is found to be needed to completely
renormalize the interaction in all channels.
This is only two more counterterms than what Weinberg's dimensional
power counting dictates for the contact range interaction
at the considered orders.

\begin{acknowledgments}

We thank Alvaro Calle Cord\'on for discussions, Evgeny Epelbaum for a
critical and careful reading of the manuscript,
and Ulf-G. Mei{\ss}ner for correcting some references.
M.P.V. is supported 
by the Helmholtz Association fund provided to the young investigator
group ``Few-Nucleon Systems in Chiral Effective Field Theory'' (grant
VH-NG-222) and the virtual institute ``Spin and strong QCD'' (VH-VI-231).
The work of E.R.A.  is supported in part by funds provided by the
Spanish DGI and FEDER funds with grant no. FIS2008-01143/FIS, and the
Junta de Andaluc{\'\i}a grant no. FQM225-05.  M.P.V. and E.R.A. are
supported by the EU HadronPhysics2 Project.

\end{acknowledgments}

\appendix
\section{Partial Wave Correlations with the One Pion Exchange Potential}
\label{app:OPE}

In this appendix we review the correlated renormalization for the one pion
exchange potential case, which corresponds to the leading order piece of
the chiral potential.
The OPE potential can be decomposed as
\begin{eqnarray}
V_{\rm OPE}(\vec{r}) = \sigma\,\tau\,W_S(r) + S_{12}(\hat{r})\,\tau\,W_T(r)
\, , 
\end{eqnarray}
where the operators $\sigma$, $\tau$ and $S_{12}$ were defined
in Eq.~(\ref{eq:S12-def}) and $W_S(r)$ and $W_T(r)$ are given by
\begin{eqnarray}
W_S(r) &=& \frac{m_{\pi}^2 g_A^2}{48 \pi f_{\pi}^2}\,
\frac{e^{-m_{\pi} r}}{r} \, ,\\
W_T(r) &=& \frac{m_{\pi}^2 g_A^2}{48 \pi f_{\pi}^2}\,
\left( 1 + \frac{3}{m_{\pi} r} + \frac{3}{(m_{\pi} r)^2}\right)\,
\frac{e^{-m_{\pi} r}}{r} 
\, .
\end{eqnarray}
As can be seen, the only singular component of the OPE potential is
the tensor piece.
Therefore partial wave correlations only arise
between attractive triplet partial waves.
Specifically, there are three sets of correlated waves:
(i) the $^3C_1-{}^3C_3-{}^3C_5$ case, which happens between coupled waves,
(ii) the $^3P_0-{}^3C_2-{}^3C_3$ case, in which there is one uncoupled wave
(the $^3P_0$) and the rest are coupled,
and (iii) $^3D_2-{}^3G_4$ in which all waves are uncoupled triplets.
All the coupled waves are of the attractive-repulsive type, and
in total only three counterterms are needed in order to obtain finite
scattering amplitudes for the OPE potential.
Nonetheless it should be noted that in usual EFT computations
a fourth counterterm will be added to renormalize the $^1S_0$ wave.
In any case, we will only consider those waves which can be related.

\begin{figure*}[htb]
\begin{center}
\epsfig{figure=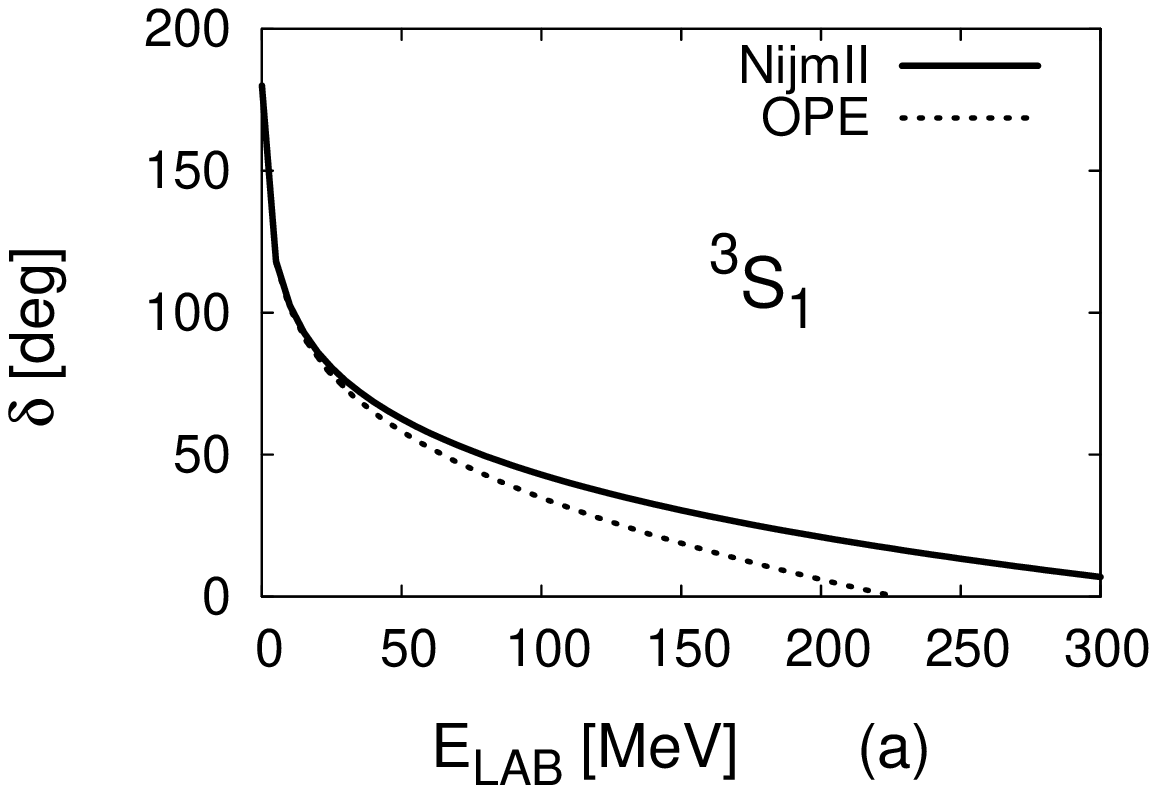, 
	height=5.0cm, width=5.5cm}
\epsfig{figure=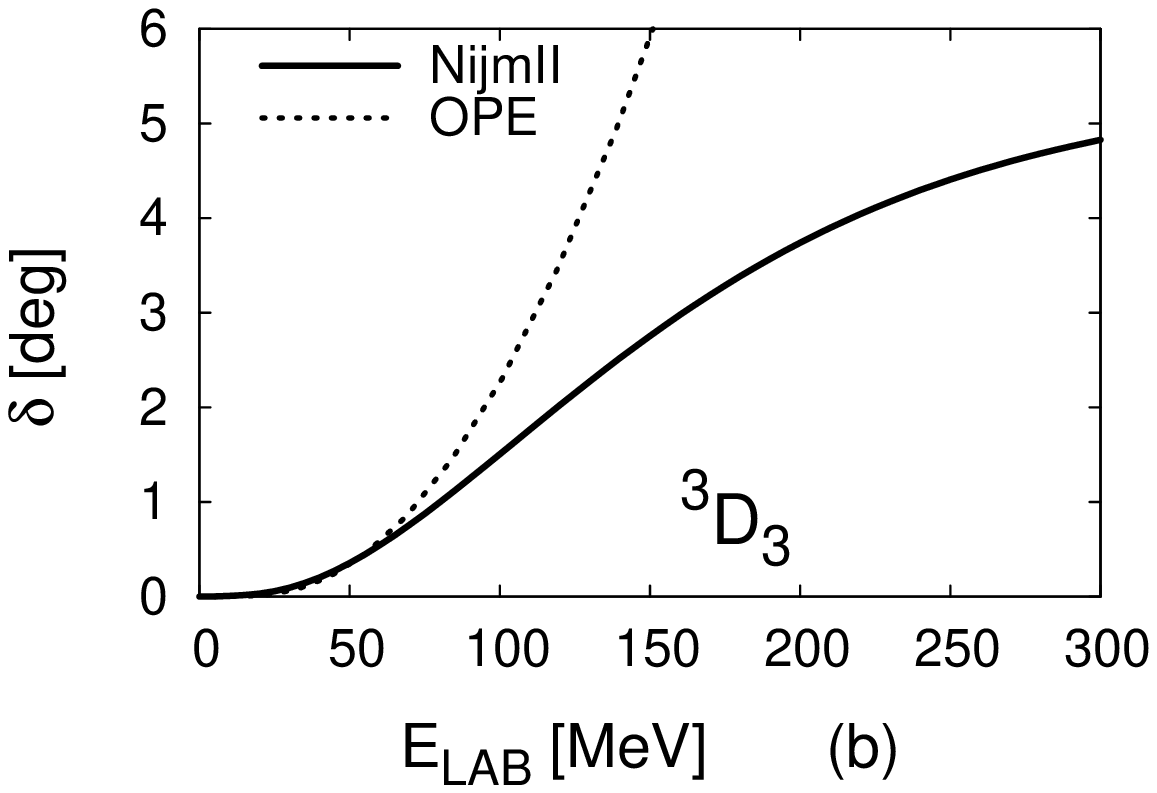, 
	height=5.0cm, width=5.5cm}
\epsfig{figure=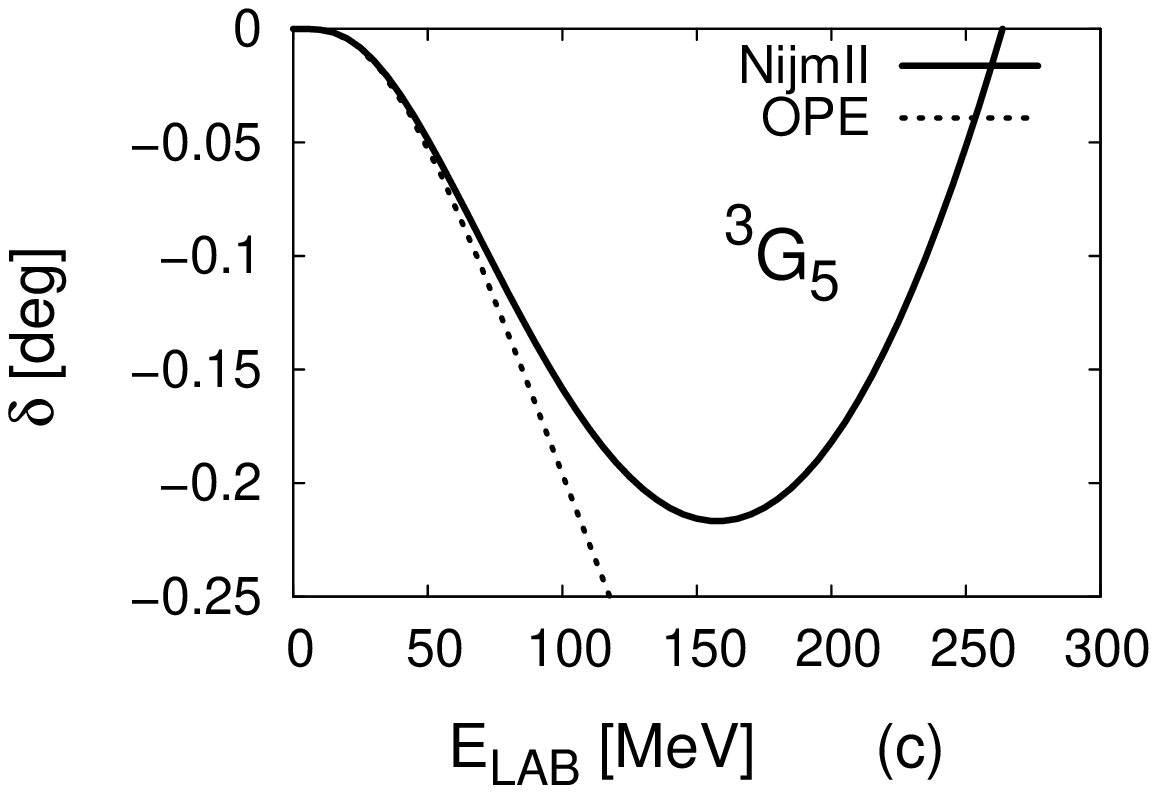, 
	height=5.0cm, width=5.5cm}
\epsfig{figure=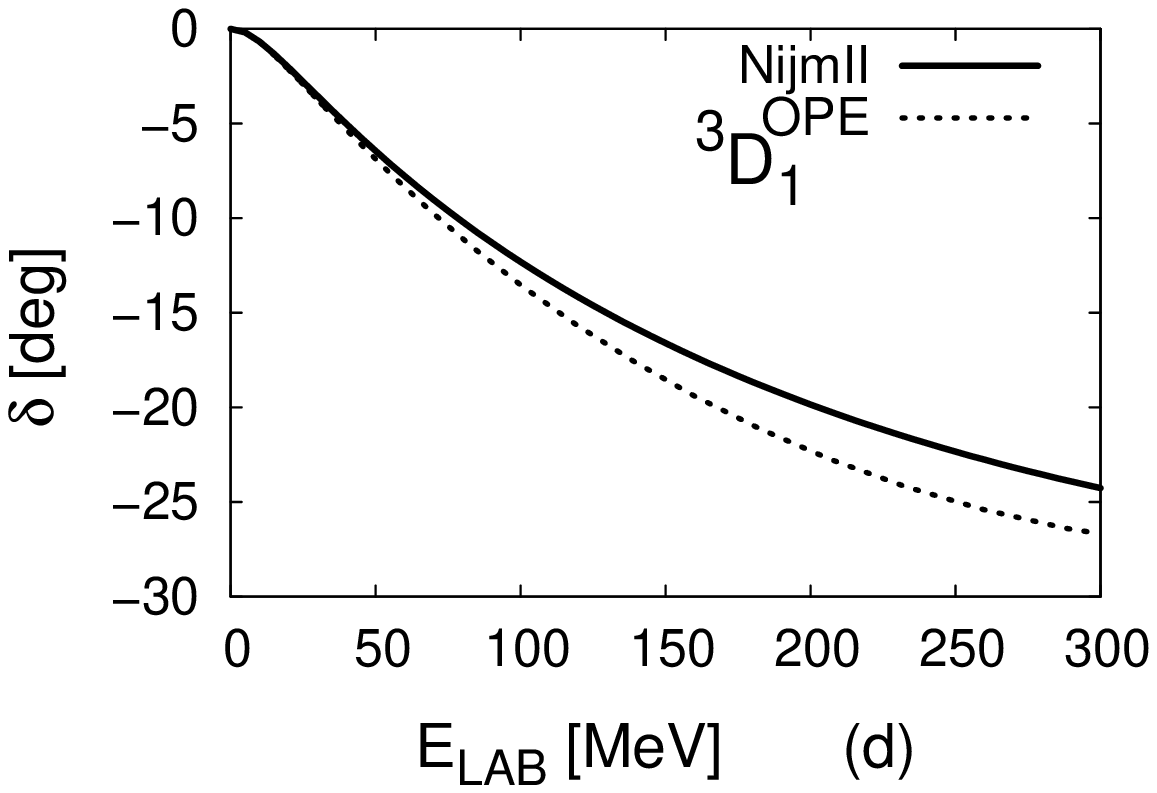, 
	height=5.0cm, width=5.5cm}
\epsfig{figure=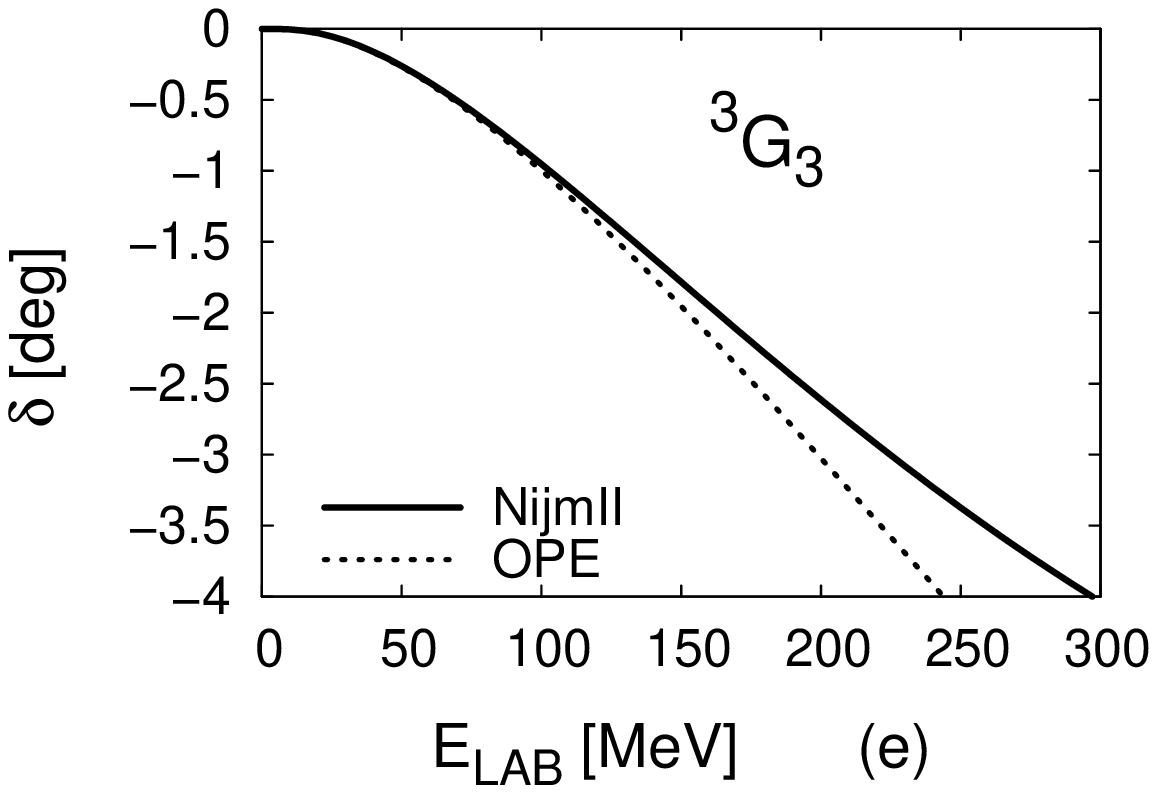, 
	height=5.0cm, width=5.5cm}
\epsfig{figure=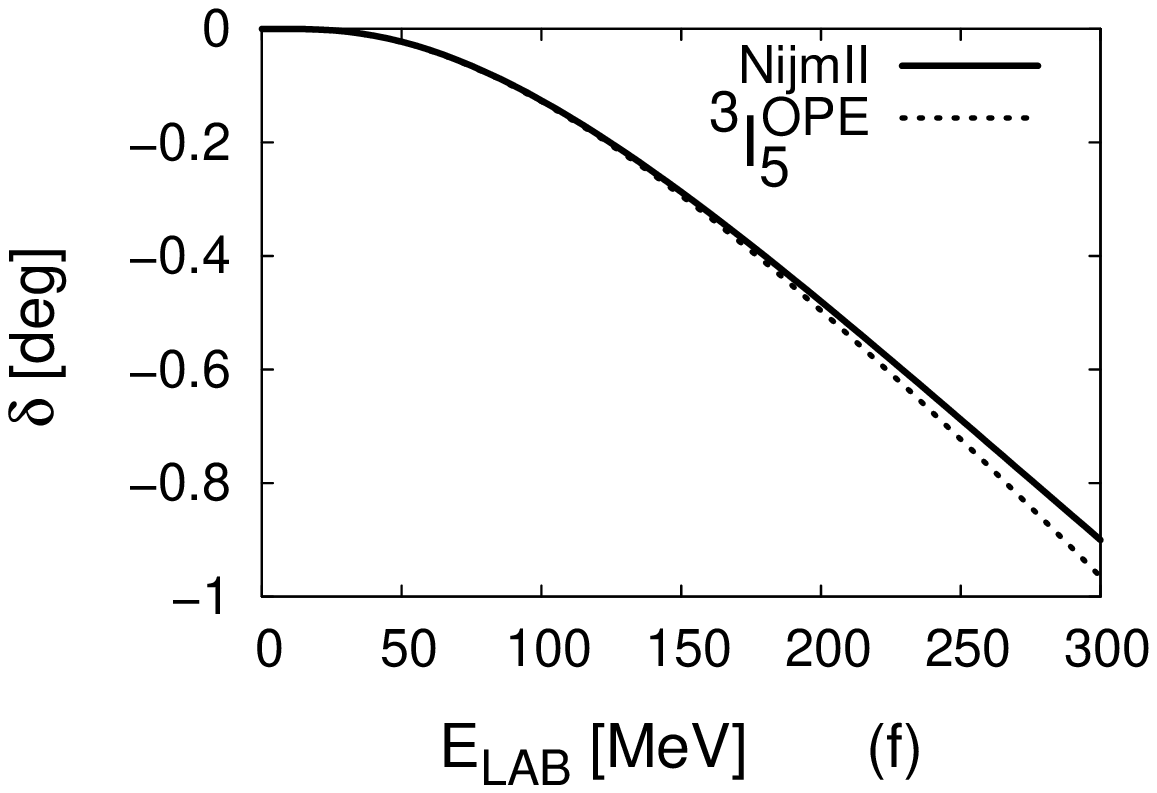, 
	height=5.0cm, width=5.5cm}
\epsfig{figure=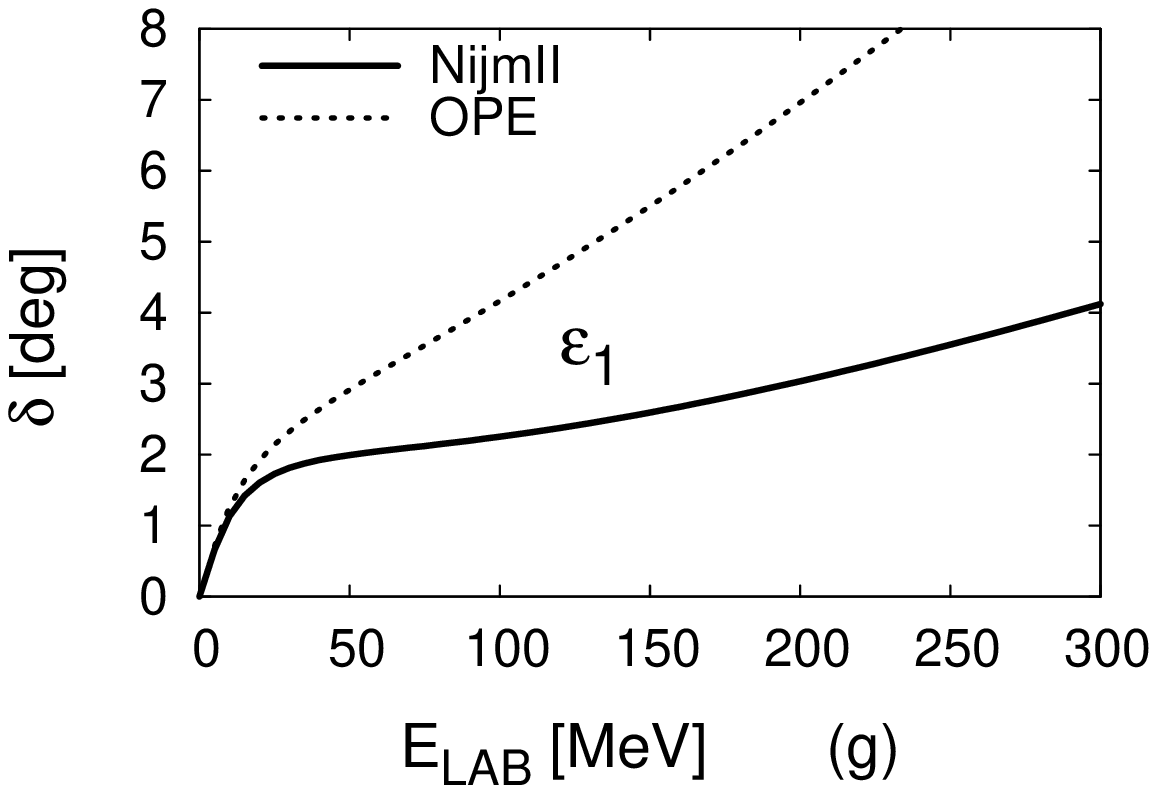, 
	height=5.0cm, width=5.5cm}
\epsfig{figure=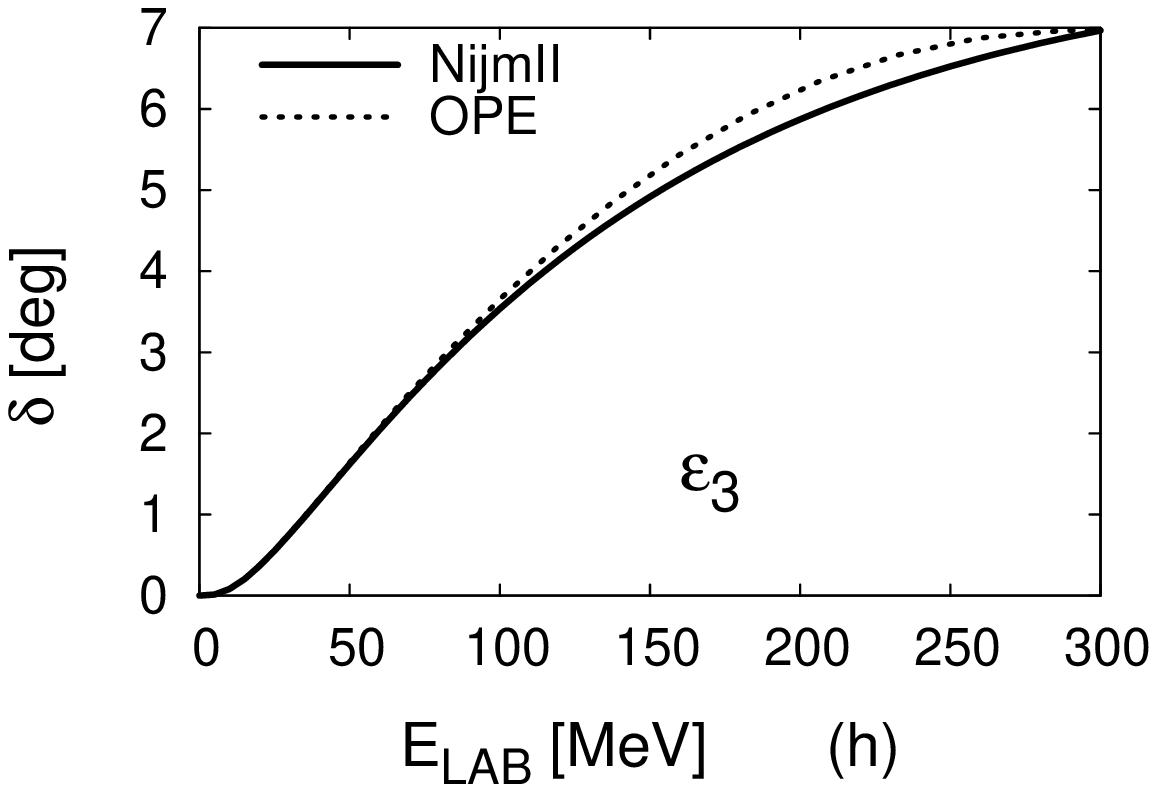, 
	height=5.0cm, width=5.5cm}
\epsfig{figure=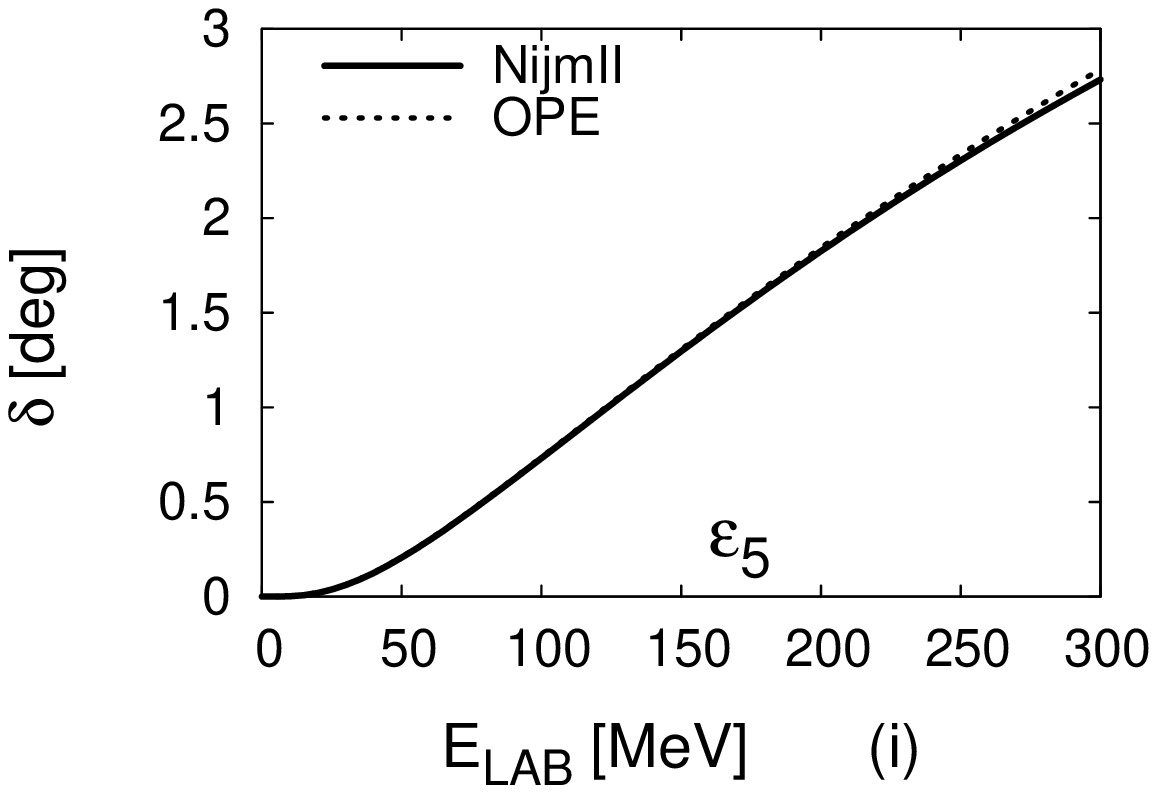, 
	height=5.0cm, width=5.5cm}
\end{center}
\caption{$^3S_1-{}^3D_1$, $^3D_3-{}^3G_3$ and $^3G_5-{}^3I_5$ OPE
coupled channel phase shifts.
The $^3S_1-{}^3D_1$ wave is computed from orthogonality to the deuteron
bound state and from the triplet scattering length
$a_{^3S_1} = 5.419\,{\rm fm}$.
The $^3D_3-{}^3G_3$ and $^3G_5-{}^3I_5$ coupled channels are
computed from the partial wave correlation given by
Eqs.~(\ref{eq:AR-corr-1}-\ref{eq:AR-corr-4})
{\it without} introducing new counterterms.
The coordinate space cut-off is taken to be $r_c = 0.15\,{\rm fm}$.
}
\label{fig:coupled-triplet-deuteron-LO}
\end{figure*}

\begin{figure*}[htb]
\begin{center}
\epsfig{figure=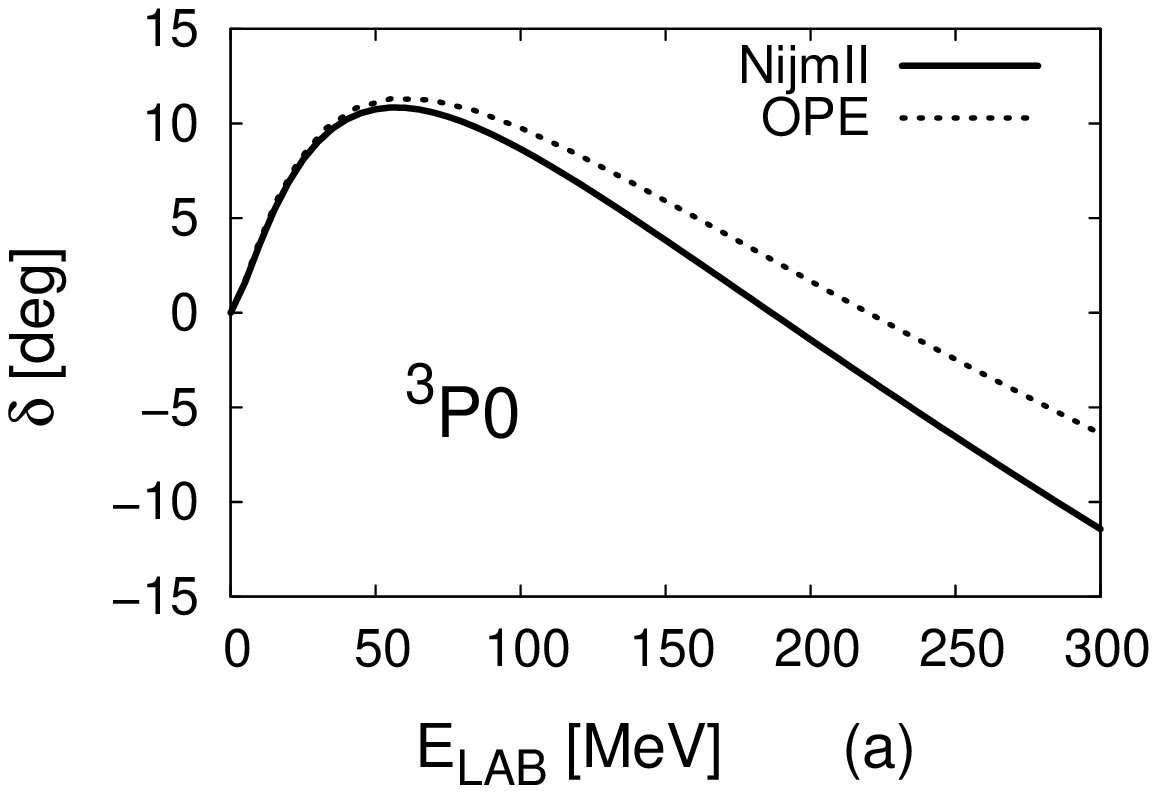, 
	height=5.0cm, width=5.5cm}
\epsfig{figure=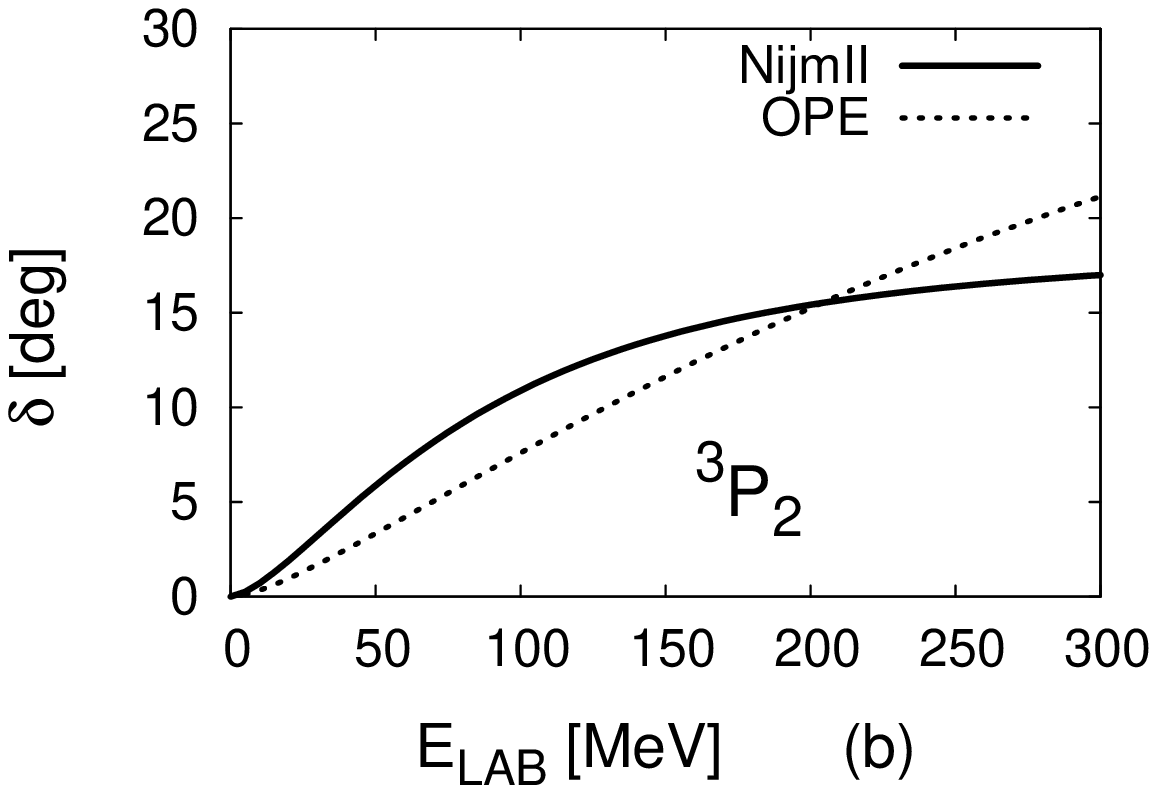, 
	height=5.0cm, width=5.5cm}
\epsfig{figure=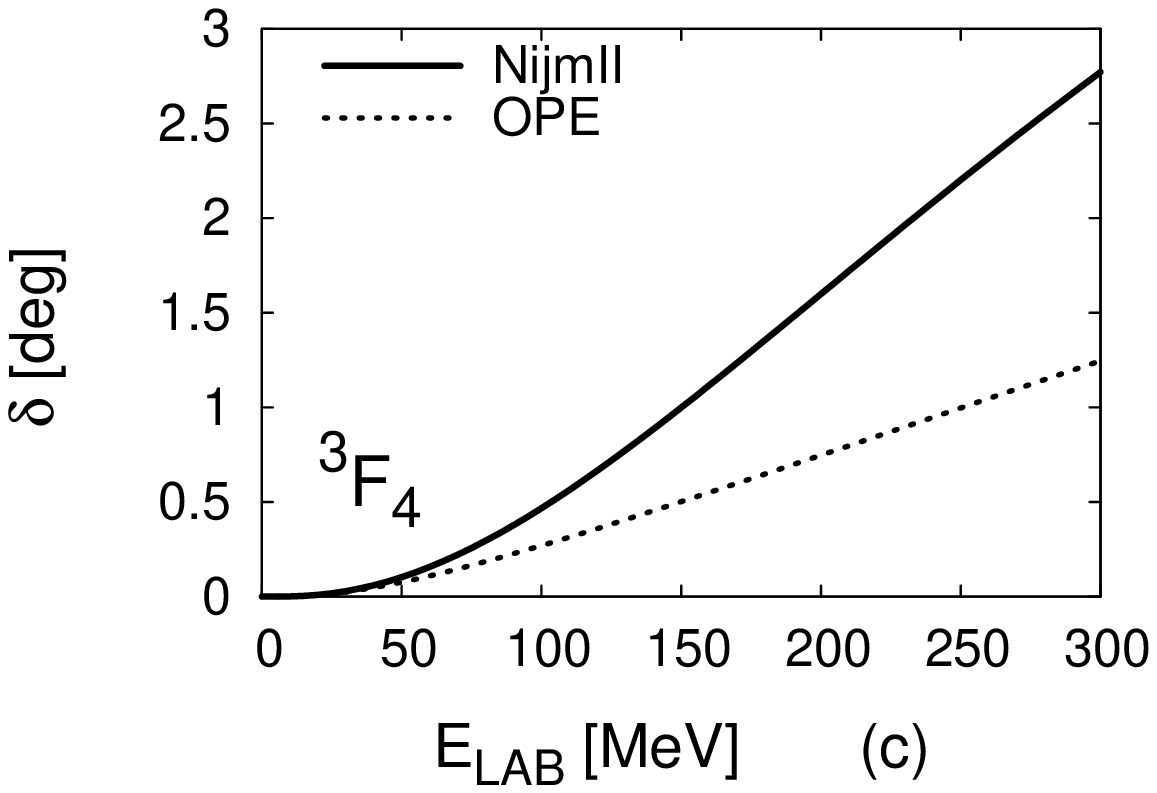, 
	height=5.0cm, width=5.5cm} \\
\epsfig{figure=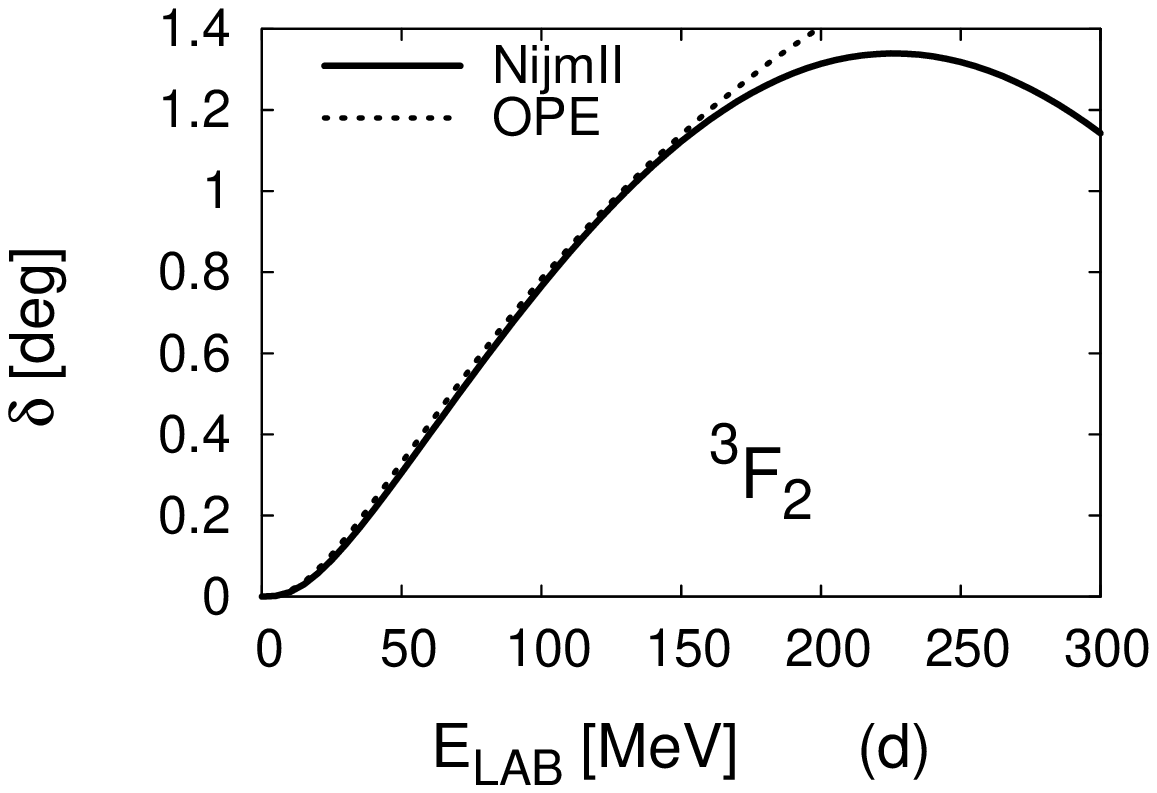, 
	height=5.0cm, width=5.5cm}
\epsfig{figure=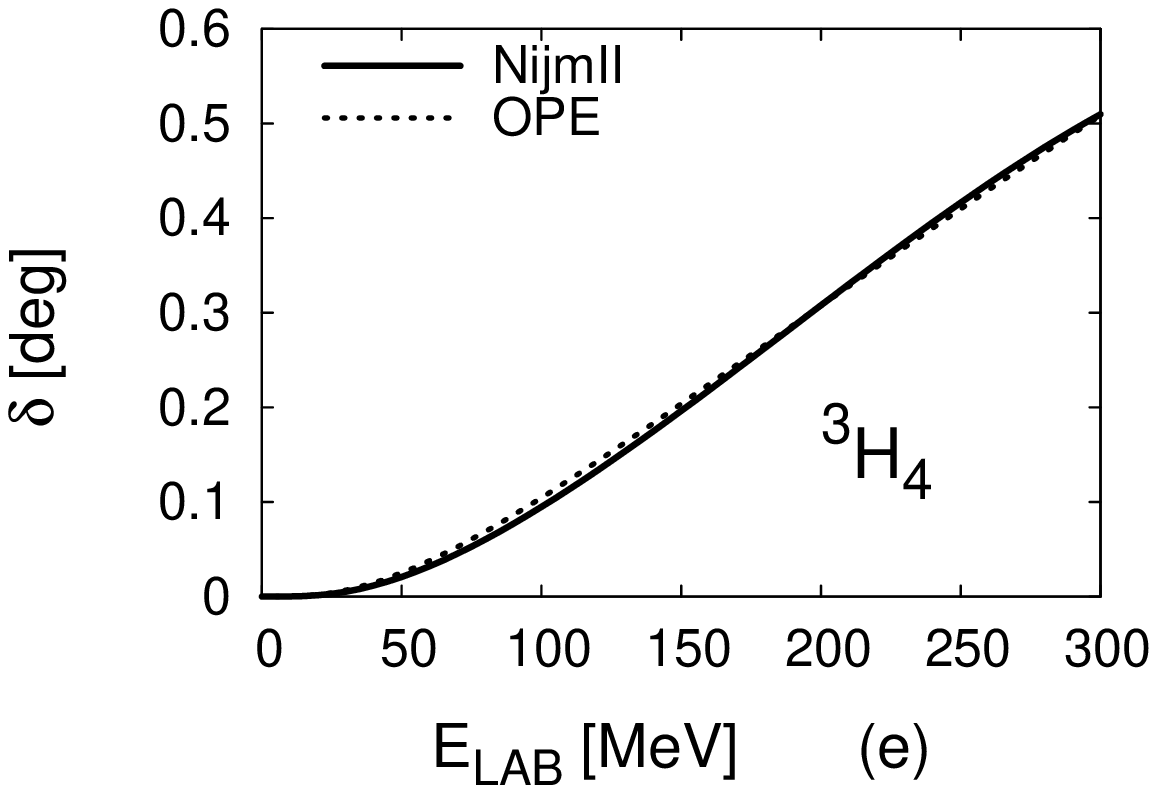, 
	height=5.0cm, width=5.5cm} \\
\epsfig{figure=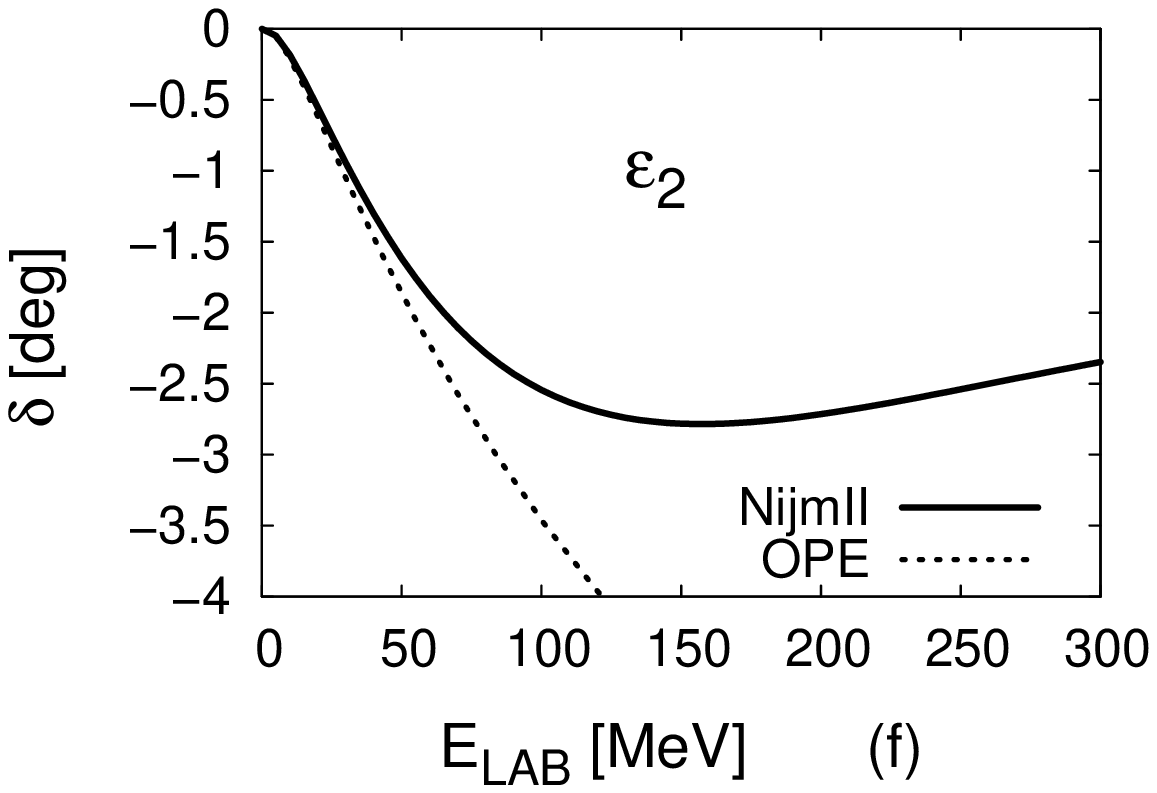, 
	height=5.0cm, width=5.5cm}
\epsfig{figure=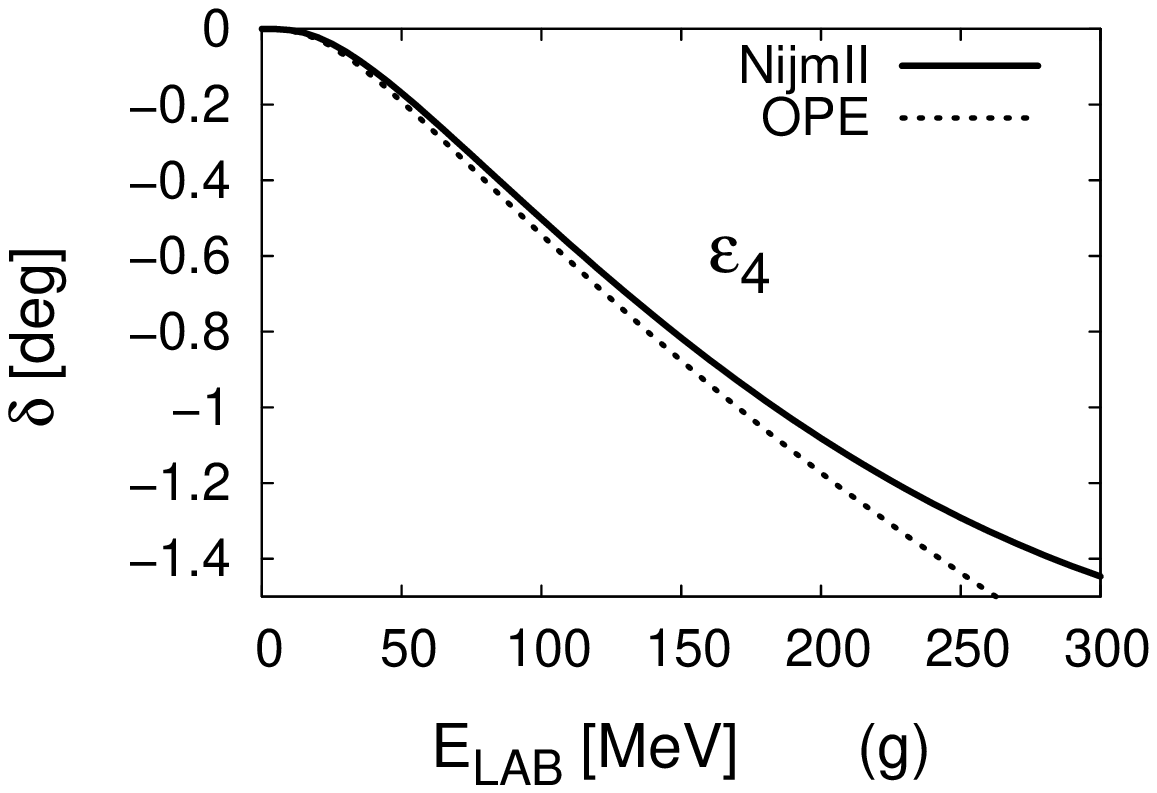, 
	height=5.0cm, width=5.5cm}
\end{center}
\caption{
$^3P_0$, $^3P_2-{}^3F_2$ and $^3F_4-{}^3H_4$ OPE phase shifts.
The $^3P_0$ wave is constructed by fixing the scattering length to the
value $a_{^3P_0} = -2.71\,{\rm fm}^3$,
while the $^3P_2-{}^3F_2$ and $^3F_4-{}^3H_4$ wave are obtained
from the partial wave correlation described in 
Eqs.~(\ref{eq:AR-corr-1}-\ref{eq:AR-corr-4})
{\it without} introducing any new counterterm.
We take the coordinate space cut-off $r_c = 0.15\,{\rm fm}$.
}
\label{fig:coupled-triplet-LO}
\end{figure*}

\begin{figure*}[htb]
\begin{center}
\epsfig{figure=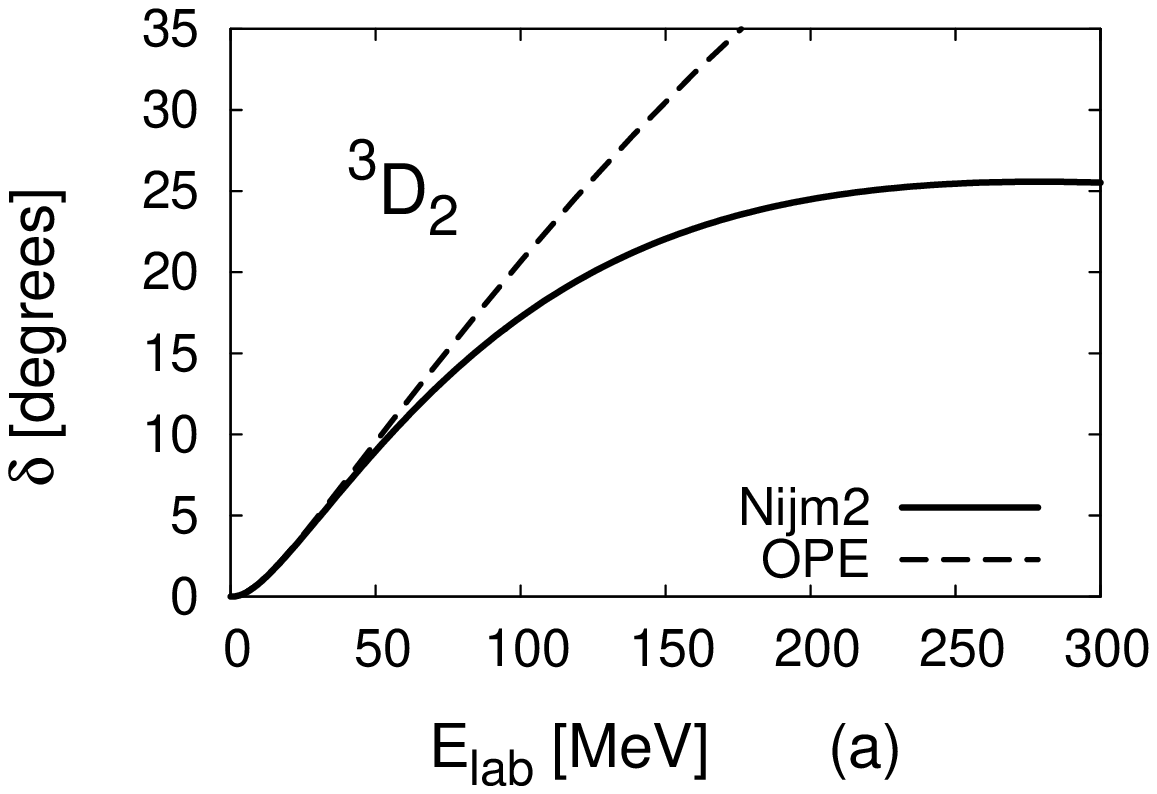, 
 	height=5.0cm, width=5.5cm}
\epsfig{figure=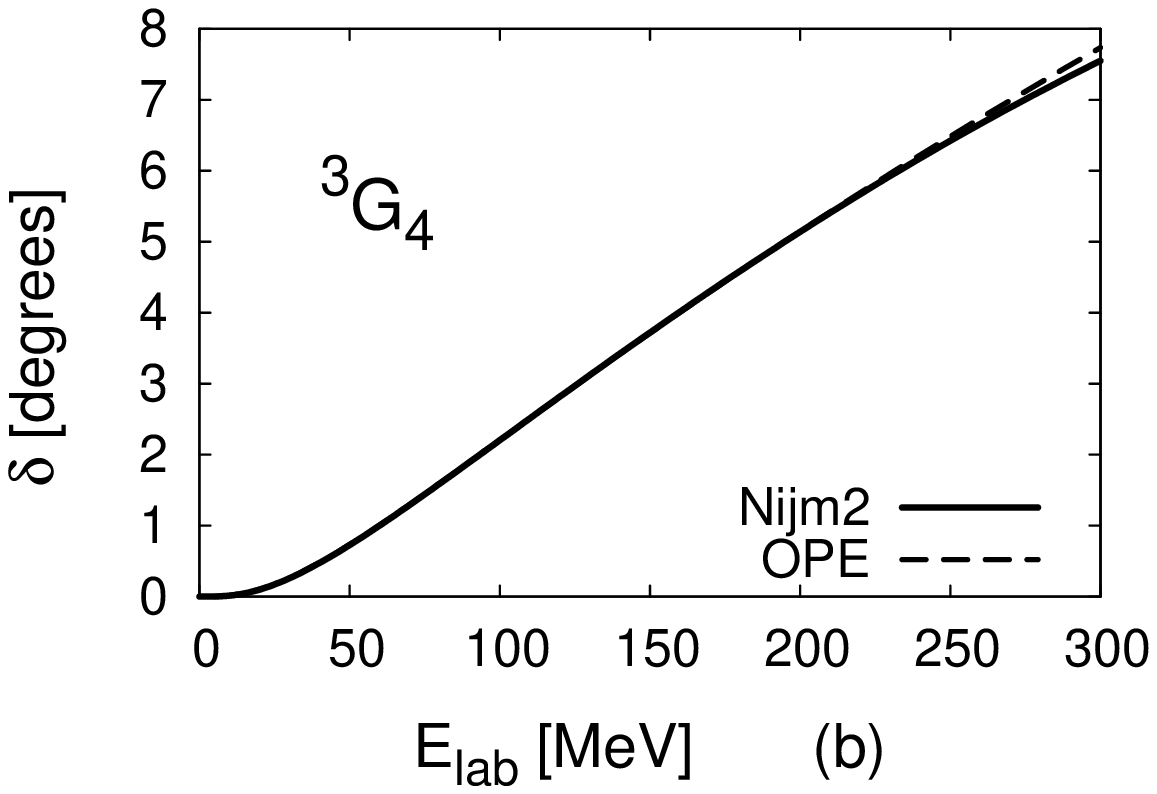, 
	height=5.0cm, width=5.5cm}
\end{center}
\caption{$^3D_2$ and $^3G_4$ phase shifts computed from
Eq.~(\ref{eq:3D2-3G4-relation}),
using the $^3D_2$ scattering length as an input parameter and the OPE
potential with a coordinate space cut-off $r_c = 0.15\,{\rm fm}$.
}
\label{fig:uncoupled-triplet-LO}
\end{figure*}

Contrary to the ${\rm NLO}-\Delta$ and ${\rm N^2LO}-\Delta$ cases,
the cut-off radius must be quite small in order
for the partial wave correlations to converge
(specially between the $^3P_0$ and $^3P_2-{}^3F_2$ waves).
In particular we take $r_c = 0.15\,{\rm fm}$.

The $^3C_1$, $^3C_3$ and $^3C_5$ correlation is shown
in Fig.~(\ref{fig:coupled-triplet-deuteron-LO}).
For this case, the $^3S_1-{}^3D_1$ wave function is renormalized by reproducing
the triplet $^3S_1$ scattering length, $a_{^3S_1} = 5.419\,{\rm fm}$,
a procedure which was described in detail in 
Ref.~\cite{PavonValderrama:2005gu}.
The other partial waves are generated by the renormalization conditions
given in Eqs.~(\ref{eq:AR-corr-1}-\ref{eq:AR-corr-4}).
As can be seen, the description of the $E_1$ and $^3D_3$ wave
is not especially good; these waves improve noticeably with
the inclusion of two-pion exchange and the $\Delta$.
The remaining $j=3$ and $j=5$ phases do not differ two much from their
${\rm NLO}$-$\Delta$ and ${\rm N^2LO}$-$\Delta$ counterparts,
as expected from the fact that peripheral waves are OPE dominated.

In Fig.~(\ref{fig:coupled-triplet-LO}) 
we show the resulting $^3P_0$, $^3P_2-{}^3F_2$ and $^3F_4-{}^3H_4$
phase shifts, which have been obtained by using the $^3P_0$ wave 
as the base wave, and where the $^3P_0$ scattering length
has been taken to be $a_{^3P_0} = -2.71\,{\rm fm}^3$.
We can see that the $^3F_2$, $E_2$ and $^3F_4$ waves are not well reproduced
with OPE alone and need the inclusion of the higher orders of the potential.

Finally, in Fig.~(\ref{fig:uncoupled-triplet-LO})
the $^3D_2$ and $^3G_4$ phase shifts are shown.
The $^3D_2$ phase has been renormalized to reproduce the Nijmegen II
value of the scattering length, $a_{^3D_2} = -7.405\,{\rm fm}^5$.
The OPE results for the $^3D_2$ are worse than those of
${\rm NLO}$-$\Delta$ and ${\rm N^2LO}$-$\Delta$ at
moderate energies of the order of $E_{\rm LAB} > 150\,{\rm MeV}$.
The $^3G_4$ phase is nicely reproduced with OPE alone.


%

\end{document}